	\documentclass[final]{siamltex}
	\usepackage[center]{subfigure}
	\usepackage{subfigure}
	\usepackage{epsfig}
	\usepackage{epstopdf}
	\usepackage{amsmath}
	\usepackage{amssymb}
	\usepackage{mathrsfs}
	\usepackage{mathtools}
	\usepackage{graphicx}
	\usepackage[usenames,dvipsnames]{xcolor}
	\usepackage{epsfig}
	\usepackage{epstopdf}	
	\usepackage[color]{showkeys}
	\usepackage{slantsc}
	\usepackage{cite}
	\usepackage{url}
	\usepackage{tikz}
	
	\usetikzlibrary{calc, decorations.markings, shapes.geometric, arrows, shapes}
	
	\newcommand{\p}[1]{\boldsymbol{#1}}

	\newcommand{\argmax}{\arg\!\max}
		
	\title{Community detection \\ in temporal multilayer networks, \\ with an application to correlation networks\thanks{This work was supported by a CASE studentship award from the EPSRC (BK/10/41), HSBC Bank, and the FET-Proactive project PLEXMATH (FP7-ICT-2011-8; grant 317614) funded by the European Commission.}} 
		
	\author{Marya Bazzi\footnotemark[2]
	\and Mason A. Porter\footnotemark[3]
	\and Stacy Williams\footnotemark[4]
	\and Mark McDonald\footnotemark[4]
	\and Daniel J. Fenn\footnotemark[4]
	\and Sam D. Howison\footnotemark[5]
	}
	
\begin{document}
	
	\maketitle
	
	\renewcommand{\thefootnote}{\fnsymbol{footnote}}
	\footnotetext[2]{Oxford Centre for Industrial and Applied Mathematics, Mathematical Institute, Oxford OX2 6GG, UK (bazzi@maths.ox.ac.uk).}
	\footnotetext[3]{Oxford Centre for Industrial and Applied Mathematics, Mathematical Institute, Oxford OX2 6GG, UK; and CABDyN Complexity Centre, University of Oxford, Oxford OX1 1HP, UK (porterm@maths.ox.ac.uk).}
	\footnotetext[4]{Global Research, HSBC Bank, London E14 5HQ, UK.}
	\footnotetext[5]{Oxford Centre for Industrial and Applied Mathematics, Mathematical Institute, Oxford OX2 6GG, UK; and Oxford-Man Institute of Quantitative Finance, University of Oxford, Oxford OX2 6ED, UK (howison@maths.ox.ac.uk).}
	\renewcommand{\thefootnote}{\arabic{footnote}}
	

		\begin{abstract} 
	
		Networks are a convenient way to represent complex systems of interacting entities. Many networks contain ``communities'' of nodes that are more densely connected to each other than to nodes in the rest of the network. In this paper, we investigate the detection of communities in temporal networks represented as multilayer networks. As a focal example, we study time-dependent financial-asset correlation networks. We first argue that the use of the ``modularity'' quality function---which is defined by comparing edge weights in an observed network to expected edge weights in a ``null network''---is application-dependent. We differentiate between ``null networks'' and ``null models'' in our discussion of modularity maximization, and we highlight that the same null network can correspond to different null models. We then investigate a multilayer modularity-maximization problem to identify communities in temporal networks. Our multilayer analysis only depends on the form of the maximization problem and not on the specific quality function that one chooses. We introduce a diagnostic to measure \emph{persistence} of community structure in a multilayer network partition. We prove several results that describe how the multilayer maximization problem measures a trade-off between static community structure within layers and larger values of persistence across layers. We also discuss some computational issues that the popular ``Louvain'' heuristic faces with temporal multilayer networks and suggest ways to mitigate them. 
		
		\end{abstract}
		
		
		
		
		\pagestyle{myheadings}
		\thispagestyle{plain}
		


		\section{Introduction}\label{Section1}
		
		In its simplest form, a network is simply a graph: it consists of a set of \emph{nodes} that represent entities and a set of \emph{edges} between pairs of nodes that represent interactions between those entities. One can consider \emph{weighted graphs} (in which each edge has an associated edge weight that quantifies the interaction of interest) or \emph{unweighted graphs} (weighted graphs with binary edge weights). Networks provide useful representations of complex systems across many disciplines \cite{Newman2010}. Common types include social networks (which arise via offline and/or online interactions), information networks (e.g., hyperlinks between webpages in the World Wide Web), infrastructure networks (e.g., transportation routes between cities), and biological networks (e.g., metabolic interactions between cells or proteins, food webs, etc.). 
		
		Given a network representation of a system, it can be useful to apply a coarse-graining technique in order to investigate features that lie between features at the ``microscale'' (e.g., nodes and pairwise interactions) and the ``macroscale'' (e.g., total edge weight and degree distribution\footnote{A node's ``degree'' is the number of edges attached to it; degree is the special case of ``strength'' \eqref{EQ2.1} for unweighted networks.}) \cite{Porter2009,Newman2011}.  One thereby studies ``mesoscale'' features such as core-periphery structure and (especially) community structure. Loosely speaking, a \emph{community} (or \emph{cluster}) in a network is a set of nodes that are ``more densely'' connected to each other than they are to nodes in the rest of the network \cite{Porter2009,Fortunato2010}. Giving a precise definition of ``densely connected'' is, of course, necessary to have a method for community detection. It is important to recognize at the outset that this definition is subjective and in particular, may depend on the application in question. Correspondingly, community detection methods may need to be tailored. We restrict ourselves to \emph{hard partitions}, in which each node is assigned to exactly one community, and we use the term ``partition'' to mean ``hard partition''.  It is also important, but beyond the scope of this paper, to consider ``soft partitions'', in which communities can overlap \cite{Porter2009,Fortunato2010,Palla2007,Jeub2014}.
		
		Analysis of community structure has been very useful in a wide range of applications; many of which are described in \cite{Porter2009, Fortunato2010, Girvan2002, Newman2011}. In social networks, communities can reveal groups of people with common interests, places of residence, or other similarities \cite{Traud2011,Newman2004}. In biological systems, communities can reveal functional groups that are responsible for synthesizing or regulating an important chemical product \cite{Guimera2004,Lewis2010}. In the present paper, we use financial-asset correlation networks as examples\cite{Catanzaro2013,Battiston2010}. Despite the diversity of markets, financial products, and geographical locations, financial assets can exhibit strong time-dependent correlations, both within and between asset classes. It is a primary concern for market practitioners (e.g., for portfolio diversification) to estimate the strengths of these correlations and to identify sets of assets that are highly correlated \cite{Williams2011, Montegna1999}. 
		
			Most methods for detecting communities are designed for static networks. However, in many applications, entities and/or interactions between entities evolve in time. In such applications, one can use the formalism of \emph{temporal networks}, where nodes and/or their edges weights vary in time \cite{Holme2012, Holme2015}. This is important for numerous applications, including person-to-person communication \cite{Wu2010}, one-to-many information dissemination (e.g., Twitter networks \cite{Bailon2011} and Facebook networks \cite{Zhao2010}), cell biology \cite{Holme2012}, neuroscience \cite{Bassett2011}, ecology \cite{Holme2012}, finance\cite{Onnela2003, Fenn2011, Fenn2009, Fenn2012}, and more. 
			
			Two main approaches have been adopted to detect communities in time-dependent networks. The first entails constructing a static network by aggregating snapshots of the evolving network at different points in time into a single network (e.g., by taking the mean or total edge weight for each edge across all time points, which can be problematic if the set of nodes varies in time and which also makes restrictive assumptions on the interaction dynamics between entities \cite{hoffmann2012}). One can then use standard network techniques. The second approach entails using static community-detection techniques on each element of a time-ordered sequence of networks at different times or on each element of a time-ordered sequence of network aggregations\footnote{One needs to distinguish between this kind of aggregation and the averaging of a set of time series over a moving window to construct a correlation matrix, which one can then interpret as a fixed-time snapshot of a time-evolving network. Although both involve averaging over a time window, the former situation entails averaging a network, and the latter situation entails averaging over a collection of time series (one for each node) with no directly observable edge weights.} (computed as above) over different time intervals (which can be either overlapping or nonoverlapping) and then tracking the communities across the sequence \cite{Hopcroft2004,Palla2007,Fenn2009,Aynaud2010,Fenn2012,Macon2012}.
			
			A third approach consists of embedding a time-ordered sequence of networks in a larger network \cite{Domenico2014, Pietri2014} (and related ideas are also available in other contexts \cite{Mucha2010,Stanley2015}). Each element of the sequence is a network \emph{layer}, and nodes at different time points are joined by \emph{inter-layer} edges. This approach was introduced in \cite{Mucha2010} and the resulting network is a type of \emph{multilayer network} \cite{Kivela2014, Boccaletti2014}. The main difference between this approach and the previous approach is that the presence of nonzero inter-layer edges introduces a dependence between communities identified in one layer and connectivity patterns in other layers. Thus far, most computations that have used a multilayer representation of temporal networks have assumed that inter-layer connections are ``diagonal'' (i.e., they exist only between copies of the same node) and ``ordinal'' (i.e., they exist only between consecutive layers) \cite{Kivela2014}. Diagonal is a natural model of the persistence of node identity in time, while ordinal preserves the time ordering.

			The authors of \cite{Mucha2010} derived a generalization of \emph{modularity maximization}, a popular clustering method for static networks, to multilayer networks. Modularity is a function that measures the ``quality'' of a network partition into disjoint sets of nodes by computing the difference between the total edge weight in sets in the observed network and the total expected edge weight in the same sets in a ``null network'' generated from some ``null model''\cite{Porter2009, Fortunato2010}. Modularity maximization consists of maximizing the modularity quality function over the space of network partitions. (In practice, given the combinatorial complexity of this maximization problem, one uses some computational heuristic and finds a local maximum \cite{Good2010}.) Intuitively, the null model controls for connectivity patterns that one anticipates finding in a network, and one uses modularity maximization to identify connectivity patterns in an observed network that are stronger than anticipated. 
			
			In this paper, we address two main issues: (1) the choice of null network and (2) the role of inter-layer edges in multilayer modularity maximization. We discuss the first issue in Section \ref{Section4} and the second issue in Section \ref{Section5}. Most of our conclusions from Sections \ref{Section4} and \ref{Section5} are applicable to an arbitrary choice of single-layer networks within layers, and we use financial correlation networks as illustrative examples. In Sections \ref{Section2} and \ref{Section3}, we give an overview of existing results.

			We give a precise definition of the modularity function for single-layer networks in Section \ref{Section2}, where (importantly) we distinguish between a ``null network'' and a ``null model'' in modularity maximization. In Section \ref{Section4}, we discuss the choice of null network for a given application. In Section \ref{Section3}, we describe the generalization of single-layer modularity to multilayer networks proposed in \cite{Mucha2010}. To date, almost no theory has explained how a multilayer partition obtained with zero inter-layer coupling (which reduces to single-layer modularity maximization on each layer independently) differs from a multilayer partition obtained with nonzero inter-layer coupling. In Section \ref{Section5}, we prove several theoretical properties of an optimal solution for the multilayer maximization problem to better understand how such partitions differ and how one can exploit this difference in practice. We also describe two computational issues that arise when using the popular Louvain heuristic \cite{Blondel2008} to solve the multilayer maximization problem, and we suggest ways to mitigate them. The results of Section \ref{Section5} depend only on the form of the maximization problem and still hold if one uses a quality function other than the modularity quality function, provided it has the same form (e.g., it is valid for ``stability'' \cite{Lambiotte2009, Delvenne2010}). We conclude in Section \ref{Section6}. 
	
		
	\section{Single-layer modularity maximization}\label{Section2}
		
			
		\subsection{The modularity function}\label{Subsection2.1}
		
		Consider an $N$-node network $\mathcal{G}$ and let the edge weights between pairs of nodes be $\left\{A_{ij} | i,j \in \{1, \ldots, N\} \right\}$, so that $\p{A} = (A_{ij}) \in\mathbb{R}^{N\times N}$ is the \emph{adjacency matrix} of $\mathcal{G}$. In this paper, we only consider symmetric adjacency matrices (and hence undirected networks), so $A_{ij} = A_{ji}$ for all $i$ and $j$. The \emph{strength} of a node $i$ is
		\begin{equation}
			k_i = \sum_{j=1}^{N}A_{ij} = \sum_{j=1}^{N} A_{ji}\,,
			\label{EQ2.1}
		\end{equation}
		 and it is given by the $i^{\text{th}}$ row (or column) sum of $\p{A}$.
	
		When studying the structure of a network, it is useful to compare what is observed with what is anticipated. We define a \emph{null model} to be a probability distribution on the set of adjacency matrices and a \emph{null network} to be the expected adjacency matrix under a specified null model. In a loose sense, null models play the role of \emph{prior models}, as they control for features that one anticipates to find in the system under investigation. One can thereby take into account known (or suspected) connectivity patterns that might obscure unknown connectivity patterns that one hopes to discover via processes like community detection. For example, in social networks, one often takes the strength of a node in a null network to be its observed strength $k_i$ \cite{Porter2009, Newman2011,Newman2006}. We discuss the use of this null network for financial-asset correlation networks in Section \ref{Section4}. In spatial networks that represent the spread of a disease or information between different locations, some authors have used null networks in which edge weights between two locations scales inversely with the distance between them \cite{Expert2011,marta2014}.	
		
		As we discussed in Section \ref{Section1}, one uses modularity maximization to partition a network into sets of nodes called ``communities'' that have a larger total internal edge weight than the expected total internal edge weight in the same sets in a null network, generated from some null model \cite{Porter2009, Fortunato2010, Newman2011, Newman2004}. \emph{Modularity maximization} consists of finding a partition that maximizes this difference \cite{Porter2009,Fortunato2010}. (As we mentioned earlier, in practice, one uses some computational heuristic and finds a local maximum \cite{Good2010}). In the present paper, we do not restrict ourselves to the usual choice of null network (i.e., the ``Newman--Girvan'' null network in subsection \ref{ng-sub}) for the modularity quality function, and we ignore any normalization constant that depends on the choice of null network but does not affect the solution of the modularity-maximization problem for a given null network. Modularity thus acts as a ``quality function'' $Q: \mathcal{C} \rightarrow \mathbb{R}$, where the set $\mathcal{C}$ is the set of all possible $N$-node network partitions.
		
		Suppose that we have a partition $C$ of a network into $K$ disjoint sets of nodes $\{C_1, \ldots, C_K\}$. We can then define a map $c(\cdot)$ from the set of nodes $\{1,\ldots,N\}$ to the set of integers $\{1,\ldots,K\}$ such that $c(i) = c(j) = k$ if and only if nodes $i$ and $j$ lie in $C_k$. We use the term \emph{global maximum} to refer to a solution of the modularity-maximization problem and the term \emph{local maximum} to refer to a solution that one obtains with a computational heuristic. We call $c(i)$ the \emph{set assignment} (or \emph{community assignment} when $C$ is a global or local maximum) of node $i$ in partition $C$. The value of modularity for a given partition $C$ is then
		\begin{equation}
			Q(C\vert \p{A};\p{P}) := \sum_{i,j=1}^{N}(A_{ij} - P_{ij})\delta(c_i,c_j)\,,
			\label{EQ2.2}
		\end{equation}
		where $\p{P} = (P_{ij})\in\mathbb{R}^{N\times N}$ is the adjacency matrix of the null network, $c_i$ is shorthand notation for $c(i)$, 
		and $\delta(c_i,c_j)$ is the Kronecker delta function.
		We state the modularity-maximization problem as follows:
		\begin{equation}
			\max_{C\in\mathcal{C}} \sum_{i,j=1}^{N}(A_{ij} - P_{ij})\delta(c_i,c_j)\,,
			\label{EQ2.3}
		\end{equation}
		which we can also write as $\max_{C\in\mathcal{C}}Q(C\vert\p{B})$ or $\max_{C\in\mathcal{C}}\sum_{i,j}^{N}B_{ij}\delta(c_i,c_j)$, where $\p{B} = \p{A} - \p{P}$ is the so-called \emph{modularity matrix} \cite{Newman2006}. The number $K$ of sets in a partition is free in the optimization problem \eqref{EQ2.3}. (In other words, one maximizes over the set of \emph{all} $N$-node partitions.) We only consider a fixed value of $K$ when we consider a particular partition $C\in\mathcal{C}$, with $\vert C\vert = K$. We allow self-edges in all numerical experiments with financial data of Sections \ref{Section4} and \ref{Section5}. (In particular, $\p{A}$ is a Pearson correlation matrix with $A_{ii} = 1$.) We assume in the rest of the paper that each of the partitions in the set $\mathcal{C}$ contains sets that do not have multiple connected components in the graph with adjacency matrix $\p{B}$. It is clear from \eqref{EQ2.3} that pairwise contributions to modularity are only counted when two nodes are assigned to the same set. These contributions are positive (respectively, negative) when the observed edge weight $A_{ij}$ between nodes $i$ and $j$ is larger (respectively, smaller) than the expected edge weight $P_{ij}$ between them. If ${A}_{ij} < P_{ij}$ for all $i$ and $j$, then the optimal solution is $N$ singleton communities. Conversely, if ${A}_{ij} > P_{ij}$ for all $i$ and $j$, then the optimal solution is a single $N$-node community. To obtain a partition of a network with a high value of modularity, one hopes to have many edges within sets that satisfy $A_{ij} > P_{ij}$ and few edges within sets that satisfy $A_{ij} < P_{ij}$. As is evident from equation \eqref{EQ2.3}, what one regards as ``densely connected'' in this setting depends fundamentally on the choice of null network.
		
		It can be useful to write the modularity-maximization problem using the trace of matrices \cite{Newman2006}. As before, we consider a partition $C$ of a network into $K$ sets of nodes $\{C_1,\ldots,C_K\}$. We define the \emph{partition matrix} $\p{S}\in\{0,1\}^{N\times K}$ as
		\begin{equation}
			S_{ij} = \delta(c_i,j)\,,
		\label{EQ2.4}
		\end{equation} 
		where $j\in\{1,\ldots,K\}$ and $c_i = j$ means that node $i$ lies in $ C_j$. The columns of $\p{S}$ are orthogonal, and the $j^{\text{th}}$ column sum of $\p{S}$ gives the number of nodes in $C_j$. This yields
		\begin{equation*}
			\sum_{i,j=1}^{N}B_{ij}\delta(c_i,c_j) = \sum_{i,j=1}^{N}\sum_{k=1}^{K}S_{ik}B_{ij}S_{jk} = \text{Tr}(\p{S}^{T}\p{B}\p{S})\,,
		\end{equation*}
		where the $(i,i)^{\text{th}}$ term of $\p{S}^{T}\p{B}\p{S}$ is twice the sum of modularity-matrix entries between pairs of nodes in $C_i$. (The $(i,j)^{\text{th}}$ off-diagonal term is the sum of modularity-matrix entries between pairs of nodes with one node in $C_i$ and one node in $C_j$.) It follows that one can restate the modularity-maximization problem in \eqref{EQ2.3} as 
		\begin{equation}
			\max_{S\in\mathcal{S}} \text{Tr}(\p{S}^{T}\p{B}\p{S})\,,
		\label{EQ2.5}
		\end{equation}
		where $\mathcal{S}$ is the set of all partition matrices in $\{0,1\}^{N\times K}$ (with $K \leq N$). 
		
		Modularity maximization is one of myriad community-detection methods \cite{Fortunato2010}, and it has many limitations (e.g., a resolution limit on the size of communities \cite{Fortunato2007} and a huge number of nearly degenerate local maxima \cite{Good2010}). Nevertheless, it is a popular method (which has been used successfully in numerous applications \cite{Fortunato2010, Porter2009}), and the ability to specify explicitly what one anticipates is a useful (and under-exploited) feature for users working on different applications. In Section \ref{Section4}, we make some observations on one's choice of null network when using the modularity quality function.  
		
	
	\subsection{The Louvain computational heuristic}\label{Subsection2.2}
		
		For a given modularity matrix $\p{B}$, a solution to the modularity-maximization problem is guaranteed to exist in any network with a finite number of nodes. However, the number of possible partitions in an $N$-node network, given by the Bell number \cite{Bell1934}, grows at least exponentially with $N$, so an exhaustive search of the space of partitions is infeasible. Modularity maximization was proven in \cite{Brandes2008} to be an NP-hard problem (at least for the null networks which we consider in this paper), so solving it requires the use of computational heuristics. In the present paper, we focus on the Louvain heuristic, which is a locally-greedy modularity-increasing sampling process over the set of partitions \cite{Blondel2008}.
		
		The Louvain heuristic consists of two phases, which are repeated iteratively. Initially, each node in the network constitutes a set, which gives an initial partition that consists of $N$ singletons. During phase 1, one considers the nodes one by one (in some order), and one places each node in a set (possibly where it already is) that results in the largest increase of modularity. This phase is repeated until one reaches a local maximum (i.e., until one obtains a partition in which the move of a single node cannot increase modularity). Phase 2 consists of constructing a reduced network $\mathcal{G'}$ from the sets of nodes in $\mathcal{G}$ that one obtains after the convergence of phase 1. We denote the sets in $\mathcal{G}$ at the end of phase 1 by $\{\hat{C}_1,\ldots,\hat{C}_{\hat{N}}\}$ (where $\hat{N}\leq N$) and the set assignment of node $i$ in this partition by $\hat{c}_i$. Each set $\hat{C}_k$ in $\mathcal{G}$ constitutes a node $k$ in $\mathcal{G'}$, and the reduced modularity matrix of $\mathcal{G'}$ is 
	\begin{displaymath}
		  \p{B}' = \p{\hat{S}^{T}}\p{B}\p{\hat{S}}\,,
	\end{displaymath}
	where $\p{\hat{S}}$ is the partition matrix of $\{\hat{C}_1,\ldots,\hat{C}_{\hat{N}}\}$. This ensures that the all-singleton partition in $\mathcal{G'}$ has the same value of modularity as the partition of $\mathcal{G}$ that we identified at the end of phase 1. One then repeats phase 1 on the reduced network and continues iterating until the heuristic converges (i.e., until phase 2 induces no further changes). 
		
		Because we use a nondeterministic implementation of the Louvain heuristic---in particular, the node order is randomized at the start of each iteration of phase 1---the network partitions that we obtain for a fixed modularity matrix can differ across runs.\footnote{The implementation \cite{NetWiki, LouvainCode} of the heuristic that we use in this paper is a generalized version of the implementation in \cite{Blondel2008}. It is independent of the null network---it takes the modularity matrix as an input to allow an arbitrary choice of null network---and it randomizes the node order at the start of each iteration of phase 1 of the heuristic to increase the search space of the heuristic. When one chooses the same null network that was assumed in \cite{Blondel2008} and uses a node order fixed to $\{1,\ldots,N\}$ at each iteration of phase 1 (the value of $N$ can change after each iteration of the heuristic's phase 2), then the implementation in \cite{LouvainCode} and the implementation described in \cite{Blondel2008} return the same output.} To account for this, one can compute the frequency of co-classification of nodes into communities for a given modularity matrix $\p{B}$ across multiple runs of the heuristic instead of using the output partition of a single run. (See \cite{Lancichinetti2012} for an application of such an approach to ``consensus clustering'' and \cite{Pardo2007} for an application of such an approach to hierarchical clustering.) We use the term \emph{association matrix} for a matrix that stores the mean number of times that two nodes are placed in the same community across multiple runs of a heuristic, and we use the term \emph{co-classification index} of nodes $i$ and $j$ to designate the  $(i,j)^{\text{th}}$ entry of an association matrix. 
		
There are many other heuristics that one can employ to maximize modularity \cite{Porter2009,Fortunato2010,Newman2010}, but the Louvain heuristic is a popular choice in practice \cite{Lancichinetti2009}. It is very fast \cite{Lancichinetti2009,Fortunato2010}, which is an important consideration in multilayer networks, for which the total number of nodes is the number of nodes in each layer multiplied by the number of layers. In Section \ref{Section5}, we point out two issues that the Louvain heuristic (independently of how it is implemented) faces with temporal multilayer networks.

		
		\subsection{Multiscale community structure}\label{Subsection2.3}
		
		Many networks include community structure at multiple scales \cite{Porter2009, Fortunato2010}, and some systems even have a hierarchical community structure of ``parts-within-parts''\cite{Simon1962}. In such a situation, although there are dense interactions within communities of some size (e.g., friendship ties between students in the same school), there are even denser interactions in subsets of nodes that lie inside of these communities (e.g., friendship ties between students in the same school and in the same class year). Some variants of the modularity function have been proposed to detect communities at different scales. A popular choice is to scale the null network using a \emph{resolution parameter} $\gamma \geq 0$ to yield a \emph{multiscale modularity-maximization} problem \cite{Reichardt2006}:
		\begin{equation} 
			\max_{C\in\mathcal{C}} \sum_{i,j=1}^{N}(A_{ij} - \gamma P_{ij})\delta(c_i,c_j)\,.
			\label{EQ2.6}
		\end{equation}
	In some sense, the value of the parameter $\gamma$ determines the importance that one assigns to the null network relative to the observed network. The corresponding modularity matrix and modularity function evaluated at a partition $C$ are $\p{B} = \p{A} - \gamma\p{P} \quad\text{ and }\quad Q(C\vert \p{A};\p{P};\gamma) = \sum_{i,j=1}^{N}(A_{ij} - \gamma P_{ij})\delta(c_i,c_j)$. The special case $\gamma = 1$ yields the modularity matrix and modularity function in the modularity-maximization problem \eqref{EQ2.3}. This formulation of multiscale modularity has a dynamical interpretation \cite{Lambiotte2009,Lambiotte2010} that we will discuss in the next subsection. 
	
	In most applications of community detection, the adjacency matrix of the observed and null networks have nonnegative entries. In these cases, the solution to \eqref{EQ2.6} when $0 \leq \gamma \leq \gamma^{-} = \min_{i\neq j,P_{ij}\neq 0}\left(A_{ij}/P_{ij}\right)$ is a single community regardless of any structure, however clear, in the observed network, because then
		\begin{equation*}
			B_{ij} = A_{ij} - \gamma P_{ij} \leq 0\quad\text{for all}\quad i,j\in\{1,\ldots,N\}\,.
		\end{equation*}
		(We exclude diagonal terms because a node is always in its own community.) However, the solution to \eqref{EQ2.6} when $\gamma > \gamma^{+} = \max_{i\neq j,P_{ij}\neq 0}\left(A_{ij}/P_{ij}\right)$ is $N$ singleton communities because
		\begin{equation*}
			B_{ij} = A_{ij} - \gamma P_{ij} < 0\quad\text{for all}\quad i,j\in\{1,\ldots,N\}\,.
		\end{equation*}
	Partitions at these boundary values of $\gamma$ correspond to the coarsest and finest possible partitions of a network, and varying the resolution parameter between these bounds makes it possible to examine a network's community structure at intermediate scales.
	
	For an observed and/or null network with signed edge weights, the intuition behind the effect of varying $\gamma$ in \eqref{EQ2.6} on an optimal solution is not straightforward. A single community and $N$ singleton communities do not need to be optimal partitions for any value of $\gamma \geq 0$. In particular, $B_{ij}$ has the same sign as $A_{ij}$ for sufficiently small values of $\gamma$, and $B_{ij}$ has the opposite sign to $P_{ij}$ for sufficiently large values of $\gamma$. For further discussion, see Section \ref{Section4}, where we explore the effect of varying the resolution parameter on an optimal partition for an observed and null network with signed edge weights. We vary the resolution parameter in the interval $[0,\gamma^+]$ instead of $[\gamma^-,\gamma^+]$ in numerical experiments with signed networks because globally optimal partitions can be different when $\gamma\in[0,\gamma^-]$.\footnote{For $\gamma\geq\gamma^{+}$, one can show that all modularity contributions no longer change signs. They are negative (respectively, positive) between pairs of nodes with $P_{ij}\geq 0$ (respectively, $P_{ij}\leq 0$).}

	It is important to differentiate between a ``resolution limit'' on the smallest community size that is imposed by a community-detection method \cite{Fortunato2007} and inherently multiscale community structure in a network \cite{Simon1962,Porter2009,Fortunato2010}. For the formulation of multiscale modularity in \eqref{EQ2.6}, the resolution limit described in \cite{Fortunato2007} applies to any fixed value of $\gamma$. By varying $\gamma$, one can identify communities that are smaller than the limit for any particular $\gamma$ value. In this sense, multiscale formulations of modularity help ``mitigate'' the resolution limit, though there remain issues \cite{Good2010,Arenas2008, Lambiotte2010}.  In this paper, we do not address the issue of how to identify communities at different scales, though we note in passing that the literature includes variants of multiscale modularity (e.g., see \cite{Arenas2008}).
We make observations on null networks in Section \ref{Section4}, and we illustrate how our observations can manifest in practice using the formulation of multiscale modularity in \eqref{EQ2.6}. (Our observations hold independently of the formulation of multiscale modularity that one adopts, but the precise manifestation can be different for different variants of multiscale modularity.)

	We use the term \emph{multiscale community structure} to refer to a set $\mathcal{C}_{\text{local}}(\p{\gamma})$ of local optima that we obtain with a computational heuristic for a set of (not necessarily all distinct) resolution-parameter values $\p{\gamma} = \{\gamma_1,\ldots,\gamma_l\}$, where $\gamma^{-} = \gamma_1 \leq \ldots \leq \gamma_l = \gamma^{+}$. We use the term \emph{multiscale association matrix} for an association matrix $\p{\hat{A}}\in [0,1]^{N\times N}$ that stores the co-classification index of all pairs of nodes for partitions in this set:
	\begin{equation}
	\hat{A}_{ij} = \frac{\sum_{C\in\mathcal{C}_{\text{local}}(\p{\gamma})}\delta(c_i,c_j)}{\vert\mathcal{C}_{\text{local}}(\p{\gamma})\vert}\,.
	\label{Asso}
	\end{equation}
	For each partition $C\in\mathcal{C}_{\text{local}}(\p{\gamma})$, nodes $i$ and $j$ either are (i.e., $\delta(c_i,c_j) = 1$) or are not (i.e., $\delta(c_i,c_j) = 0$) in the same community. It follows that $\hat{A}_{ij}$ is the mean number of partitions in $\mathcal{C}_{\text{local}}(\p{\gamma})$ for which nodes $i$ and $j$ are in the same community. The value of $\vert\mathcal{C}_{\text{local}}(\p{\gamma})\vert$ is the number of distinct values of $\gamma$ that we consider multiplied by the number of runs of the Louvain algorithm performed for each value of $\gamma$. In Section \ref{Section4}, we use a discretization of $[\gamma^{-},\gamma^{+}]$ (respectively, $[0,\gamma^{+}]$) for unsigned (respectively, signed) networks and perform a single run for each resolution-parameter value. The number of partitions in $\mathcal{C}_{\text{local}}(\p{\gamma})$ is then precisely the number of distinct values of $\gamma$ that we consider (i.e., one partition per value of $\gamma$). We use the matrix $\p{\hat{A}}$ repeatedly in our computational experiments in Section \ref{Section4}.

		
		\subsection{Null models and null networks}\label{Subsection2.4}

	In this section, we describe three null networks. We make several observations on the interpretation of communities that we obtain from Pearson correlation matrices using each of these null networks in the computational experiments of Section \ref{Section4}. 
	
		
		\subsubsection{Newman--Girvan (NG) null network}\label{ng-sub}
	
		A popular choice of null network for networks with positive edge weights is the Newman--Girvan (NG) null network, whose adjacency-matrix entries are $P_{ij} = k_ik_j/(2m)$, where $k_i$ are the observed node strengths \cite{Newman2004,Newman20042}. This yields the equivalent maximization problems
		\begin{equation}
			\max_{C\in\mathcal{C}} \sum_{i,j=1}^{N}\Bigg(A_{ij} - \frac{k_ik_j}{2m}\Bigg)\delta(c_i,c_j)
			\,\Leftrightarrow\,\max_{\p{S}\in\mathcal{S}}\text{Tr}\Bigg[\p{S}^{T}\Bigg(\p{A} - 
			\frac{\p{k}\p{k}^T}{2m}\Bigg)\p{S}\Bigg]\,,
			\label{EQ2.7}
		\end{equation}
		where $\p{k} = \p{A}\p{1}$ is the $N\times 1$ vector of node strengths and $2m = \p{1}^T\p{A}\p{1}$ is the total edge weight of the observed network. This null network can be derived from a variety of null models. One way to generate an unweighted network with expected adjacency matrix $\p{k}\p{k}^T/(2m)$ is to generate each of its edges and self-edges with probability $k_ik_j/(2m)$ (provided $k_ik_j \leq 2m$ for all $i,j$). That is, the presence and absence of edges and self-edges is a Bernoulli random variable with probability $k_ik_j/(2m)$ \cite{Bollobas2009, Bollobas2007}. More generally, any probability distribution on the set of adjacency matrices that satisfies $\mathbb{E}\big(\sum_{j=1}^{N}W_{ij}\big) = k_i$ (i.e., the expected strength equals the observed strength, see for e.g., \cite{Chung2002}) and $\mathbb{E}(W_{ij}) = f(k_i)f(k_j)$ for some real-valued function $f$ has an expected adjacency matrix of $\mathbb{E}(\p{W}) = \p{k}\p{k}^T/(2m)$.\footnote{The linearity of the expectation and the assumptions $\mathbb{E}\big(\sum_{j=1}^{N}W_{ij}\big) = k_i$ and $\mathbb{E}(W_{ij}) = f(k_i)f(k_j)$ imply that $f(k_i) = k_i/\sum_{j=1}^{N}f(k_j)$ and $\sum_{j=1}^{N}f(k_j) = \sqrt{2m}$. Combining these equations gives the desired result.} The adjacency matrix of the NG null network is symmetric and positive semidefinite.
			
		We briefly describe a way of deriving the NG null network from a model on time-series data (in contrast to a model on a network). The \emph{partial correlation} $\text{corr}(a,b\,\vert\, c)$ between $a$ and $b$ given $c$ is the Pearson correlation between the residuals that result from the linear regression of $a$ with $c$ and $b$ with $c$, and it is given by
	\begin{equation}
		\text{corr}(a,b\,\vert\, c) = \frac{\text{corr}(a,b) - \text{corr}(a,c)\text{corr}(b,c)}{\sqrt{1 - \text{corr}^2(a,c)}\sqrt{1 - \text{corr}^2(b,c)}}\,.
		\label{EQ2.8}
		\end{equation}
Suppose that the data used to construct the observed network is a set of time series $\{z_{i}\vert i \in\{1,\ldots,N\}\}$, where $ z_i = \{z_{i}(t)\vert t\in T\}$ and $T$ is a discrete set of time points. The authors of \cite{MacMahon2013} pointed out that $A_{ij} = \text{corr}(z_{i},z_{j})$ implies that $k_i = \text{cov}(\hat{z}_i,\hat{z}_{\text{tot}})$ and thus that
	\begin{equation}
		\frac{k_ik_j}{2m} = \text{corr}(\hat{z}_i,\hat{z}_{\text{tot}})\text{corr}(\hat{z}_j,\hat{z}_{\text{tot}})\,,
	\label{EQ2.9}
	\end{equation}
	where $\hat{z}_i(t) = (z_i(t) - \langle z_i\rangle)/\sigma(z_i)$ is a standardized time series and $\hat{z}_{\text{tot}}(t) = \sum_{i=1}^{N} \hat{z}_i(t)$ is the sum of the standardized time series.\footnote{The equality \eqref{EQ2.9} holds for networks constructed using Pearson correlations between time series \cite{MacMahon2013}. Such networks are examples of signed networks with edge weights in $[-1,1]$. The strength of a node $i$ is given by the $i^{\text{th}}$ (signed) column or row sum of the correlation matrix.} Taking $a = \hat{z}_i$, $b = \hat{z}_j$, and $c = \hat{z}_{\text{tot}}$, equation  \eqref{EQ2.8} implies that if $\text{corr}(\hat{z}_i,\hat{z}_j\,\vert\, \hat{z}_{\text{tot}}) = 0$ then $\text{corr}(\hat{z}_i,\hat{z}_j) = k_ik_j/(2m)$. That is, Pearson correlation coefficients between pairs of time series that satisfy $\text{corr}(\hat{z}_i,\hat{z}_j\,\vert\,\hat{z}_{\text{tot}}) = 0$ are precisely the adjacency-matrix entries of the NG null network. One way of generating a set of time series in which pairs of distinct time series satisfy this condition is to assume that each standardized time series depends linearly on the mean time series and that residuals are mutually uncorrelated (i.e., $\hat{z}_i = \alpha_i \hat{z}_{\text{tot}}/N + \beta_i + \epsilon_i$ for some $\alpha_i,\beta_i\in\mathbb{R}$ and $\text{corr}(\epsilon_i,\epsilon_j) = 0$ for $i\neq j$).
			
		The multiscale modularity-maximization problem in \eqref{EQ2.6} was initially introduced in \cite{Reichardt2006} using an \emph{ad hoc} approach. Interestingly, one can derive this formulation of the maximization problem for sufficiently large values of $\gamma$ by considering a quality function based on a continuous-time Markov process $X(t)$ on an observed network \cite{Lambiotte2009,Lambiotte2010}. The probability density of a continuous-time Markov process with exponentially distributed waiting times at each node parametrized by $\lambda(i)$ satisfies
	\begin{equation}
		\p{\dot{p}} = \p{p}\p{\Lambda}\p{M} - \p{p}\p{\Lambda}\,,
	\label{EQ2.10}
	\end{equation}
	where the vector $\p{p}(t)\in [0,1]^{1\times N}$ is the probability density of a random walker at each node [i.e., $p_i(t) := \mathbb{P}(X(t) = i)$ for each $i$], $\p{\Lambda}$ is a diagonal matrix with the rate $\lambda(i)$ on its $i^{\text{th}}$ diagonal entry, and $\p{M}$ is the transition matrix of a random walker (i.e., $M_{ij} :=  A_{ij}/k_i$). The solution to equation \eqref{EQ2.10} is $\p{p}(t) = \p{p}_0e^{\p{\Lambda(M-\p{I})}t}$ and its stationary distribution is unique (provided the network is connected) and given by $\p{\pi} = c\p{k}^T\p{\Lambda}^{-1}/(2m)$, where the normalization constant $c$ ensures that $\sum_{i=1}^{N} \pi_i = 1$. The \emph{stability} of a partition is a quality function defined by \cite{Lambiotte2009, Lambiotte2010, Delvenne2010}
	\begin{displaymath}
		r(\p{S},t) = \text{Tr}\left[\p{S}^{T}\left(\p{\Pi}e^{\p{\Lambda(M-\p{I})}t} - \p{\pi}^{T}\p{\pi}\right)\p{S}\right]\,,
	\end{displaymath}
	where $\Pi_{ij} = \delta(i,j)\pi_i$. Equivalently, the stability is
	\begin{equation}
		r(C,t) = \sum_{i,j=1}^{N} \left[\pi_i\left(e^{\p{\Lambda(M-\p{I})}t}\right)_{ij} - \pi_i\pi_j\right]\delta(c_i,c_j)\,.
		\label{EQ2.11}
	\end{equation}
	Taking $\p{p}_0 = \p{\pi}$, the term in brackets on the left-hand side of \eqref{EQ2.11} is $\mathbb{P}(X(0) = i\cap X(t)=j )$ and the term in brackets on the right-hand side is $\mathbb{P}(X(0) = i\cap X(t\to\infty) = j)$ (provided the system is ergodic). The intuition behind the stability quality function is that a good partition at a given time before reaching stationarity corresponds to one in which the time that a random walker spends within communities is large compared with the time that it spends transiting between communities. In other words, a random walker that starts out at a community ends up there again in the early stages of the random walk, long before stationarity.
	The resulting maximization problem is $\max_{\p{S}\in\mathcal{S}}r(\p{S},t)$, or equivalently $\max_{C\in\mathcal{C}}r(C,t)$. By linearizing $e^{\p{\Lambda(M-\p{I})}t}$ at $t=0$ and taking $\p{\Lambda} = \p{I}$, one obtains the multiscale modularity-maximization problem in \eqref{EQ2.6} at short timescales with $\gamma = 1/t$ and $P_{ij} = k_ik_j/(2m)$. This approach provides a dynamical interpretation of the resolution parameter $\gamma$ as the inverse (after linearization) of the time used to explore a network by a random walker.

		
	\subsubsection{Generalization of Newman--Girvan null network to signed networks (NGS)}
			
	In \cite{Gomez2009}, G\'omez et al. proposed a generalization of the NG null network to signed networks. They separated $\p{A}$ into its positive and negative edge weights:
	\begin{equation*}
		\p{A} = \p{A}^{+} - \p{A}^{-}\,,
	\end{equation*}
	where $\p{A}^{+}$ denotes the positive part of $\p{A}$ and $-\p{A}^{-}$ denotes its negative part. Their generalization of the NG null network to signed networks (NGS) is $P_{ij} = k_i^{+}k_j^{+}/(2m^+) - k_i^{-}k_j^{-}/(2m^-)$. This yields the maximization problem
		\begin{equation}
			\max_{C\in\mathcal{C}} \sum_{i,j=1}^{N}\Bigg[\Bigg(A_{ij}^{+} - \frac{k_i^{+}k_j^{+}}{2m^+}\Bigg) -
			\Bigg(A_{ij}^{-}-\frac{k_i^{-}k_j^{-}}{2m^-}\Bigg)\Bigg]\delta(c_i,c_j)\,,
			\label{EQ2.12}
		\end{equation}
	where $k_i^{+}$ and $2m^{+}$ (respectively, $k_i^{-}$ and $2m^{-}$) are the strengths and total edge weight in $\p{A}^{+}$ (respectively, $\p{A}^{-}$). The intuition behind this generalization is to use an NG null network on both unsigned matrices $\p{A}^{+}$ and $\p{A}^{-}$ but to count contributions to modularity from negative edge weights (i.e., the second group of terms in \eqref{EQ2.12}) in an opposite way to those from positive edge weights (i.e., the first group of terms in \eqref{EQ2.12}). Negative edge weights that exceed their expected edge weight are penalized (i.e., they decrease modularity) and those that do not are rewarded (i.e., they increase modularity). One can generate a network with edge weights $0$, $1$, or $-1$ and expected edge weights $k_i^+k_j^+/(2m^+) - k_i^-k_j^-/(2m^-)$ by generating one network with expected edge weights $W_{ij}^{+} = k_i^+k_j^+/(2m^+)$ and a second network with expected edge weights $W_{ij}^{-} = k_i^-k_j^-/(2m^-)$ using the procedure described for the NG null network in Section \ref{ng-sub}, and then defining a network whose edge weights are given by the difference between the edge weights of these two networks. More generally, by Footnote 5 and linearity of expectation, any probability distribution on the set $\{\p{W} \in\mathbb{R}^{N\times N}\}$ of signed adjacency matrices that satisfies the two conditions
\begin{align}
	&\quad\mathbb{E}\bigg(\sum_{j=1}^{N}W_{ij}^{+}\bigg) = k_i^{+} \,\text{ and }\, \mathbb{E}(W_{ij}^{+}) = f(k_i^{+})f(k_j^{+})\,,\\
	&\quad\mathbb{E}\bigg(\sum_{j=1}^{N}W_{ij}^{-}\bigg) = k_i^{-} \,\text{ and }\, \mathbb{E}(W_{ij}^{-}) = g(k_i^{-})g(k_j^{-})\,,
\end{align}
where $f$ and $g$ are real-valued functions and $W_{ij} = W_{ij}^{+} - W_{ij}^{-}$, has an expected edge weight of $\mathbb{E}(W_{ij}) = k_i^+k_j^+/(2m^+) - k_i^-k_j^-/(2m^-)$ for all $i,j\in\{1,\ldots,N\}$.
		
	The authors of \cite{Mucha2010} derived a variant of the multiscale formulation of modularity in \eqref{EQ2.6} for the NGS null network at short time scales by building on the random-walk approach used to derive the NG null network.\footnote{In particular, they derived the multiscale formulation of modularity obtained using a Potts-model approach in \cite{Traag2009}. This multiscale formulation results in one resolution parameter $\gamma_1$ for the term $(k_i^{+}k_j^{+})/(2m^{+})$ and a second resolution parameter $\gamma_2$ for the term $(k_i^{-}k_j^{-})/(2m^{-})$ in \eqref{EQ2.12} (see \cite{Macon2012} for an application of this multiscale formulation to the United Nations General Assembly voting networks). Without an application-driven justification for how to choose these parameters, this increases the parameter space substantially, so we only consider the case $\gamma_1 = \gamma_2$ in this paper.} 
	They considered the function 
		\begin{equation}
			\hat{r}(C,t) = \sum_{i,j=1}^{N} \bigg(\pi_i\big[\delta_{ij} + t\Lambda_{ii}(M_{ij} - \delta_{ij})\big] - \pi_i\rho_{i\vert j}\bigg)\delta(c_i,c_j)\,,
			\label{EQ2.13}
		\end{equation}
	 where the term in brackets on the left-hand side is a linearization of the exponential term in \eqref{EQ2.11}, $\p{M}$ and $\pi_i$ are as defined in \eqref{EQ2.11} on a network with adjacency matrix $\vert \p{A} \vert:=\p{A}^{+} + \p{A}^{-}$, and $\rho_{i\vert j}$ is the probability of jumping from node $i$ to node $j$ at stationarity in one step conditional on the network structure \cite{Mucha2010}. If the network is non-bipartite, unsigned, and undirected, then $\rho_{i\vert j}$ reduces to the stationary probability $\pi_j$.

		
		\subsubsection{Uniform (U) null network}
			
		A third null network that we consider is a uniform (U) null network, with adjacency-matrix entries  $P_{ij} = \langle k\rangle^2/(2m)$, where $\langle k\rangle := \bigl(\sum_{i=1}^{N}k_i\bigr)/N$ denotes the mean strength in a network. We thereby obtain the equivalent maximization problems
		\begin{equation}
			\max_{C\in\mathcal{C}} \sum_{i,j=1}^{N}\Bigg(A_{ij} - \frac{\langle k\rangle^2}{2m}\Bigg)\delta(c_i,c_j) \Leftrightarrow \max_{\p{S}\in\mathcal{S}} \text{Tr}\bigg[\p{S}^{T}\bigg(\p{A} - \frac{\langle k\rangle^2}{2m}\p{1}_{N}\bigg)\p{S}\bigg]\,,
			\label{EQ2.14}
		\end{equation}
		where $\p{A}$ is an unsigned adjacency matrix and $\p{1}_{N}$ is an $N\times N$ matrix in which every entry is 1.\footnote{For a network in which all nodes have the same strength, the uniform and Newman--Girvan null networks are equivalent because $k_i = k_j$ for all $i,j \Leftrightarrow k_i = 2m/N = \langle k\rangle$ for all $i$. This was pointed out for an application to foreign exchange markets in \cite{Fenn2009}.} The expected edge weight in \eqref{EQ2.14} is constant and satisfies
		\begin{equation*}
			\frac{\langle k\rangle^2}{2m} = \frac{\left(\sum_{i=1}^{N}k_i\big/N\right)^2}{\sum_{i=1}^{N}k_i} = \frac{2m}{N^2} = \langle \p{A}\rangle\,,
		\end{equation*}
		where $\langle \p{A}\rangle$ denotes the mean value of the adjacency matrix.\footnote{Although we use the uniform null network on unsigned adjacency matrices in this paper, the expected edge weight in the uniform null network is always nonnegative for correlation matrices, as positive semidefiniteness guarantees that $\langle\p{A}\rangle = \p{1}^T\p{A}\p{1}/(N^2) \geq 0$.} One way to generate an unweighted network with adjacency matrix $\langle \p{A}\rangle\p{1}_N$ is to generate each edge with probability $\langle \p{A}\rangle$ (provided $\langle \p{A}\rangle\leq 1$). That is, the presence and absence of an edge (including self-edges) are independent and identically distributed (i.i.d.) Bernoulli random variables with probability $\langle \p{A}\rangle$. More generally, any probability distribution on the set of adjacency matrices that satisfies $\mathbb{E}\big(\sum_{i,j=1}^{N}W_{ij}\big) = 2m$ and $\mathbb{E}(W_{ij}) = \mathbb{E}(W_{i'j'})$ for all $i,j,i',j'$ has an expected adjacency matrix $\mathbb{E}(\p{W}) = \langle \p{A}\rangle\p{1}_N$. The adjacency matrix of the U null network is symmetric and positive semidefinite. One can derive the multiscale formulation in \eqref{EQ2.6} for the U null network from the stability quality function in precisely the same way as it is derived for the NG null network, except that one needs to consider exponentially distributed waiting times at each node with rates proportional to node strength (i.e.,  $\Lambda_{ij} = \delta(i,j)k_i/\langle k\rangle$) \cite{Lambiotte2009}. 
			
		 
		
		\section{Multilayer modularity maximization}\label{Section3}
		
		
		\subsection{Multilayer representation of temporal networks}
		
		We restrict our attention to temporal networks in which only edges vary in time. (Thus, each node is present in all layers.) We use the notation $\p{A}_s$ for a $\emph{layer}$ in a sequence of adjacency matrices $\mathcal{T} = \{\p{A_1},\ldots,\p{A}_{\vert\mathcal{T}\vert}\}$, and we denote node $i$ in layer $s$ by $i_s$. We use the term \emph{multilayer network} for a network defined on the set of nodes $\{1_1,\ldots,N_1;1_2,\ldots,N_2;\ldots;1_{\vert\mathcal{T}\vert},\ldots,N_{\vert\mathcal{T}\vert}\}$ \cite{Kivela2014}.  
			
\begin{figure}
			\begin{equation*}
				\begin{tikzpicture}[baseline = -43pt, scale = 0.55]
				\draw[very thick] (-0.25cm,0cm) -- (3.75cm,0cm);
				\draw[very thick] (-0.25cm,0cm) -- (-1.25cm,-3cm);
				\draw[very thick] (-1.25cm,-3cm) -- (2.75cm,-3cm) node [pos = .25, below] {Layer 1};
				\draw[very thick] (2.75cm,-3cm) -- (3.75cm,0cm);
				\fill[black] (0.75cm, -0.75cm) circle (1mm) node [below] {\footnotesize $1_{\scalebox{0.8}{1}}$}; 
				\fill[black] (2.5cm, -0.75cm) circle (1mm) node [below] {\footnotesize $2_{\scalebox{0.8}{1}}$};
				\fill[black] (0.2cm, -2.5cm) circle (1mm) node [right] {\footnotesize $3_{\scalebox{0.8}{1}}$};
				
				\draw[very thin] (0.75cm, -0.75cm) -- (2.5cm, -0.75cm); 
				\draw[very thin] (0.75cm, -0.75cm) -- (0.2cm, -2.5cm);
				
				\draw[dashed] (0.75cm, -0.75cm) .. controls (2.25, 0).. (3.75cm, -1.75cm); 
				\draw[dashed] (2.5cm, -0.75cm) .. controls (4, 0) .. (5.5cm, -1.75cm);
				\draw[dashed] (0.2cm, -2.5cm) .. controls (1.7, -3.75) .. (3.2cm, -3.5cm);
		
				\draw[very thick] (3cm,-1cm) -- (7cm,-1cm);
				\draw[very thick] (3cm,-1cm) -- (2cm,-4cm);
				\draw[very thick] (2cm,-4cm) -- (6cm,-4cm) node [pos = .25, below] {Layer 2};
				\draw[very thick] (6cm,-4cm) -- (7cm,-1cm);
				\fill[black] (3.75cm, -1.75cm) circle (1mm) node [below] {\footnotesize $1_{\scalebox{0.8}{2}}$}; 
				\fill[black] (5.5cm, -1.75cm) circle (1mm) node [below] {\footnotesize $2_{\scalebox{0.8}{2}}$};
				\fill[black] (3.2cm, -3.5cm) circle (1mm) node [right] {\footnotesize $3_{\scalebox{0.8}{2}}$};
				
				\draw[very thin] (3.75cm, -1.75cm) -- (5.5cm, -1.75cm); 
				\draw[very thin] (3.75cm, -1.75cm) -- (3.2cm, -3.5cm);
				\draw[very thin] (5.5cm, -1.75cm) -- (3.2cm, -3.5cm);
				
				\draw[dashed] (3.75cm, -1.75cm) .. controls (5.25, -1).. (6.75cm, -2.75cm); 
				\draw[dashed] (5.5cm, -1.75cm) .. controls (7, -1) .. (8.5cm, -2.75cm);
				\draw[dashed] (3.2cm, -3.5cm) .. controls (4.7, -4.75) .. (6.2cm, -4.5cm);
				
				\draw[very thick] (6.25cm,-2cm) -- (10.25cm,-2cm);
				\draw[very thick] (6.25cm,-2cm) -- (5.25cm,-5cm);
				\draw[very thick] (5.25cm,-5cm) -- (9.25cm,-5cm) node [pos = 0.25, below] {Layer 3};
				\draw[very thick] (9.25cm,-5cm) -- (10.25cm,-2cm);
				\fill[black] (6.75cm, -2.75cm) circle (1mm) node [below] {\footnotesize $1_{\scalebox{0.8}{3}}$}; 
				\fill[black] (8.5cm, -2.75cm) circle (1mm) node [below] {\footnotesize $2_{\scalebox{0.8}{3}}$};
				\fill[black] (6.2cm, -4.5cm) circle (1mm) node [right] {\footnotesize $3_{\scalebox{0.8}{3}}$};
				
				\draw[very thin] (8.5cm, -2.75cm) -- (6.75cm, -2.75cm); 
				\draw[very thin] (8.5cm, -2.75cm) -- (6.2cm, -4.5cm);
			
			\end{tikzpicture}
			\longleftrightarrow
			\scalebox{0.8}{$\left[\begin{array}{ccc | ccc | ccc}
				0&1&1&\omega&0&0&0&0&0\\
				1&0&0&0&\omega&0&0&0&0\\
				1&0&0&0&0&\omega&0&0&0\\ \hline
				\omega&0&0&0&1&1&\omega&0&0\\
				0&\omega&0&1&0&1&0&\omega&0\\
				0&0&\omega&1&1&0&0&0&\omega\\ \hline
				0&0&0&\omega&0&0&0&1&0\\
				0&0&0&0&\omega&0&1&0&1\\
				0&0&0&0&0&\omega&0&1&0
				\end{array}\right]$}
				\end{equation*}
		\caption{Example of (left) a multilayer network with unweighted intra-layer connections (solid lines) and uniformly weighted inter-layer connections (dashed curves) and (right) its corresponding adjacency matrix. (The adjacency matrix that corresponds to a multilayer network is sometimes called a ``supra-adjacency matrix'' in the network-science literature \cite{Kivela2014}.) 
		}
		\label{FIG3.2}
		\end{figure}
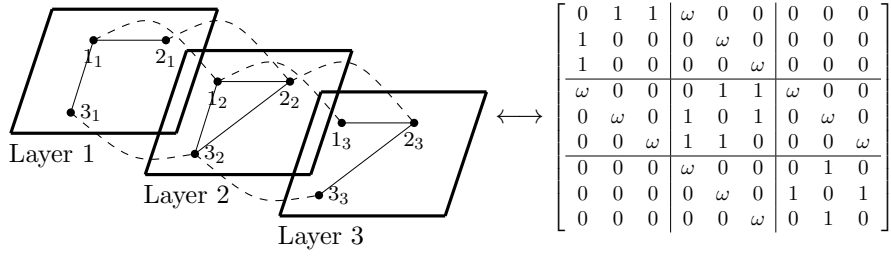

			Thus far, computations that have used a multilayer framework for temporal networks have almost always assumed (1) that inter-layer connections exist only between nodes that correspond to the same entity (i.e., between nodes $i_s$ and $i_r$ for some $i$ and $s\neq r$) and (2) that the network layers are ``ordinal'' (i.e., inter-layer edges exist only between consecutive layers) \cite{Mucha2010, Mucha20102, Bassett2011,  Radicchi2013, Kivela2014}. It is also typically assumed that (3) inter-layer connections are uniform (i.e., inter-layer edges have the same weight). In a recent review article on multilayer networks \cite{Kivela2014}, condition (1) was called ``diagonal'' coupling, and condition (2) implies that a network is ``layer-coupled''. We refer to the type of coupling defined by (1), (2), and (3) as \emph{ordinal diagonal} and \emph{uniform} inter-layer coupling and we denote the value of the inter-layer edge weight by $\omega \in\mathbb{R}$. We show a simple illustration of a multilayer network with ordinal diagonal and uniform inter-layer coupling in Fig.~\ref{FIG3.2}. One can consider more general inter-layer connections (e.g., nonuniform ones). Although we restrict our attention to uniform coupling in our theoretical and computational discussions, we give an example of a nonuniform choice of inter-layer coupling in Section \ref{Section5}. Results similar to those of subsection \ref{Subsection5.2} also apply in this more general case.

		\subsection{The multilayer modularity function}
		
		The authors of \cite{Mucha2010} generalized the single-layer multiscale modularity-maximization problem in \eqref{EQ2.6} to a multilayer network using a similar approach as the one used to derive the NGS null network from a stochastic Markov process on the observed network. For simplicity, we express intra-layer and inter-layer connections in an $N\vert\mathcal{T}\vert$-node multilayer network using a single $N\vert\mathcal{T}\vert\times N\vert\mathcal{T}\vert$ matrix. Each node $i_s$ in layer $s$ has the unique index $i':= i + (s-1)N$, and we use $\p{\mathcal{A}}$ to denote the multilayer adjacency matrix, which has entries $\mathcal{A}_{i'j'} = A_{ijs}\delta(s,r) + \omega\delta(\vert s-r\vert,1)$ when the inter-layer coupling is ordinal diagonal and uniform. (As discussed in \cite{Kivela2014}, one can use either an adjacency tensor or an adjacency matrix to represent a multilayer network.) The generalization in \cite{Mucha2010} consists of applying the function in \eqref{EQ2.13} to the $N\vert\mathcal{T}\vert$-node multilayer network:
	\begin{equation}
		\hat{r}(C,t) = \sum_{i,j=1}^{N\vert\mathcal{T}\vert} \bigg(\pi_i\big[\delta_{ij} + t\Lambda_{ii}(\mathcal{M}_{ij} - \delta_{ij})\big] - \pi_i\rho_{i\vert j}\bigg)\delta(c_i,c_j)\,,
		\label{EQ3.1}
	\end{equation}	
	where $C$ is now a multilayer partition (i.e., a partition of an $N\vert\mathcal{T}\vert$-node multilayer network), $\p{\Lambda}$ is the $N\vert\mathcal{T}\vert\times N\vert\mathcal{T}\vert$ diagonal matrix with the rates of the exponentially distributed waiting times at each node of each layer on its diagonal, $\p{\mathcal{M}}$ (with entries $\mathcal{M}_{ij} := \mathcal{A}_{ij}/\sum_{j}\mathcal{A}_{ij}$) is the $N\vert\mathcal{T}\vert\times N\vert\mathcal{T}\vert$ transition matrix for the $N\vert\mathcal{T}\vert$-node multilayer network with adjacency matrix $\p{\mathcal{A}}$, $\pi_i$ is the corresponding stationary distribution (with the strength of a node and the total edge weight now computed from the multilayer adjacency matrix $\p{\mathcal{A}}$), and $\rho_{i\vert j}$ is the probability of jumping from node $i$ to node $j$ at stationarity in one step conditional on the structure of the network within and between layers. The authors' choice of $\rho_{i\vert j}$, which accounts for the ``sparsity pattern''\footnote{The \emph{sparsity pattern} of a matrix $\p{X}$ is a matrix $\p{Y}$ with entries $Y_{ij} = 1$ when $X_{ij}\neq 0$ and $Y_{ij} = 0$ when $X_{ij} = 0$.} of inter-layer edges in the multilayer adjacency matrix, motivates the \emph{multilayer modularity-maximization problem}
	\begin{equation}
		\max_{C\in\mathcal{C}}\sum_{i,j = 1}^{N\vert\mathcal{T}\vert}\mathcal{B}_{ij}\delta(c_i,c_j)\,,
		\label{EQ3.2}
	\end{equation}	 
	which we can also write as $\max_{C\in\mathcal{C}}Q(C\vert\p{\mathcal{B}})$, where $\p{\mathcal{B}}$ is the \emph{multilayer modularity matrix}
	\begin{equation}
		\p{\mathcal{B}} = 
		\begin{bmatrix}
			\p{B}_1 & \omega\p{I} & \p{0}  & \ldots & \p{0} \\
			\omega\p{I} &\ddots  &\ddots   & \ddots  & \vdots \\
			\p{0}  & \ddots & \ddots & \ddots & \p{0}\\
			\vdots & \ddots & \ddots & \ddots & \omega\p{I}\\
			\p{0} & \ldots &\p{0} & \omega\p{I} & \p{B}_{\vert\mathcal{T}\vert}
			\end{bmatrix}\,,
		\label{EQ3.3}
	\end{equation}
	and $\p{B}_s$ is a single-layer modularity matrix computed on layer $s$. (For example, $\p{B}_s = \p{A}_s - \langle \p{A}_s\rangle\p{1}_N$ if one uses the U null network and sets $\gamma = 1$.) We rewrite the multilayer modularity-maximization problem in \cite{Mucha2010} as
		\begin{equation}
			\max_{C\in\mathcal{C}}\Bigg[\sum_{s=1}^{\vert\mathcal{T}\vert}\sum_{i,j = 1}^{N}B_{ijs}\delta(c_{i_s},c_{j_s}) + 	2\omega\sum_{s=1}^{\vert\mathcal{T}\vert-1}\sum_{i=1}^{N}\delta(c_{i_s},c_{i_{s+1}})\Bigg]\,,
			\label{EQ3.4}
		\end{equation}
	where $B_{ijs}$ denotes the $(i,j)^{\text{th}}$ entry of $\p{B}_s$. Equation \eqref{EQ3.4} clearly separates intra-layer contributions (left term) from inter-layer contributions (right term) to the multilayer quality function.
	
	In practice, one can solve this multilayer modularity-maximization problem with the Louvain heuristic in subsection \ref{Subsection2.2} by using the multilayer modularity matrix $\p{\mathcal{B}}$ instead of the single-layer modularity matrix $\p{B}$ as an input (the number of nodes in the first iteration of phase 1 becomes $N\vert\mathcal{T}\vert$ instead of $N$). In this case, the initial partition consists of $N\vert\mathcal{T}\vert$ singletons. One first places each of the $N\vert\mathcal{T}\vert$ nodes into a set (possibly where it already is) that results in the largest increase of multilayer modularity. We then iterate this procedure on a reduced network (as defined in subsection \ref{Subsection2.2}) until the heuristic converges. It is clear from \eqref{EQ3.4} that placing nodes from different layers into the same set, which we call an inter-layer \emph{merge}, decreases the value of the multilayer quality function when $\omega < 0$, so we only consider $\omega \geq 0$. As with single-layer networks, we assume that each of the partitions in the set $\mathcal{C}$ of $N\vert\mathcal{T}\vert$-node partitions contains sets that do not have multiple connected components in the graph with adjacency matrix $\mathcal{B}$.
		
	In Section \ref{Section5}, we try to gain some insight into how to interpret a globally optimal multilayer partition by proving several properties that it satisfies.  The results that we show hold for \emph{any} choice of matrices $\p{B}_1,\ldots,\p{B}_{\vert\mathcal{T}\vert}$, so (for example) they still apply when one uses the stability quality function in \eqref{EQ2.11} on each layer instead of the modularity quality function.  For ease of writing (and because modularity is the quality function that we use in our computational experiments of Section \ref{Section5}), we will continue to refer to the maximization problem \eqref{EQ3.4} as a multilayer modularity-maximization problem.
		
			
	\section{Interpretation of community structure in correlation networks with different null networks}\label{Section4}
	
		It is clear from the structure of $\p{\mathcal{B}}$ in equation \eqref{EQ3.3} that the choice of quality function within layers (i.e., diagonal blocks in the multilayer modularity matrix) and the choice of coupling between layers (i.e., off-diagonal blocks) for a given quality function affect the solution of the maximization problem in \eqref{EQ3.4}. In this section, we make some observations on the choice of null network for correlation networks when using the modularity quality function. To do this, we consider the multilayer modularity-maximization problem (\ref{EQ3.4}) with zero inter-layer coupling (i.e., $\omega = 0$), which is equivalent to performing single-layer modularity maximization on each layer independently.

		
	\subsection{Toy examples}\label{Subsection4.1}
	
	We describe two simple toy networks to illustrate some features of the NG \eqref{EQ2.7} and NGS \eqref{EQ2.12} null networks that can be misleading for asset correlation networks.

		
	\subsubsection{NG null network}\label{ng-example}
	
	Assume that the nodes in a network are divided into $K$ nonoverlapping categories (e.g., asset classes) such that all intra-category edge weights have a constant value $a>0$ and all inter-category edge weights have a constant value $b$, with $0\leq b<a$. Let $\kappa_i$ denote the category of node $i$, and rewrite the strength of node $i$ as
	\begin{equation*}
		k_{i} = \vert\kappa_i\vert a + (N - \vert\kappa_i\vert)b = \vert\kappa_i\vert(a-b) + Nb\,.
		\label{EQ4.2}
	\end{equation*}		
	The strength of a node in this network scales affinely with the number of nodes in its category. Suppose that we have two categories $\kappa_1$, $\kappa_2$ that do not contain the same number of nodes. Taking $\vert \kappa_1\vert > \vert \kappa_2 \vert$ without loss of generality, it follows that 
	\begin{equation}\label{size}
			P_{i,j\in\kappa_1} = \frac{1}{2m}\Bigg[\vert\kappa_1\vert (a - b) + Nb\Bigg]^2 > \frac{1}{2m}\Bigg[\vert\kappa_2\vert (a - b) + Nb\Bigg]^2 =P_{i,j\in\kappa_2}\,,
	\end{equation}
	where $P_{i,j\in\kappa_i}$ is the expected edge weight between pairs of nodes in $\kappa_i$ in the NG null network. That is, pairs of nodes in an NG null network that belong to larger categories have a larger expected edge weight than pairs of nodes that belong to smaller categories.

	\begin{figure}
			\center
			\subfigure[Unsigned adjacency matrix]{
			\includegraphics[width = 3cm]{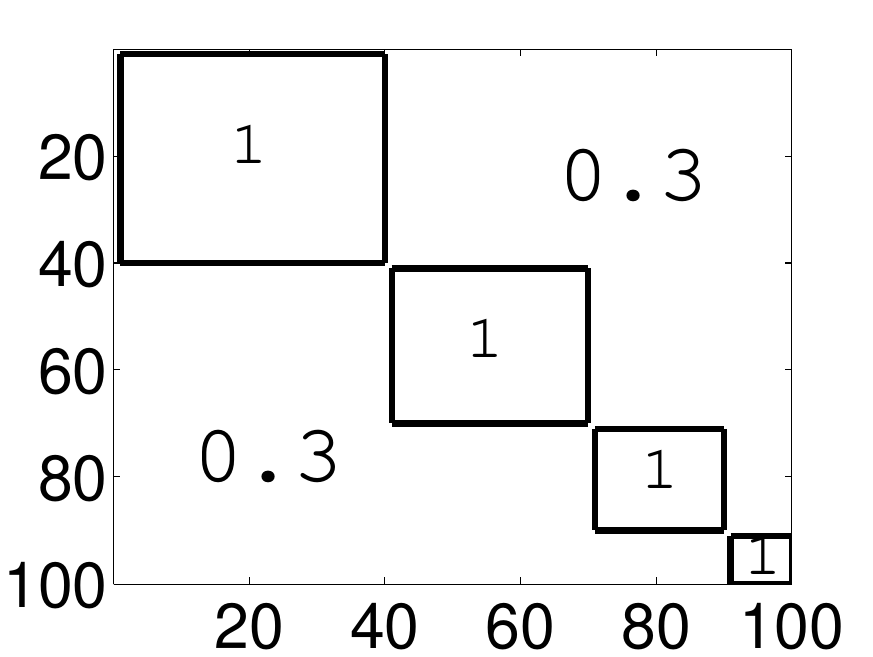}}
			\subfigure[Multiscale association matrix for the NG null network]{
			\includegraphics[width = 3cm]{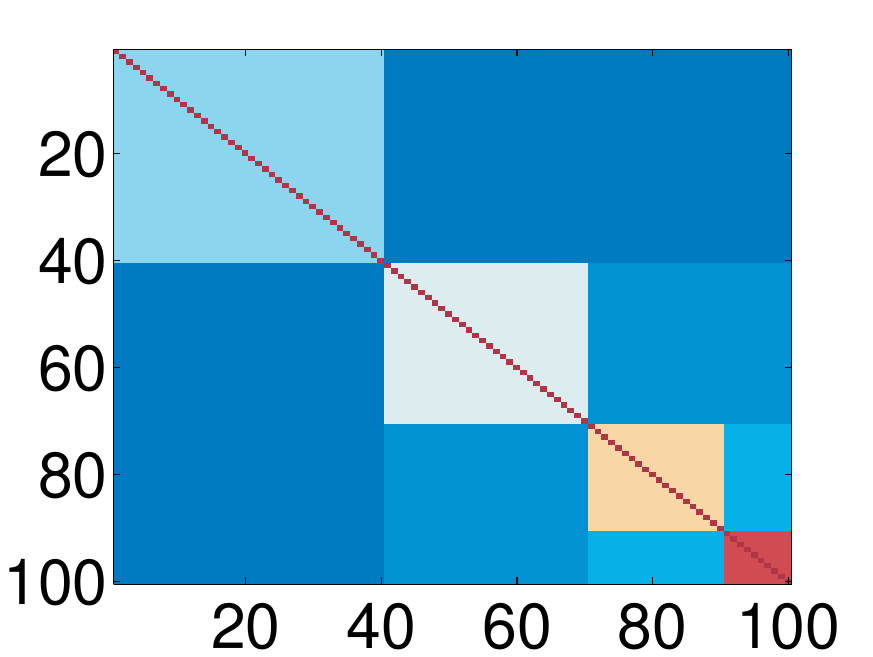}}
			\subfigure[Multiscale association matrix for the U null network]{
			\includegraphics[width = 3cm]{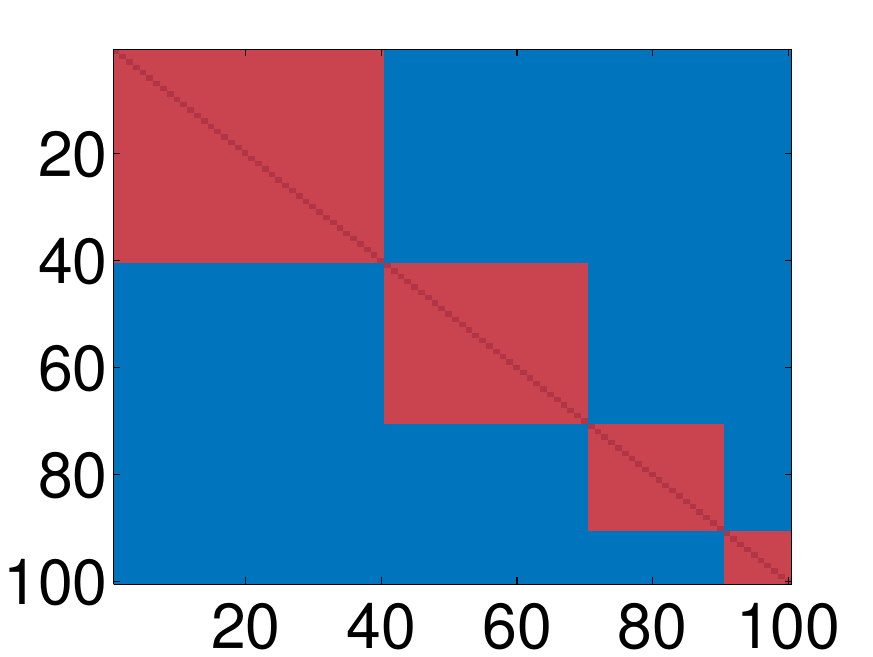}}
			  
			\hspace{3.2cm}\includegraphics[width = 7cm]{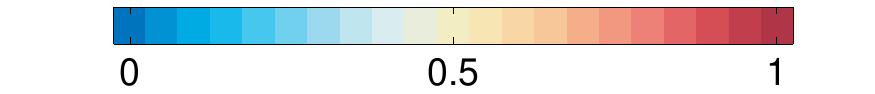}
				
			\subfigure[Signed adjacency matrix]{
			\includegraphics[width = 3cm]{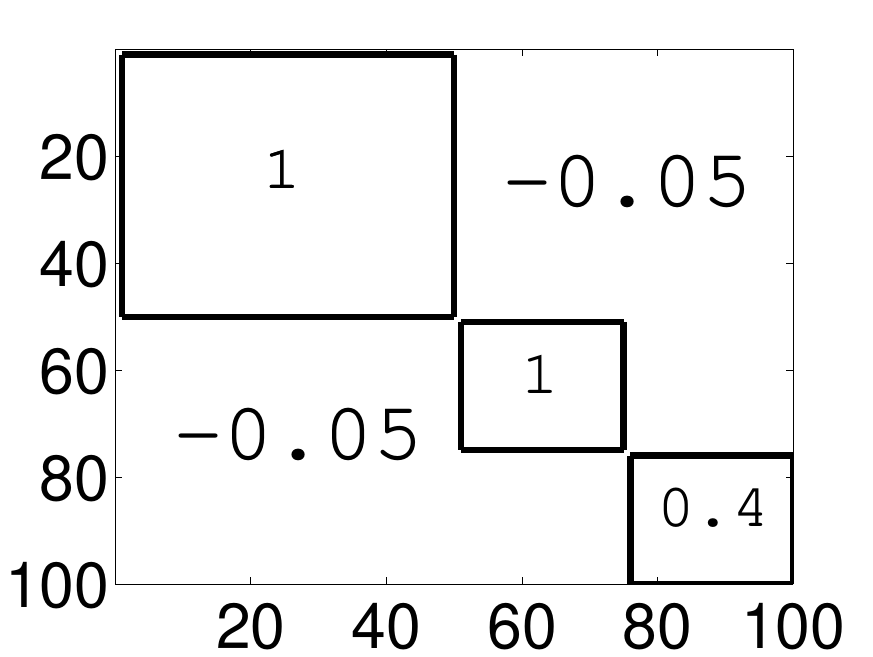}}
			\subfigure[Multiscale association matrix for the NGS null network]{
			\includegraphics[width = 3cm]{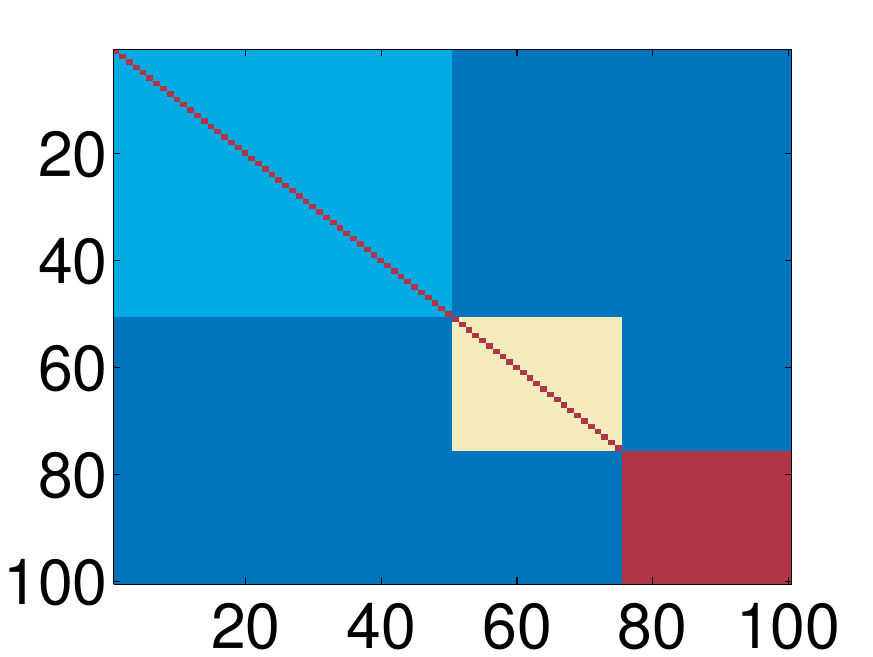}}
			\subfigure[Multiscale association matrix for the U null network]{
			\includegraphics[width = 3cm]{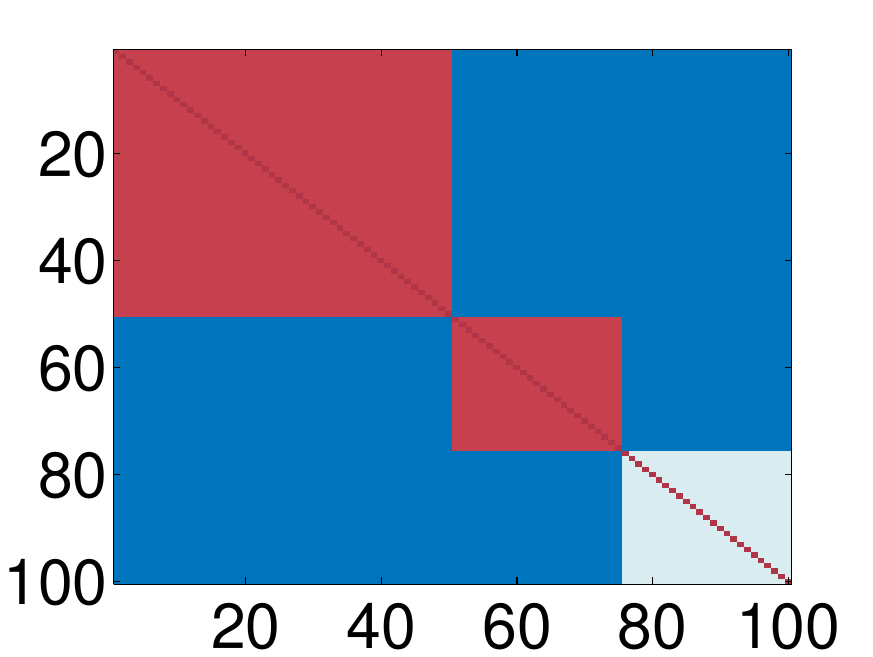}}
			    
			\caption{(a) Toy unsigned block matrix with constant diagonal and off-diagonal blocks that take the value indicated in the block. (b) Multiscale association matrix of (a) that gives the frequency of co-classification of nodes across resolution-parameter values using an NG null network. (c) Multiscale association matrix of (a) that uses a U null network. (d) Toy signed block matrix with constant diagonal and off-diagonal blocks that take the value indicated in the block. (e) Multiscale association matrix of (d) that uses an NGS null network. (f) Multiscale association matrix of (d) that uses a U null network.  For the NG and U (respectively, NGS) null networks, our sample of resolution-parameter values  is the set $\{\gamma^{-},\dots,\gamma^{+}\}$ (respectively, $\{0,\dots,\gamma^{+}\}$) with a discretization step of $10^{-3}$ between each pair of consecutive values.}
	\label{FIG4.1}
	\end{figure}

	To see how equation \eqref{size} can lead to misleading results, we perform a simple experiment. Consider the toy network in Fig.~\ref{FIG4.1}(a) that contains 100 nodes divided into four categories of sizes 40, 30, 20, and 10.  We set intra-category edge weights to 1 and inter-category edge weights to 0.3 (i.e., $a = 1$ and $b = 0.3$ in equation \eqref{size}). In  Fig.~\ref{FIG4.1}(b) (respectively, Fig.~\ref{FIG4.1}(c)), we show the multiscale association matrix defined in \eqref{Asso} using an NG null network (respectively, a U null network). Colors scale with the frequency of co-classification of pairs of nodes into the same community across resolution-parameter values. Because the nodes are ordered by category, diagonal blocks in Fig.~\ref{FIG4.1}(b,c) indicate the co-classification index of nodes in the same category, and off-diagonal blocks indicate the co-classification index of nodes in different categories. We observe in Fig.~\ref{FIG4.1}(b) that larger categories are identified as a community across a smaller range of resolution-parameter values than smaller categories when using an NG null network. In particular, category $\kappa$ is identified as a single community when $\gamma < a/P_{i,j\in\kappa}$ (with $a/P_{i,j\in\kappa_1} < a/P_{i,j\in\kappa_2}$ when $\vert\kappa_1\vert > \vert\kappa_2\vert$ by equation \eqref{size}). When $\gamma \geq  a/P_{i,j\in\kappa}$, category $\kappa$ is identified as $\vert\kappa\vert$ singleton communities. However, we observe in Fig.~\ref{FIG4.1}(c) that all four categories are identified as a single community across the same range of resolution-parameter values when using the U null network. In particular, category $\kappa$ is identified as a single community when $\gamma < a/\langle \p{A}\rangle$ and as $\vert\kappa\vert$ singleton communities when $\gamma \geq a/\langle \p{A}\rangle$.
		
	The standard interpretation of multiscale modularity maximization is that the communities that one obtains for larger values of $\gamma$ reveal ``smaller'' and ``more densely'' connected nodes in the observed network \cite{Reichardt2006, Lambiotte2010}. Although all diagonal blocks in Fig.~\ref{FIG4.1}(a) have the same internal connectivity, different ones are identified as communities for different values of $\gamma$ when using the NG null network---as $\gamma$ increases, nodes in the largest category split into singletons first, followed by those in the second largest category, etc. One would need to be cautious in using multiscale community structure to gain information about connectivity patterns in the observed network in this example.

	
		
		\subsubsection{NGS null network}
		
		A key difference between an NG null network \eqref{EQ2.7} and an NGS null network \eqref{EQ2.12} is that the expected edge weight between two nodes must be positive in the former but can be negative in the latter. Consider a signed variant of the example in Section \ref{ng-example} in which intra-category edge weights equal a constant $a>0$ and inter-category edge weights equal a constant $b<0$. The strengths of node $i$ in the $\kappa^{\text{th}}$ category are
		\begin{equation*}
			k_i^{+} = \vert\kappa\vert a\quad\text{ and }\quad k_{i}^{-} = (N - \vert\kappa\vert)b\,.
		\end{equation*}
		We consider two categories $\kappa_1$, $\kappa_2$ with different numbers of nodes. Taking $\vert\kappa_1\vert > \vert\kappa_2\vert$ without loss of generality, it follows that
		\begin{align*}
			& P_{i,j\in\kappa_1} = \frac{1}{2m^{+}}\bigg(\vert\kappa_1\vert a\bigg)^2 - \frac{1}{2m^{-}}\bigg[(N-\vert\kappa_1\vert)b\bigg]^2\\
			& > \frac{1}{2m^{+}}\bigg(\vert\kappa_2\vert a\bigg)^2 - \frac{1}{2m^{-}}\bigg[(N-\vert\kappa_2\vert)b\bigg]^2 = P_{i,j\in\kappa_2}\,,
		\end{align*}
	where $P_{i,j\in\kappa_i}$ is the expected edge weight between pairs of nodes in $\kappa_i$ in the NGS null network. As was the case for an NG null network, pairs of nodes in an NGS null network that belong to larger categories have a larger expected edge weight than pairs of nodes that belong to smaller categories. 
	
	However, the fact that the expected edge weight can be negative can further complicate interpretations of multiscale community structure. A category $\kappa$ for which $P_{i,j\in\kappa} < 0$ and $P_{i\in\kappa,j\notin\kappa} \geq 0$ is identified as a community when $A_{ij} < -\gamma P_{ij}$ for all $i,j\in\kappa$ (this inequality must hold for sufficiently large $\gamma$ because $P_{i,j\in\kappa} < 0$) and does not split further for larger values of $\gamma$. This poses a particular problem in the interpretation of multiscale community structure obtained with the NGS null network because nodes with negative expected edge weights do not need to be ``densely connected'' in the observed network to contribute positively to modularity. 
	In fact, if one relaxes the assumption of uniform edge weights across categories, one can ensure that nodes in the category with \emph{lowest} intra-category edge weight will $\emph{never}$ split. This is counterintuitive to standard interpretations of multiscale community structure \cite{Lambiotte2010}.

	In Fig.~\ref{FIG4.1}(d,e), we illustrate the above feature of the NGS null network using a simple example. The toy network in Fig.~\ref{FIG4.1}(d) contains 100 nodes divided into three categories: one of size $50$ and two of size $25$. The category of size 50 and one category of size 25 have an intra-category edge weight of $1$ between each pair of nodes. The other category of size $25$ has an intra-category edge weight of $0.4$ between each pair of nodes. All inter-category edges have weights of $-0.05$. (We choose these values so that the intra-category expected edge weight is negative for the third category but positive for the first two and so that inter-category expected edge weights are positive.) We observe in Fig.~\ref{FIG4.1}(e) that the first and second categories split into singletons for sufficiently large $\gamma$, that the smaller of the two categories splits into singletons for a larger value of the resolution parameter, and that the third category never splits. We repeat the same experiment with the U null network in Fig.~\ref{FIG4.1}(f) (after a linear shift of the adjacency matrix to the interval $[0,1]$ using $A_{ij} \mapsto \frac{1}{2}(A_{ij} + 1)$ for all $i$ and $j$), and we observe that the co-classification index of nodes reflects the value of the edge weight between them. It is largest for pairs of nodes in the first and second category, and it is smallest for pairs of nodes in the third category. 
	
		
		\subsection{Data sets}\label{datasets}
		
		We illustrate how the features that we discussed in Section \ref{Subsection4.1} can manifest in real data. We use two data sets of financial time series for our computational experiments. 
		
The first data set, which we call {\sc MultiAssetClasses}, has multiple types of assets and consists of weekly price time series for $N = 98$ financial assets during the time period $01 \text{ Jan } 99$--$01 \text{ Jan } 10$ (resulting in $574$ prices for each asset). The assets are divided into seven asset classes: 20 government bond indices (Gov.), 4 corporate bond indices (Corp.), 28 equity indices (Equ.), 15 currencies (Cur.), 9 metals (Met.), 4 fuel commodities (Fue.), and 18 commodities (Com.). This data set was studied in \cite{Fenn2011} using principal component analysis and a detailed description of the financial assets can be found in that paper.
		
		The second data set, which we call {\sc SingleAssetClass}, consists of daily price time series for $N=859$ financial assets from the Standard \& Poor's (S \& P's) $1500$ index during the time period $01 \text{ Jan } 99$--$01 \text{ Jan } 13$ (resulting in $3673$ prices for each asset).\footnote{We consider fewer than 1500 nodes because we only include nodes for which data is available at all time points to avoid issues associated with choices of data-cleaning techniques.} The financial assets are all equities and are divided into ten sectors: 64 materials, 141 industrials, 150 financials, 142 information technology, 55 utilities, 47 consumer staples, 138 consumer discretionary, 48 energy, 68 health care, and 6 telecommunication services.
		
		The precise way that one chooses to compute a measure of similarity between pairs of time series and the subsequent choices that one makes (e.g., uniform or nonuniform window length, and overlap or no overlap if one uses a \emph{rolling time window}) affect the values of the similarity measure. There are myriad ways to define similarity measures---the best choices depend on facets such as application domain, time-series resolution, and so on---and this is an active and contentious area of research \cite{Schafer2010, Zalesky2012, Potters2005,Smith2011}. Constructing a similarity matrix from a set of time series and investigating community structure in a given similarity matrix are separate problems, and we are concerned with the latter in the present paper. Accordingly, in all of our experiments, we use Pearson correlation coefficients for our measure of similarity. We compute them using a rolling time window with a uniform window length and uniform amount of overlap.

			\begin{figure}	
				\centering
				\subfigure[{\sc MultiAssetClasses}: Surface plot of correlations over all 238 time windows]{
				\includegraphics[width = 6.1cm]{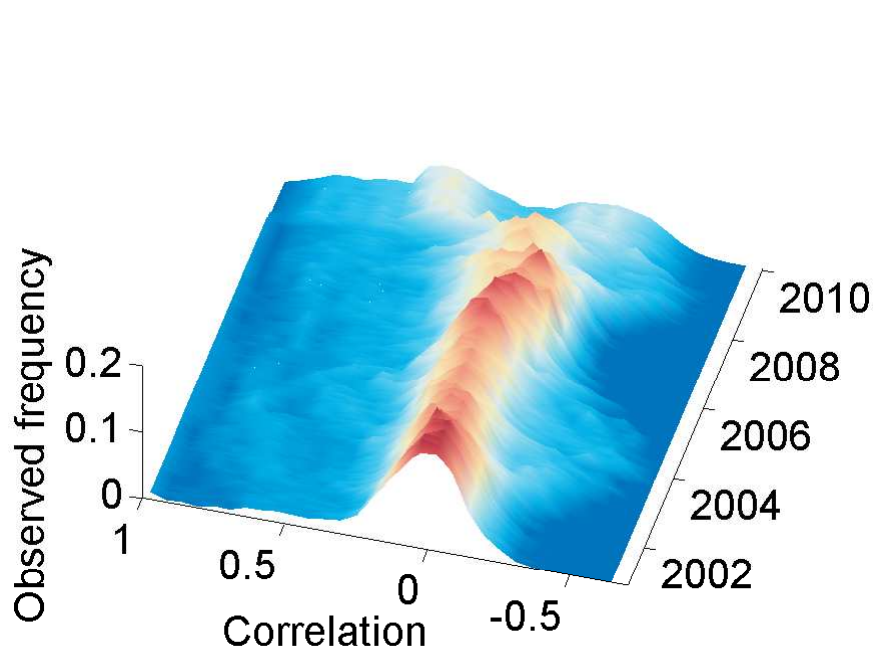}}
				\hspace{1em}
				\subfigure[{\sc SingleAssetClass}: Surface plot of correlations over all 854 time windows]{
				\includegraphics[width = 6.1cm]{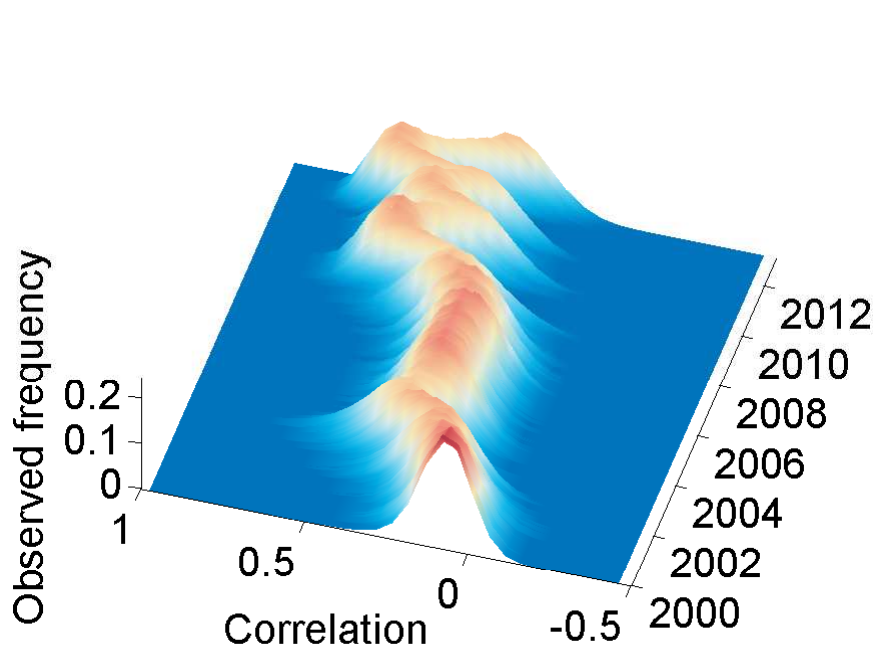}}
					
				\includegraphics[width = 5cm]{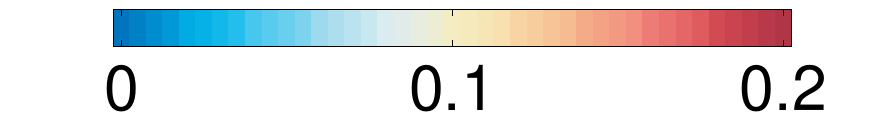}
							
				\caption{Surface plots of the correlations over all time windows for (a) {\sc MultiAssetClasses} data set and (b) {\sc SingleAssetClass} data set. The colors in each panel scale with the value of the observed frequency.		
				}
				\label{FIG3.1}
			\end{figure}
			
		We adopt the same network representation for both data sets. We use the term \emph{time window} for a set of discrete time points and divide each time series into overlapping time windows that we denote by $\mathcal{T} = \{T_s\}$. The length of each time window $\vert T\vert$ and the amount of overlap between consecutive time windows $\vert T\vert-\delta t$ are uniform. The amount of overlap determines the number of data points that one adds and removes from each time window. It thus determines the number of data points that can alter the connectivity patterns in each subsequent correlation matrix (i.e., each subsequent layer). We fix $(\vert T\vert,\delta t) = (100,2)$ for the {\sc MultiAssetClasses} data set (which amounts to roughly two years of data in each time window) and $(\vert T\vert,\delta t) = (260,4)$ for the {\sc SingleAssetClass} data set (which amounts to roughly one year of data in each time window). Every network layer with adjacency matrix $\p{A}_s$ is a Pearson correlation matrix between the time series of logarithmic returns during the time window $T_s$. We take correlations between logarithmic returns because it is standard practice \cite{Cont2001}, but one can also examine correlations between other quantities (such as arithmetic returns \cite{Groppelli2000}). For each data set, we study the sequence of matrices
		\begin{equation*}
				\left\{\p{A}_{s} \in[-1,1]^{N\times N}\vert s\in\{1,\ldots,\vert\mathcal{T}\vert\}\right\}\,.
		\end{equation*}
We show a surface plot of the observed frequency of correlations in each layer for each data set in Fig.~\ref{FIG3.1}.

		 	
	\subsection{Multiscale community structure in asset correlation networks}\label{Subsection4.2}
		
		We perform the same experiments as in Fig.~\ref{FIG4.1} on the correlation matrices of both data sets. Our resolution-parameter sample is the set $\{\gamma^{-},\ldots,\gamma^{+}\}$ (respectively, $\{0,\ldots,\gamma^{+}\}$) for the U and NG (respectively, NGS) null networks with a discretization step of the order of $10^{-3}$. We store the co-classification index of pairs of nodes averaged over all resolution-parameter values in the sample. We use the U and NG null networks for a correlation matrix that is linearly shifted to the interval $[0,1]$. For each null network, we thereby produce $\vert\mathcal{T}\vert$ multiscale association matrices with entries between $0$ and $1$ that indicate how often pairs of nodes are in the same community across resolution-parameter values. 
		\begin{figure}[t]
				
				\center
				\subfigure[Reordered correlation matrix]{
				\includegraphics[width = 2.7cm]{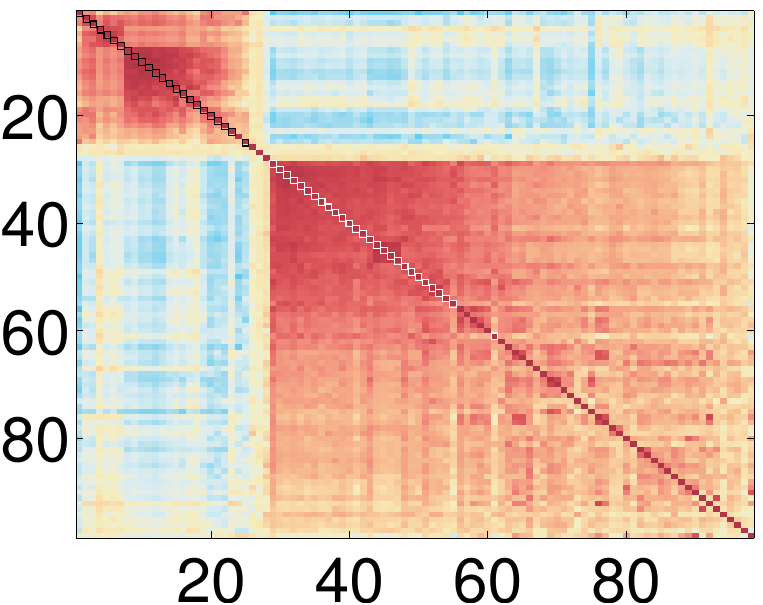}}\hspace{1em}
				\subfigure[Reordered multiscale association matrix (U)]{
				\includegraphics[width = 2.7cm]{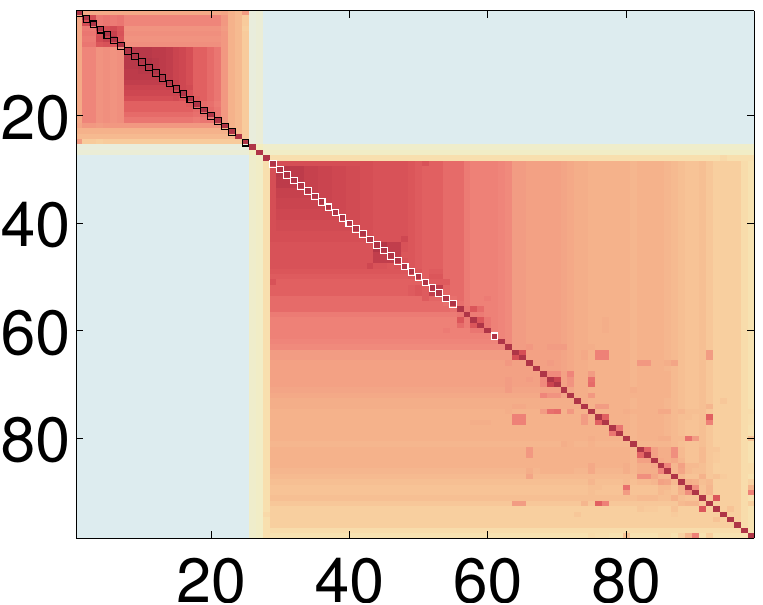}}\hspace{1em}
				\subfigure[Reordered multiscale association matrix (NG)]{
				\includegraphics[width = 2.7cm]{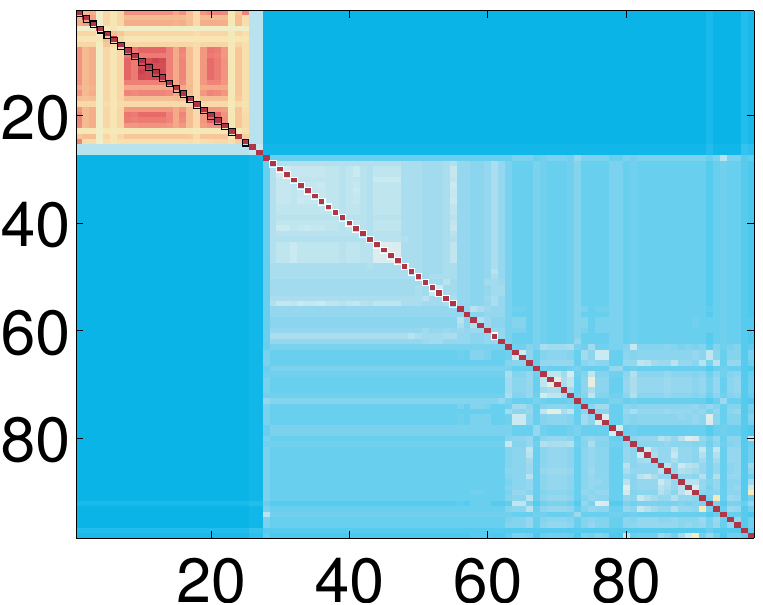}}\hspace{1em}
				\subfigure[Reordered multiscale association matrix (NGS)]{
				\includegraphics[width = 2.7cm]{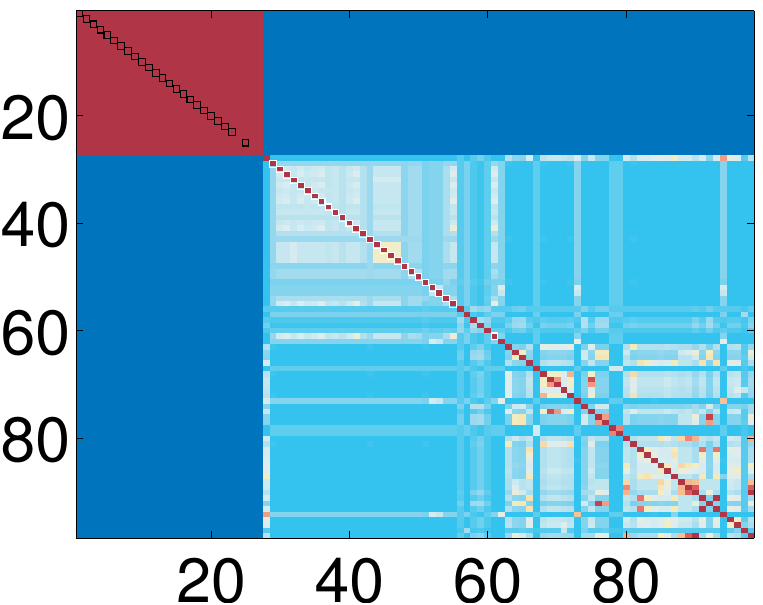}}
				  
				\center
				\subfigure[Reordered correlation matrix]{
				\includegraphics[width = 2.7cm]{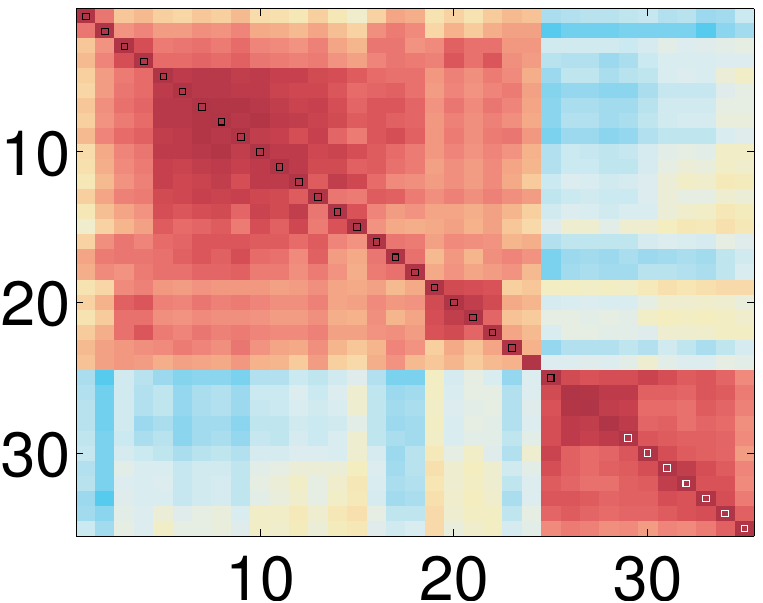}}\hspace{1em}
				\subfigure[Reordered multiscale association matrix (U)]{
				\includegraphics[width = 2.7cm]{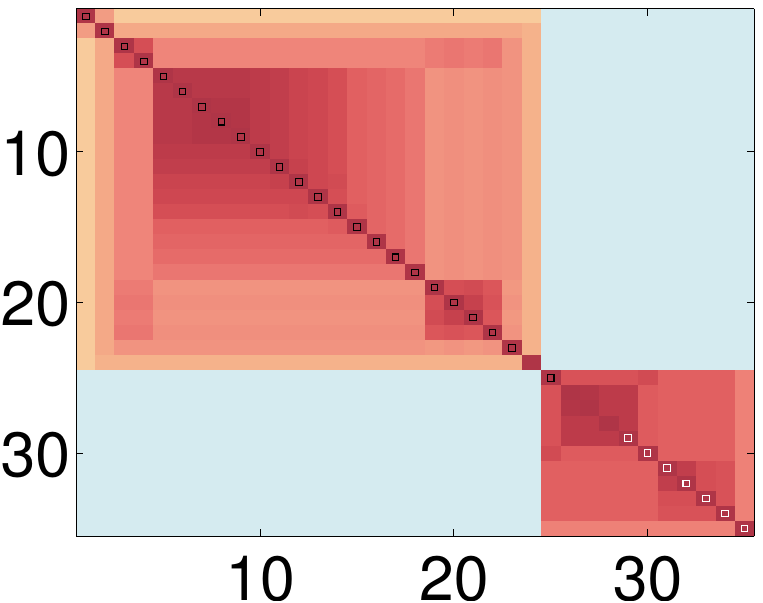}}\hspace{1em}
				\subfigure[Reordered multiscale association matrix (NG)]{
				\includegraphics[width = 2.7cm]{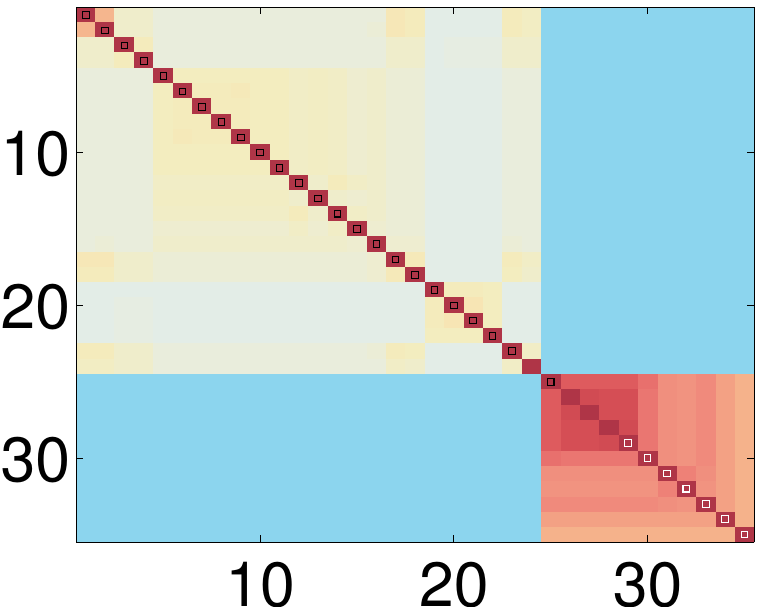}}\hspace{1em}
				\subfigure[Reordered multiscale association matrix (NGS)]{
				\includegraphics[width = 2.7cm]{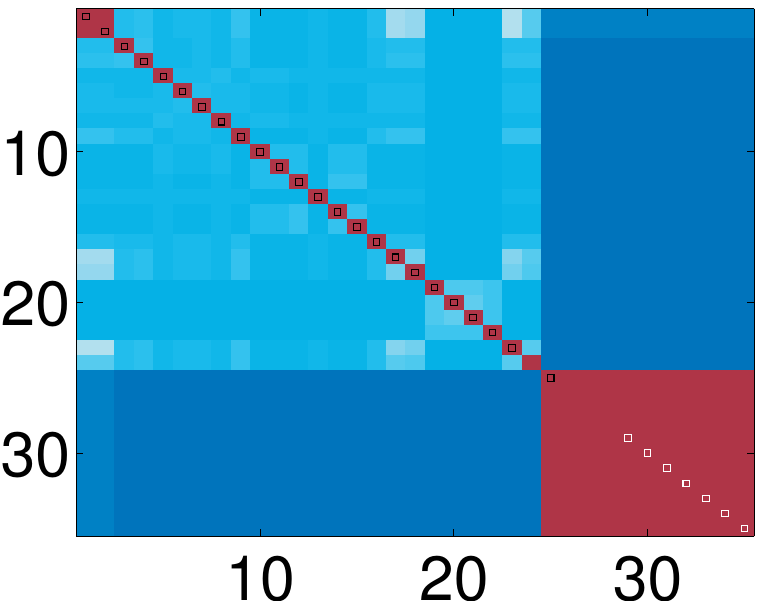}}
				  
				\center\includegraphics[width = 7cm]{figs2/colorbar.pdf}
				  
				\caption{Multiscale association matrix for the U, NG, and NGS null networks for the entire correlation matrix and a subset of the correlation in the last layer of the {\sc MultiAssetClasses} data set. In panel (a), we show the entire matrix; in panels (b,c,d), we show the multiscale association matrix that we obtain from this matrix using each of the three null networks. In panel (e), we show the first $35\times 35$ block of the correlation matrix from panel (a); and in panels (f,g,h), we show the multiscale association matrix that we obtain from this subset of the correlation matrix using each of the three null networks. The colors scale with the entries of the multiscale association and the entries of the correlation matrix. Black squares on the diagonals correspond to government and corporate bond assets, and white squares correspond to equity assets.	
				}
				\label{FIG4.2}
			\end{figure}

		We show the multiscale association matrices for a specific layer of {\sc MultiAssetClasses} in Fig.~\ref{FIG4.2}. The matrix in Fig.~\ref{FIG4.2}(a) corresponds to the correlation matrix during the interval 08 Feb 08--01 Dec 10. In accord with the results in \cite{Fenn2011}, this matrix reflects the increase in correlation between financial assets that took place after the Lehman bankruptcy in 2008 compared to correlation matrices that we compute from earlier time periods. (One can also see this feature in the surface plot of Fig.~\ref{FIG3.1}(a).) The matrices in Fig.~\ref{FIG4.2}(b,c,d) correspond, respectively, to the multiscale association matrix for the U, NG, and NGS null networks. We reorder all matrices (identically) using a node ordering based on the partitions that we obtain with the U null network that emphasizes block-diagonal structure in the correlation matrix. We observe that the co-classification indices in the multiscale association matrix of Fig.~\ref{FIG4.2}(b) are a better reflection of the strength of correlation between assets in Fig.~\ref{FIG4.2}(a) than the multiscale association matrices in Fig.~\ref{FIG4.2}(c,d). As indicated by the darker shades of red in the upper left corner in Fig.~\ref{FIG4.2}(c,d), we also observe that the government and corporate bond assets (which we represent with black squares on the diagonal) are in the same community for a larger range of resolution-parameter values than the range for which equity assets (which we represent with white squares on the diagonal) are in the same community. In fact, when we use an NGS null network, the expected weight between two government or corporate bonds is negative (it is roughly $-0.1$), and these assets are in the same community for arbitrarily large values of the resolution parameter. (In other words, they do not split into smaller communities for large $\gamma$.) One would need to be cautious in using the multiscale association matrices in Fig.~\ref{FIG4.2}(c,d) to gain insight about the connectivity between assets in Fig.~\ref{FIG4.2}(a).

		When studying correlation matrices of multi-asset data sets, one may wish to vary the size of the asset classes included in the data (e.g., by varying the ratio of equity and bond assets). We show how doing this can lead to further misleading conclusions. By repeating the same experiment using only a subset of the correlation matrix (the first 35 nodes), we consider an example where we have inverted the relative sizes of the bond asset class and the equity asset class. As indicated by the darker shades of red in the lower right corner in Fig.~\ref{FIG4.2}(g,h), equity assets now have a larger co-classification index than government and corporate bond assets when using the NG or NGS null networks. If one uses the co-classification index in the multiscale association matrices of Fig.~\ref{FIG4.2}(c,d) (respectively, Fig.~\ref{FIG4.2}(g,h)) to gain information about the observed correlation between equity and bond assets in Fig.~\ref{FIG4.2}(a) (respectively, Fig.~\ref{FIG4.2}(e)), one may draw different conclusions despite the fact that these have not changed. However, the multiscale association matrix with a U null network in Fig.~\ref{FIG4.2}(f) reflects the observed correlation between equity and bond assets in Fig.~\ref{FIG4.2}(e).\footnote{The authors of \cite{Traag2011} showed that a globally optimal partition for a null network called the ``constants Potts model'' (CPM), in which the edge weights are given by a constant that is \emph{independent} of the network, is ``sample-independent''. Their result can be generalized as follows for the U null network (in which expected edge weights are constant but are not independent of the observed network). Suppose that $C_{\mathrm{max}}$ is a partition that maximizes $Q(C\vert\p{A};\p{P};\gamma_1)$ and consider the subgraph induced by the network on a set of communities $C_1,\ldots, C_l \in C_{\mathrm{max}}$. It then follows that $\{C_1\cup C_2 \ldots \cup C_l\}$ maximizes $Q(C\vert\p{\hat{A}};\p{P};\gamma_2)$, where $\p{\hat{A}}$ is the adjacency matrix of the induced subgraph and $\gamma_2 = \gamma_1\langle \p{A}\rangle/\langle \p{\hat{A}}\rangle$. For the CPM null network, the same result holds with $\gamma_1 = \gamma_2$.}

		\begin{figure}[t!]
			\center
			\subfigure[Correlation between multiscale association matrix and adjacency matrix for the {\sc MultiAssetClasses} data set]{\includegraphics[width = 4.5cm]{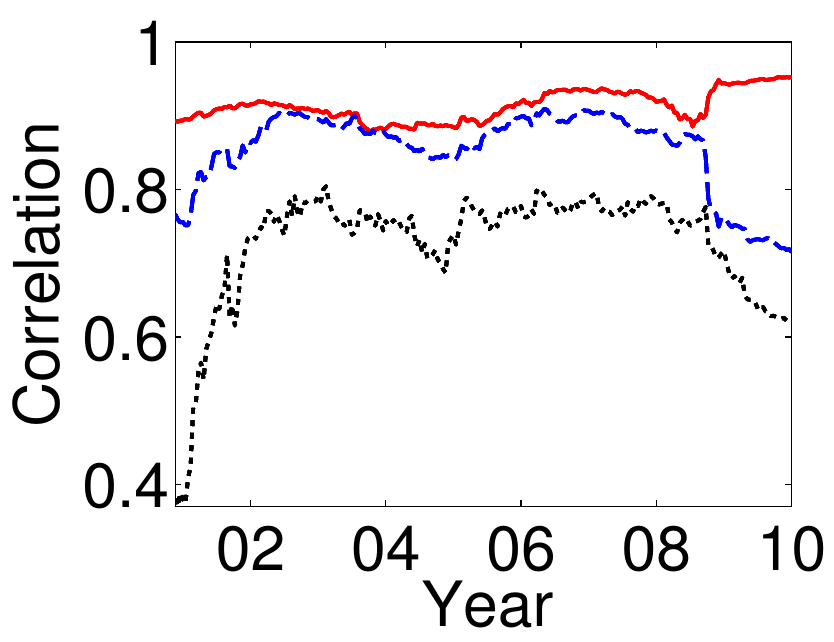}}\hspace{2em}
			\includegraphics[width = 1cm,height = 1.5cm]{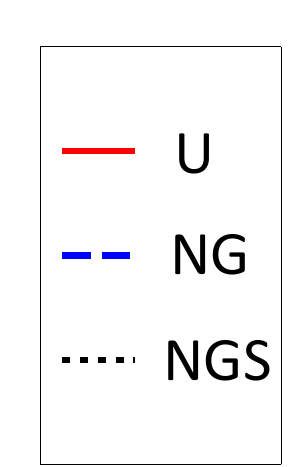}\hspace{2em}
			\subfigure[Correlation between multiscale association matrix and adjacency matrix for the {\sc SingleAssetClass} data set]{\includegraphics[width = 4.5cm]{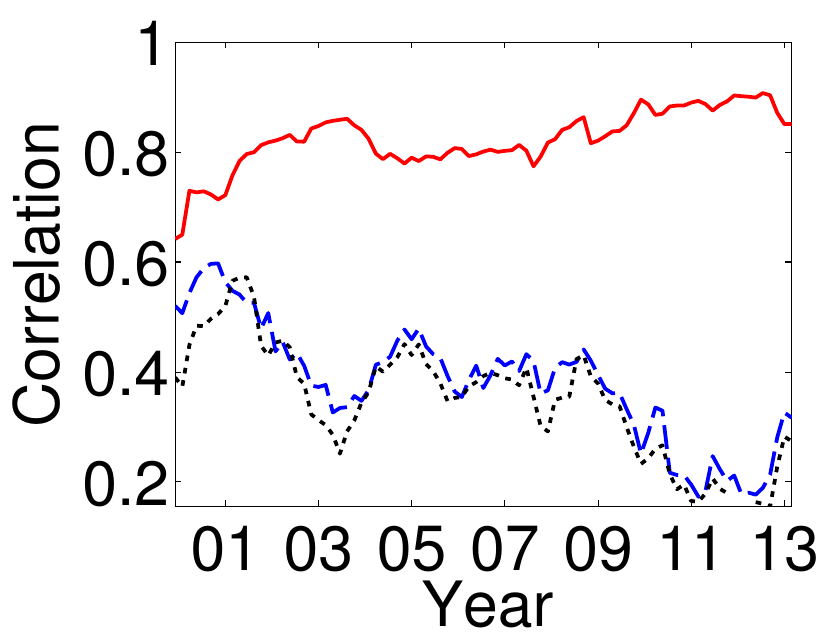}}
					  
			\caption{Correlation between the adjacency matrix and the multiscale association matrix for the U (solid curve), NG (dashed curve), and NGS (dotted curve) null networks over all time layers for (a) the {\sc MultiAssetClasses} data set and (b) the {\sc SingleAssetClass} data set. We compute the Pearson correlation coefficients between entries in the upper diagonal blocks in each matrix (to avoid double counting, as the matrices are symmetric), and we exclude diagonal entries (which, by construction, are equal to 1 in both matrices).}
			  
			\label{FIG4.3}
		\end{figure}

		To quantify the sense in which a multiscale association matrix of one null network ``reflects'' the values in the correlation matrix, we compute the Pearson correlation between the upper triangular part of each multiscale association matrix and its corresponding adjacency matrix across all time layers of both data sets for the U, NG, and NGS null networks. We show these correlation plots in Fig.~\ref{FIG4.3}. Observe that the correlation between the adjacency and multiscale association matrix in Fig.~\ref{FIG4.3}(a,b) is strongest in each layer for the U null network and weakest in (almost) each layer for the NGS null network.
		
		The above observation can be explained as follows. Recall from \eqref{EQ2.5} that we can write the modularity-maximization problem as $\max_{S\in\mathcal{S}}\text{Tr}(\p{S^{T}}\p{B}\p{S})$, where $\mathcal{S}$ is the set of partition matrices. When one uses a U null network, the entries of the modularity matrix are the entries of the adjacency matrix shifted by a constant $\gamma\langle\p{A}\rangle$, and the quality function reduces to
		\begin{equation}
			\max_{\p{S}\in\mathcal{S}}\left[\text{Tr}(\p{S^{T}}\p{A}\p{S}) - \gamma \langle \p{A}\rangle||c(\p{S})||_2 \right]\,,
			\label{EQ4.10}
		\end{equation}
		where $||c(\p{S})||_2 =  ||\text{Tr}(\p{S^{T}}\p{1}_{N}\p{S})||_2$ is the 2-norm of the vector of set sizes in $\p{S}$ (i.e., $c(\p{S})$ is the vector whose $k^{\text{th}}$ entry is $\sum_{i=1}^{N}S_{ik}$).
	It follows that modularity maximization with a U null network is equivalent to a block-diagonalization of the adjacency matrix $\p{A}$ (the first term in \eqref{EQ4.10}) with a penalty on the size of communities (the second term). As one increases the resolution parameter, one favors smaller sets of nodes with stronger internal connectivity. Note that one could also apply equation \eqref{EQ4.10} on adjusted adjacency matrices $\p{A}^{\prime} = \p{A} - \p{\tilde{A}}$. For example, one can let $\p{\tilde{A}}$ be a matrix that controls for random fluctuations in a correlation matrix $\p{A}$ (e.g., the ``random component'' $\p{C}^{r}$ in \cite{MacMahon2013}).

	For a general null network, equation \eqref{EQ4.10} takes the form 
		\begin{equation*}
			\max_{S\in\mathcal{S}}\left[\text{Tr}(\p{S^{T}}\p{A}\p{S}) - \text{Tr}(\p{S^{T}}(\gamma\p{P})\p{S})\right]\,,
		\end{equation*}
	where $\p{P}$ is the adjacency matrix of the null network. That is, modularity maximization finds block-diagonal structure in $\p{A}$ (first term) that is not in $\gamma\p{P}$ (second term). It is common to avoid using the U null network in applications because ``it is not a good representation of most real-world networks'' \cite{Newman2006}. The extent to which one wants a null network to be a good representation of an observed network depends on the features that one wants to take as given. We argue that whether an NG null network is more appropriate than a U null network for a given situation depends at least in part on one's interpretation of node strength for that application. As we discussed in Section \ref{Subsection2.4}, the strength of a node in correlation matrices is given by the covariance between its standardized time series and the mean time series. When using the NG null network, it thus follows that pairwise differences $B_{ij} - B_{i^{'}j^{'}}$ in the modularity quality function depend on $\text{corr}(\hat{z}_i,\hat{z}_j), \text{corr}(\hat{z}_{i^{'}},\hat{z}_{j^{'}})$, and $\text{corr}(\hat{z}_k,\hat{z}_\text{tot})$, where $k \in \{i,j,i^{'},j^{'}\}$, the quantity $\hat{z}_i$ is the standardized time series of asset $i$ defined in subsection \ref{ng-sub}, and $\hat{z}_\text{tot} = \sum_{i=1}^{N}\hat{z}_i$. When using the U null network, pairwise differences in the modularity quality function depend only on the observed edge weights $\text{corr}(\hat{z}_i,\hat{z}_j)$ and $\text{corr}(\hat{z}_{i^{'}},\hat{z}_{j^{'}})$. The term $\text{corr}(\hat{z}_k,\hat{z}_\text{tot})$ introduces a dependence between the communities that one finds using the NG null network and the extent to which nodes in those communities are representative of the mean time series for the sample [as measured by $\text{corr}(\hat{z}_k,\hat{z}_\text{tot})$]. In situations in which one may wish to vary one's node sample (e.g., by changing the size of asset classes), one needs to bear such dependencies in mind when interpreting the communities that one obtains.

		 
		 	 
	\section{Effect of inter-layer coupling on a multilayer partition}\label{Section5}
	
		In Section \ref{Section4}, we set the inter-layer connection weights to $0$ in the multilayer network. The solution to the multilayer modularity-maximization problem \eqref{EQ3.4} then depends solely on the values in the modularity matrix of each time layer, and the multilayer modularity-maximization problem reduces to performing single-layer modularity maximization on each layer independently.
			
		Recall the multilayer modularity-maximization problem 
		\begin{equation*}
			\max_{C\in\mathcal{C}}\Bigg[\sum_{s=1}^{\vert\mathcal{T}\vert}\sum_{i,j = 1}^{N}B_{ijs}\delta(c_{i_s},c_{j_s}) + 	2\omega\sum_{s=1}^{\vert\mathcal{T}\vert-1}\sum_{i=1}^{N}\delta(c_{i_s},c_{i_{s+1}})\Bigg]\,.
		\end{equation*}
		 A solution to this problem is a partition of an $N\vert\mathcal{T}\vert$-node multilayer network. Its communities can contain nodes from the same layer and nodes from different layers. Nodes from different layers can be the same node at different times (($i_s, i_r$) with $s\neq r$) or different nodes at different times (($i_s, j_r$) with $i\neq j$ and $s\neq r$). We say that a node $i_s$ remains in the same community (respectively, changes communities) between consecutive layers $s$ and $s+1$ if $\delta(c_{i_{s}},c_{i_{s+1}}) = 1$ (respectively, $\delta(c_{i_{s}},c_{i_{s+1}}) = 0$). 
		 	 
		Positive ordinal, diagonal, and uniform inter-layer connections favor nodes remaining in the same community between consecutive layers. Every time a node does not change communities between two consecutive layers (i.e., $\delta(c_{i_{s}},c_{i_{s+1}}) = 1$), a positive contribution of $2\omega$ is added to the multilayer quality function. One thereby favors communities that do not to change in time because community assignments are transitive: if $\delta(c_{i_s},c_{j_s}) = 1$ and $\delta(c_{i_s},c_{i_{s+1}}) = \delta(c_{j_s},c_{j_{s+1}}) = 1$, then $\delta(c_{i_{s+1}},c_{j_{s+1}}) = 1$. We define the \emph{persistence} of a multilayer partition to be the total number of nodes that do not change communities between layers:
		\begin{equation}
			\text{Pers}(C) := \sum_{s=1}^{\vert\mathcal{T}\vert-1}\sum_{i=1}^{N}\delta(c_{i_s},c_{i_{s+1}}) \in\{0,\ldots,N(\vert\mathcal{T}\vert-1)\}\,.
			\label{EQ5.1}
		\end{equation}
	As indicated in equation \eqref{EQ5.1}, $\text{Pers}(C)$ is an integer between 0, which occurs when no node ever remains in the same community across layers, and $N(\vert\mathcal{T}\vert-1)$, which occurs when every node always remains in the same community. (See \cite{Bassett2011} for a closely related measure called ``flexibility'' that has been applied to functional brain networks.) Let $\text{Pers}(C)\vert_s$ denote the number of nodes that remain in the same community between two consecutive layers $s$ and $s+1$:
		\begin{equation}
			\text{Pers}(C)\vert_s := \sum_{i=1}^{N}\delta(c_{i_s},c_{i_{s+1}})\in\{0,\ldots,N\}\,,
			\label{EQ5.2}
		\end{equation}
		so that $\text{Pers}(C) = \sum_{s=1}^{\vert\mathcal{T}\vert-1}\text{Pers}(C)\vert_s$. Persistence provides an insightful way of rewriting the multilayer modularity-maximization problem:
		\begin{equation}
				\max_{C\in\mathcal{C}}\left[\sum_{s=1}^{\vert\mathcal{T}\vert}\sum_{i,j = 1}^{N}B_{ijs}\delta(c_{i_s},c_{j_s}) + 2\omega\text{Pers}(C)\right]\,.
				\label{EQ5.3}
		\end{equation}
		The multilayer maximization problem thus measures a trade-off between static community structure within layers (the first term in \eqref{EQ5.3}) and temporal persistence across layers (the second term in \eqref{EQ5.3}). 
		
		To better understand how the partitions that one obtains with nonzero inter-layer coupling (i.e., using $\omega > 0$) can differ from the partitions that one obtains without inter-layer coupling (i.e., using $\omega = 0$), we define notation that helps to compare a multilayer partition to single-layer partitions. Let $\mathcal{N}_s:= \{1_s,\ldots,N_s\}$ denote the set of nodes in layer $s$. The restriction of a set of nodes $C_l\subseteq\{1_1,\ldots,N_1;1_2,\ldots,N_2;\ldots;1_{\vert\mathcal{T}\vert},\ldots,N_{\vert\mathcal{T}\vert}\}$ to a layer $s$ is 
$C_l\vert_s := C_l\cap\mathcal{N}_s$, and we define the \emph{partition induced by a multilayer partition} $C\in\mathcal{C}$ on layer $s$ by
	\begin{equation*}
		C\vert_s := \{C_l\vert_s,  C_l\in C\}\,.
	\end{equation*}
We refer to a ``globally optimal partition'' as an ``optimal partition'' in this section for ease of writing and we refer to  $\sum_{s=1}^{\vert\mathcal{T}\vert}\sum_{i,j = 1}^{N}B_{ijs}\delta(c_{i_s},c_{j_s})$ (i.e., the first term in \eqref{EQ5.3}) as \emph{intra-layer modularity}. In the next two subsections, we illustrate how the set of partitions induced by a multilayer partition with $\omega > 0$ on individual layers can differ from intra-layer partitions obtained with $\omega = 0$.

		
	\subsection{Toy examples}\label{Subsection5.1}
	
	\subsubsection{Changes in connectivity patterns}
\begin{figure}
			\centering
			\subfigure[Layer $s\neq 3,6$]	{
				\begin{tikzpicture}[scale = 1.3]
			
				\fill[black] (0,0) circle (0.75mm) node [left] {\scriptsize $1_{\scalebox{0.8}{s}}$};
				\fill[black] (0.5,0) circle (0.75mm) node [right] {\scriptsize $2_{\scalebox{0.8}{s}}$};		
				\fill[black] (0,-0.5) circle (0.75mm) node [left] {\scriptsize $3_{\scalebox{0.8}{s}}$};
				\fill[black] (0.5,-0.5) circle (0.75mm) node [right] {\scriptsize $4_{\scalebox{0.8}{s}}$};
							
				\fill[black] (0,-1) circle (0.75mm) node [left] {\scriptsize $5_{\scalebox{0.8}{s}}$};
				\fill[black] (0.5,-1) circle (0.75mm) node [right] {\scriptsize $6_{\scalebox{0.8}{s}}$};		
				\fill[black] (0,-1.5) circle (0.75mm) node [left] {\scriptsize $7_{\scalebox{0.8}{s}}$}; 
				\fill[black] (0.5,-1.5) circle (0.75mm) node [right] {\scriptsize $8_{\scalebox{0.8}{s}}$};
						
				\draw[very thin] (0,0) -- (0.5,0); \draw[very thin] (0,0) -- (0,-0.5); \draw[very thin] (0,0) -- (0.5,-0.5);
				\draw[very thin] (0.5,0) -- (0,-0.5); \draw[very thin] (0.5,0) -- (0.5,-0.5);
				\draw[very thin] (0,-0.5) -- (0.5,-0.5);
				\draw[very thin, color=white] (0.6,0) -- (0.8,0); 
				\draw[very thin, color=white] (-0.1,0) -- (-0.2,0);
				
				\draw[very thin] (0,-1) -- (0.5,-1); \draw[very thin] (0,-1) -- (0,-1.5); \draw[very thin] (0,-1) -- (0.5,-1.5);
				\draw[very thin] (0.5,-1) -- (0,-1.5); \draw[very thin] (0.5,-1) -- (0.5,-1.5);
				\draw[very thin] (0,-1.5) -- (0.5,-1.5);
				
				\end{tikzpicture}}\hspace{1cm}
			\subfigure[Layer 3]{
				\begin{tikzpicture}[scale = 1.3]
				
				\fill[black] (0,0) circle (0.75mm) node [left] {\scriptsize $1_{\scalebox{0.8}{3}}$};
				\fill[black] (0.5,0) circle (0.75mm) node [right] {\scriptsize $2_{\scalebox{0.8}{3}}$};		
				\fill[black] (0,-0.5) circle (0.75mm) node [left] {\scriptsize $3_{\scalebox{0.8}{3}}$};
				\fill[black] (0.5,-0.5) circle (0.75mm) node [right] {\scriptsize $4_{\scalebox{0.8}{3}}$};
							
				\fill[black] (0,-1) circle (0.75mm) node [left] {\scriptsize $5_{\scalebox{0.8}{3}}$};;
				\fill[black] (0.5,-1) circle (0.75mm) node [right] {\scriptsize $6_{\scalebox{0.8}{3}}$};;		
				\fill[black] (0,-1.5) circle (0.75mm) node [left] {\scriptsize $7_{\scalebox{0.8}{3}}$};;
				\fill[black] (0.5,-1.5) circle (0.75mm) node [right] {\scriptsize $8_{\scalebox{0.8}{3}}$};;
							
				\draw[very thin] (0,0) -- (0.5,0); \draw[very thin] (0,0) -- (0,-0.5); \draw[very thin] (0,0) -- (0.5,-0.5);
				\draw[very thin] (0.5,0) -- (0,-0.5); \draw[very thin] (0.5,0) -- (0.5,-0.5);
				\draw[very thin] (0,-0.5) -- (0.5,-0.5);
				\draw[very thin] (0,-1) .. controls (-0.5, -0.5).. (0,0);
				\draw[very thin] (0,-1) .. controls (1, -0.5).. (0.5,0);
				
				\draw[very thin] (0.5,-1) -- (0,-1.5); \draw[very thin] (0.5,-1) -- (0.5,-1.5);
				\draw[very thin] (0,-1.5) -- (0.5,-1.5);
				
				\end{tikzpicture}}\hspace{0.7cm}
			\subfigure[Layer 6]{
				\begin{tikzpicture}[scale = 1.3]
			
				\fill[black] (0,0) circle (0.75mm) node [left] {\scriptsize $1_{\scalebox{0.8}{6}}$};
				\fill[black] (0.5,0) circle (0.75mm) node [right] {\scriptsize $2_{\scalebox{0.8}{6}}$};		
				\fill[black] (0,-0.5) circle (0.75mm) node [left] {\scriptsize $3_{\scalebox{0.8}{6}}$};
				\fill[black] (0.5,-0.5) circle (0.75mm) node [right] {\scriptsize $4_{\scalebox{0.8}{6}}$};
							
				\fill[black] (0,-1) circle (0.75mm) node [left] {\scriptsize $5_{\scalebox{0.8}{6}}$};
				\fill[black] (0.5,-1) circle (0.75mm) node [right] {\scriptsize $6_{\scalebox{0.8}{6}}$};		
				\fill[black] (0,-1.5) circle (0.75mm) node [left] {\scriptsize $7_{\scalebox{0.8}{6}}$};
				\fill[black] (0.5,-1.5) circle (0.75mm) node [right] {\scriptsize $8_{\scalebox{0.8}{6}}$};
							
				\draw[very thin] (0,0) -- (0.5,0); \draw[very thin] (0,0) -- (0,-0.5); \draw[very thin] (0,0) -- (0.5,-0.5);
				\draw[very thin] (0.5,0) -- (0,-0.5); \draw[very thin] (0.5,0) -- (0.5,-0.5);
				\draw[very thin] (0,-0.5) -- (0.5,-0.5);
				\draw[very thin] (0,-1) .. controls (-0.5, -0.5).. (0,0);
				\draw[very thin] (0,-1) .. controls (1, -0.5).. (0.5,0);
				\draw[very thin] (0,-1) .. controls (-0.2, -0.75).. (0,-0.5);
				\draw[very thin] (0,-1) .. controls (0.25, -0.75).. (0.5,-0.5);
				
				\draw[very thin] (0.5,-1) -- (0,-1.5); \draw[very thin] (0.5,-1) -- (0.5,-1.5);
				\draw[very thin] (0,-1.5) -- (0.5,-1.5);
				\end{tikzpicture}}
				
		\subfigure[Partition $\mathcal{C}_0$]{
			\begin{tikzpicture}[scale = 0.9]
				\foreach \i in {1,...,10}{ 
					\foreach \j in {1,...,8}{ 
						{\pgfmathsetmacro{\x}{0.3*(\i-1)} 
						\pgfmathsetmacro{\y}{-0.23*(\j-1)} 
					 	\fill[black] (\x,\y) circle (0.07);}}}
				\foreach \i in {1,2,4,5,7,8,9,10}{
						{\pgfmathsetmacro{\first}{1}
						\pgfmathsetmacro{\last}{4}
						\pgfmathsetmacro{\a}{0.3*(\i-1) - 0.06 - 0.06} 
						\pgfmathsetmacro{\b}{0.3*(\i-1) + 0.06 + 0.06}
						\pgfmathsetmacro{\c}{-0.23*(\first-1)} 
						\pgfmathsetmacro{\A}{0.3*(\i-1)}
						\pgfmathsetmacro{\B}{-0.23*(\first-1) + 0.06 + 0.06} 
						\pgfmathsetmacro{\d}{-0.23*(\last-1)} 
						\pgfmathsetmacro{\C}{0.3*(\i-1)}
						\pgfmathsetmacro{\D}{-0.23*(\last-1) - 0.06 - 0.06} 
						
						\draw[thin] (\a, \c) .. controls (\A, \B).. (\b, \c);
						\draw[thin] (\a, \d) .. controls (\C, \D).. (\b, \d);
						\draw[thin] (\a, \c) -- (\a, \d);
						\draw[thin] ((\b, \c) -- (\b, \d);}
						
						{\pgfmathsetmacro{\first}{5}
						\pgfmathsetmacro{\last}{8}
						\pgfmathsetmacro{\a}{0.3*(\i-1) - 0.06 - 0.06} 
						\pgfmathsetmacro{\b}{0.3*(\i-1) + 0.06 + 0.06}
						\pgfmathsetmacro{\c}{-0.23*(\first-1)} 
						\pgfmathsetmacro{\A}{0.3*(\i-1)}
						\pgfmathsetmacro{\B}{-0.23*(\first-1) + 0.06 + 0.06} 
						\pgfmathsetmacro{\d}{-0.23*(\last-1)} 
						\pgfmathsetmacro{\C}{0.3*(\i-1)}
						\pgfmathsetmacro{\D}{-0.23*(\last-1) - 0.06 - 0.06} 
						
						\draw[thin] (\a, \c) .. controls(\A,\B) .. (\b, \c);
						\draw[thin] (\a, \d) .. controls (\C, \D) .. (\b, \d);
						\draw[thin] (\a, \c) -- (\a, \d);
						\draw[thin] ((\b, \c) -- (\b, \d);}}
						
						\foreach \i in {3,6}{
						{\pgfmathsetmacro{\first}{1}
						\pgfmathsetmacro{\last}{5}
						\pgfmathsetmacro{\a}{0.3*(\i-1) - 0.06 - 0.06} 
						\pgfmathsetmacro{\b}{0.3*(\i-1) + 0.06 + 0.06}
						\pgfmathsetmacro{\c}{-0.23*(\first-1)} 
						\pgfmathsetmacro{\A}{0.3*(\i-1)}
						\pgfmathsetmacro{\B}{-0.23*(\first-1) + 0.06 + 0.06} 
						\pgfmathsetmacro{\d}{-0.23*(\last-1)} 
						\pgfmathsetmacro{\C}{0.3*(\i-1)}
						\pgfmathsetmacro{\D}{-0.23*(\last-1) - 0.06 - 0.06} 
						
						\draw[thin] (\a, \c) .. controls (\A, \B).. (\b, \c);
						\draw[thin] (\a, \d) .. controls (\C, \D).. (\b, \d);
						\draw[thin] (\a, \c) -- (\a, \d);
						\draw[thin] ((\b, \c) -- (\b, \d);}
						
						{\pgfmathsetmacro{\first}{6}
						\pgfmathsetmacro{\last}{8}
						\pgfmathsetmacro{\a}{0.3*(\i-1) - 0.06 - 0.06} 
						\pgfmathsetmacro{\b}{0.3*(\i-1) + 0.06 + 0.06}
						\pgfmathsetmacro{\c}{-0.23*(\first-1)} 
						\pgfmathsetmacro{\A}{0.3*(\i-1)}
						\pgfmathsetmacro{\B}{-0.23*(\first-1) + 0.06 + 0.06} 
						\pgfmathsetmacro{\d}{-0.23*(\last-1)} 
						\pgfmathsetmacro{\C}{0.3*(\i-1)}
						\pgfmathsetmacro{\D}{-0.23*(\last-1) - 0.06 - 0.06} 
						
						\draw[thin] (\a, \c) .. controls(\A,\B) .. (\b, \c);
						\draw[thin] (\a, \d) .. controls (\C, \D) .. (\b, \d);
						\draw[thin] (\a, \c) -- (\a, \d);
						\draw[thin] ((\b, \c) -- (\b, \d);}}	
					\end{tikzpicture}}
		\subfigure[Partition $\mathcal{C}_1$]{
			\begin{tikzpicture}[scale = 1.2]
				\foreach \i in {1,...,10}{ 
					\foreach \j in {1,...,4}{ 
						{\pgfmathsetmacro{\x}{0.22*(\i-1)} 
						\pgfmathsetmacro{\y}{-0.17*(\j-1)} 
					 	\fill[RedOrange] (\x,\y) circle (0.8mm);}}}			 	
					\foreach \i in {1,...,10}{ 
						\foreach \j in {5,...,8}{ 
							{\pgfmathsetmacro{\x}{0.22*(\i-1)} 
							\pgfmathsetmacro{\y}{-0.17*(\j-1)} 
					 		\fill[SkyBlue] (\x,\y) circle (0.8mm);}}}	
					 \foreach \i in {3,6}{ 
						\foreach \j in {5}{ 
							{\pgfmathsetmacro{\x}{0.22*(\i-1)} 
							\pgfmathsetmacro{\y}{-0.17*(\j-1)} 
					 		\fill[RedOrange] (\x,\y) circle (0.8mm);}}}	
					 	\end{tikzpicture}}
		\subfigure[Partition $\mathcal{C}_2$]{
		\begin{tikzpicture}[scale = 1.2]
				\foreach \i in {1,...,10}{ 
					\foreach \j in {1,...,4}{ 
						{\pgfmathsetmacro{\x}{0.22*(\i-1)} 
						\pgfmathsetmacro{\y}{-0.17*(\j-1)} 
					 	\fill[RedOrange] (\x,\y) circle (0.8mm);}}}			 	
					\foreach \i in {1,...,10}{ 
						\foreach \j in {5,...,8}{ 
							{\pgfmathsetmacro{\x}{0.22*(\i-1)} 
							\pgfmathsetmacro{\y}{-0.17*(\j-1)} 
					 		\fill[SkyBlue] (\x,\y) circle (0.8mm);}}}	
					\foreach \i in {6}{ 
						\foreach \j in {5}{ 
							{\pgfmathsetmacro{\x}{0.22*(\i-1)} 
						\pgfmathsetmacro{\y}{-0.17*(\j-1)} 
				 		\fill[RedOrange] (\x,\y) circle (0.8mm);}}}	
				 	\end{tikzpicture}}
	\subfigure[Partition $\mathcal{C}_3$]{
	\begin{tikzpicture}[scale = 1.2]
			\foreach \i in {1,...,10}{ 
				\foreach \j in {1,...,4}{ 
					{\pgfmathsetmacro{\x}{0.22*(\i-1)} 
					\pgfmathsetmacro{\y}{-0.17*(\j-1)} 
				 	\fill[RedOrange] (\x,\y) circle (0.8mm);}}}			 	
				\foreach \i in {1,...,10}{ 
					\foreach \j in {5,...,8}{ 
						{\pgfmathsetmacro{\x}{0.22*(\i-1)} 
						\pgfmathsetmacro{\y}{-0.17*(\j-1)} 
				 		\fill[SkyBlue] (\x,\y) circle (0.8mm);}}}	
				 	\end{tikzpicture}}
				 		
		\hspace{-1.5cm}\subfigure[Change in multilayer modularity value with respect to partition $\mathcal{C}_1$]{	
		\includegraphics[width = 5cm]{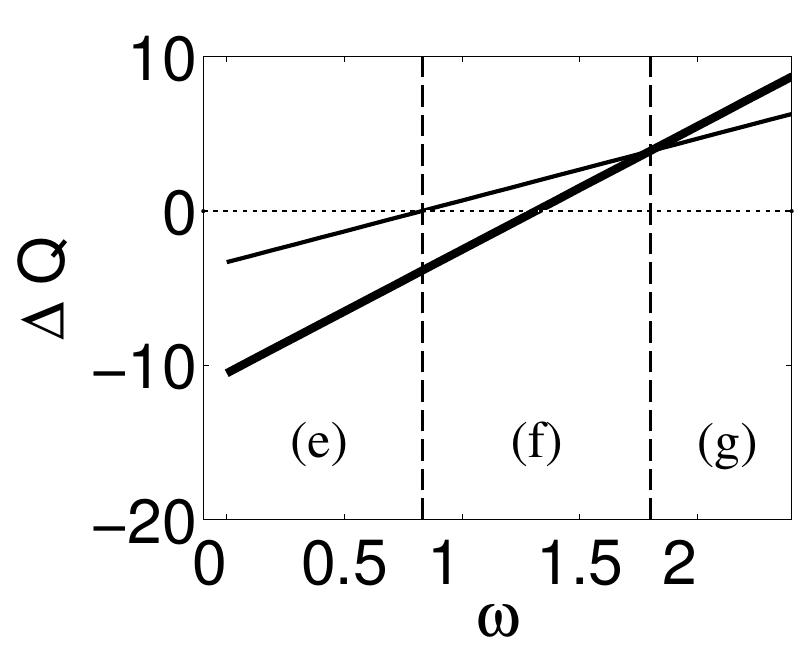}}
		\caption{Toy example illustrating the use of ordinal diagonal and uniform inter-layer coupling for detecting changes in community structure across layers. We consider ten layers ($\vert\mathcal{T}\vert=10$) with eight nodes ($N=8$) in each layer. We show the network structures in (a) layers $s\not\in \{3,6\}$, (b) layer $3$, and (c) layer $6$. Panels (d)--(g) illustrate four different multilayer partitions. In each panel, the $s^{\text{th}}$ column of circles represents the nodes in the $s^{\text{th}}$ layer, which we order from 1 to 8. We show sets of nodes in the same community using solid curves in panel (d) (to avoid having to use $20$ distinct colors) and using colors in panels (e)--(g). In panel (h), we show the difference between the multilayer modularity value between the partition in panels (f) (thin line) and (g) (thick line) and the partition in panel (e) for different values of $\omega$. We include the horizontal dotted line to show the point at which the thin line intercepts the horizontal axis. The panel labels in the regions defined by the area between two consecutive vertical lines in panel (h) indicate which of the multilayer partitions in panels (e), (f), and (g) has a higher value of modularity.}
		 \label{FIG5.1}
\end{figure}
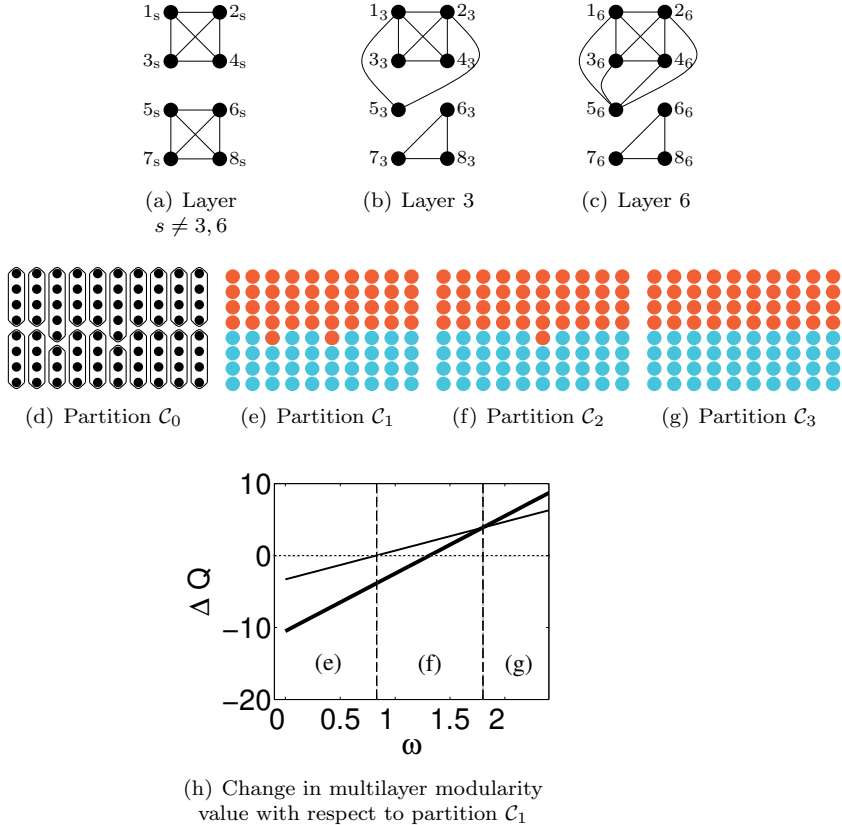

	This toy example illustrates how inter-layer coupling can enable us to detect and differentiate between changes in connectivity patterns across layers. In Fig.~\ref{FIG5.1}, we show an unweighted multilayer network with $\vert\mathcal{T}\vert = 10$ layers and $N = 8$ nodes in each layer. Every layer except for layers $3$ and $6$ contains two 4-node cliques. In layer $3$, node $5_3$ is connected to nodes $\{1_3,2_3\}$ instead of nodes $\{6_3,7_3,8_3\}$.  In layer $6$, node $5_6$ is connected to nodes $\{1_6,2_6,3_6,4_6\}$ instead of nodes $\{6_6,7_6,8_6\}$. We show the layers of the multilayer network in panels (a)--(c) of Fig.~\ref{FIG5.1}. We examine its communities using a U null network with a resolution parameter $\gamma = 1$. Layer $s$ then has the following single-layer modularity matrix:
	\begin{equation*}
		B_{ijs} = \left\{
		\begin{array}{l l}
		1 - \langle \p{A}_s\rangle\,, & \text{if $i$ is connected to $j$}\\
		- \langle \p{A}_s\rangle\,, & \text{otherwise\,.}
		\end{array}
		\right.
	\end{equation*}
	The optimal partition in each layer is unique and is $C_s = \{\{1_s,2_s,3_s,4_s\},\{5_s,6_s,7_s,8_s\}\}$ in layer $s$ for $s\notin \{3, 6\}$ and is $C_s = \{\{1_s,2_s,3_s,4_s,5_s\},\{6_s,7_s,8_s\}\}$ in layers $3$ and $6$. 
	When the value of inter-layer coupling is $0$, the optimal multilayer partition is the union of $\vert\mathcal{T}\vert$ disconnected optimal single-layer partitions. The resulting multilayer partition $\mathcal{C}_0 = \bigcup_{i=1}^{10}C_s$, which we show in panel (d) of Fig.~\ref{FIG5.1}, has a persistence of $\text{Pers}(\mathcal{C}_0) = 0$.
For any $\omega>0$, any partition with the same intra-layer partitions as $\mathcal{C}_0$ and a nonzero value of persistence yields a higher value of multilayer modularity than $\mathcal{C}_0$. This follows immediately from the expression of the multilayer quality function:
	\begin{equation*}
			Q(C\vert\p{\mathcal{B}}) = \sum_{s=1}^{\vert\mathcal{T}\vert}\sum_{i,j = 1}^{N}B_{ijs}\delta(c_{i_s},c_{j_s}) + 2\omega\text{Pers}(C)\,.
	\end{equation*}
	Increasing persistence without changing intra-layer partitions  increases the last term of $Q(C\vert\p{\mathcal{B}})$ without changing the other terms. (In Section \ref{Subsection5.2}, we will prove that $\omega > 0$ is both necessary and sufficient for an optimal partition to have a positive value of persistence.) To obtain the multilayer partition in panel (e), we combine all of the sets in panel (d) that contain $1_s$ into one set and all of the sets that contain $N_s$ into another set. This partition has a persistence equal to  $N(\vert\mathcal{T}\vert-1) - 4$, and any other way of combining the sets in $\mathcal{C}_0$ yields a lower value of persistence.

	We now examine Fig.~\ref{FIG5.1} further. We consider the multilayer partitions in panels (e)--(g). The example in panel (e) shows the structural changes from both layer 3 [see panel (b)] and layer 6 [see panel (c)], the example in panel (f) shows only the change from layer 6, and panel (g) does not show either change. As we quantify shortly below in terms of modularity cost, the change in layer 6 is the ``stronger'' of the two changes. We let $\mathcal{C}_1$ denote the multilayer partition in panel (e), $\mathcal{C}_2$ denote the multilayer partition in panel (f), and $\mathcal{C}_3$ denote the multilayer partition in panel (g). Additionally, note that $\text{Pers}(\mathcal{C}_1) < \text{Pers}(\mathcal{C}_2)< \text{Pers}(\mathcal{C}_3)$.  The value $\omega$ of inter-layer coupling determines which partition of these three has the largest value of multilayer modularity. To see this, we compute the modularity cost of changing static community structure within layers in partition $\mathcal{C}_1$ in favor of persistence. (Such a computation is a multilayer version of the calculations for static networks in \cite{Good2010}.) The intra-layer modularity cost in $\mathcal{C}_1$ of moving node $5_s$ from community $\{1_s,2_s,3_s,4_s,5_s\}$ to community $\{6_s,7_s,8_s\}$  in layers $s\in\{3,6\}$ is
	\begin{align*}	
		\Delta Q(s) &= 2\Bigg(\sum_{j\in \{6,7,8\}}B_{5js} - \sum_{j\in \{1,2,3,4\}}B_{5js}\Bigg) \\ 
			&= \left\{
		\begin{array}{l l}
		-4 + 2\langle \p{A}\rangle_3 \approx -3.3\,, & \text{if $s=3$}\,,\\
		-8 + 2\langle \p{A}\rangle_6 \approx -7.2\,, & \text{if $s=6$}\,.
			\end{array}
		\right.
	\end{align*}
	The inter-layer modularity cost from this move is $+4\omega$ in both cases; the first $+2\omega$ contribution of the $+4\omega$ follows by symmetry of $\p{\mathcal{B}}$, and the second $+2\omega$ contribution of the $+4\omega$ follows from the fact that either move increases persistence by $+2$. Consequently, for $0 < 4\omega <  \vert\Delta Q(3)\vert$, the partition in panel (e) yields a larger value of multilayer modularity than the partitions in (f) and (g). When $\vert\Delta Q(3)\vert < 4\omega < \vert\Delta Q(6)\vert$, the multilayer modularity value of the partition in (f) is larger than those of (e) or (g). Finally, when $4\omega > \vert\Delta Q(6)\vert$, the partition in panel (g) has the largest value of multilayer modularity of the three. When $4\omega = \vert\Delta Q(3)\vert$ (respectively, $4\omega = \vert\Delta Q(6)\vert$), the multilayer partition in panels (e) and (f) (respectively, (f) and (g)) have the same value of multilayer modularity. We illustrate these results in Fig.~\ref{FIG5.1}(h) by plotting $Q(\mathcal{C}_2\vert\p{\mathcal{B}}) - Q(\mathcal{C}_1\vert\p{\mathcal{B}})$ and $Q(\mathcal{C}_3\vert\p{\mathcal{B}}) - Q(\mathcal{C}_1\vert\p{\mathcal{B}})$ against $\omega$. This example is a simple illustration of how inter-layer connections can help distinguish between changes in connectivity patterns: stronger changes (in terms of modularity cost) persist across larger values of inter-layer coupling (see \cite{Peel2014, Onella2014} for other approaches to ``change point detection'' in temporal networks).

	
	\subsubsection{Shared connectivity patterns}
	
	In the previous toy example, the intra-layer partitions induced on each layer by the multilayer partitions in Fig.~\ref{FIG5.1}(e,f,g) are optimal for at least one layer when $\omega = 0$ (see Fig.~\ref{FIG5.1}(d)). This second example illustrates how inter-layer coupling can identify intra-layer partitions that are not optimal for any individual layer when $\omega = 0$ but which reflect connectivity patterns that are shared across layers.

	\begin{figure}
	\centering
				
		\subfigure[Layer 1]	{
		\begin{tikzpicture}[scale = 1.2]
		\fill[black] (0,0) circle (0.8mm) node [left] {\textcolor{black}{\scriptsize $4_{\scalebox{0.8}{1}}$}};
		\fill[black] (0,-0.5) circle (0.8mm) node [left] {\textcolor{black}{\scriptsize $5_{\scalebox{0.8}{1}}$}};		
		\fill[black] (0.5,-0.5) circle (0.8mm) node [right] {\textcolor{black}{\scriptsize $6_{\scalebox{0.8}{1}}$}};
		\draw[very thin] (0,0) -- (0,-0.5);\draw[very thin] (0,-0.5) -- (0.5,-0.5);\draw[very thin] (0.5,-0.5) -- (0,0);

		\fill[black] (0.5,-1.5) circle (0.8mm) node [right] {\textcolor{black}{\scriptsize $7_{\scalebox{0.8}{1}}$}};
		\fill[black] (0,-1.5) circle (0.8mm) node [left] {\textcolor{black}{\scriptsize $8_{\scalebox{0.8}{1}}$}};
		\fill[black] (0,-2) circle (0.8mm) node [left] {\textcolor{black}{\scriptsize $9_{\scalebox{0.8}{1}}$}};	
		\draw[very thin] (0.5,-1.5) -- (0,-1.5);\draw[very thin] (0,-1.5) -- (0,-2);\draw[very thin] (0,-2) -- (0.5,-1.5);
		
		\fill[black] (-1,0) circle (0.8mm) node [right] {\textcolor{black}{\scriptsize $1_{\scalebox{0.8}{1}}$}};
		\fill[black] (-1,-0.5) circle (0.8mm) node [right] {\textcolor{black}{\scriptsize $2_{\scalebox{0.8}{1}}$}};
		\fill[black] (-1.5,-0.5) circle (0.8mm) node [left] {\textcolor{black}{\scriptsize $3_{\scalebox{0.8}{1}}$}};
		\draw[very thin] (-1,0) -- (-1,-0.5);\draw[very thin] (-1,-0.5) -- (-1.5,-0.5);\draw[very thin] (-1.5,-0.5) -- (-1,0);
		
		\fill[black] (-1,-1.5) circle (0.8mm) node [right] {\scriptsize \textcolor{black}{$11_{\scalebox{0.8}{1}}$}};
		\fill[black] (-1.5,-1.5) circle (0.8mm) node [left] {\scriptsize \textcolor{black}{$10_{\scalebox{0.8}{1}}$}};
		\fill[black] (-1,-2) circle (0.8mm) node [right] {\scriptsize \textcolor{black}{$12_{\scalebox{0.8}{1}}$}};
		\draw[very thin] (-1,-1.5) -- (-1.5,-1.5);\draw[very thin] (-1.5,-1.5) -- (-1,-2);\draw[very thin] (-1,-2) -- (-1,-1.5);
		
		\fill[black] (-0.5,-1) circle (0.8mm) node [right] {\scriptsize \textcolor{black}{$13_{\scalebox{0.8}{1}}$}};
		\draw[dashed, very thin] (-0.5,-1) -- (-1,-0.5); \draw[dashed, very thin] (-0.5,-1) .. controls(-0.6,-0.5).. (-1,0);\draw[dashed, very thin] (-0.5,-1) .. controls(-1.2,-0.75) ..(-1.5,-0.5);
		\draw[very thin] (-0.5,-1) .. controls(-1.2,-1.25).. (-1.5,-1.5); \draw[very thin] (-0.5,-1) .. controls(-0.48,-1.5).. (-1,-2);
		\end{tikzpicture}}
		\subfigure[Layer 2]	{
		\begin{tikzpicture}[scale = 1.2]
		\fill[black] (0,0) circle (0.8mm) node [left] {\scriptsize \textcolor{black}{$4_{\scalebox{0.8}{2}}$}};
		\fill[black] (0,-0.5) circle (0.8mm) node [left] {\scriptsize \textcolor{black}{$5_{\scalebox{0.8}{2}}$}};		
		\fill[black] (0.5,-0.5) circle (0.8mm) node [right] {\scriptsize \textcolor{black}{$6_{\scalebox{0.8}{2}}$}};
		\draw[very thin] (0,0) -- (0,-0.5);\draw[very thin] (0,-0.5) -- (0.5,-0.5);\draw[very thin] (0.5,-0.5) -- (0,0);

		\fill[black] (0.5,-1.5) circle (0.8mm) node [right] {\scriptsize \textcolor{black}{$7_{\scalebox{0.8}{2}}$}};
		\fill[black] (0,-1.5) circle (0.8mm) node [left] {\scriptsize \textcolor{black}{$8_{\scalebox{0.8}{2}}$}};
		\fill[black] (0,-2) circle (0.8mm) node [left] {\scriptsize \textcolor{black}{$9_{\scalebox{0.8}{2}}$}};	
		\draw[very thin] (0.5,-1.5) -- (0,-1.5);\draw[very thin] (0,-1.5) -- (0,-2);\draw[very thin] (0,-2) -- (0.5,-1.5);
		
		\fill[black] (-1,0) circle (0.8mm) node [right] {\scriptsize \textcolor{black}{$1_{\scalebox{0.8}{2}}$}};
		\fill[black] (-1,-0.5) circle (0.8mm) node [right] {\scriptsize \textcolor{black}{$2_{\scalebox{0.8}{2}}$}};
		\fill[black] (-1.5,-0.5) circle (0.8mm) node [left] {\scriptsize \textcolor{black}{$3_{\scalebox{0.8}{2}}$}};
		\draw[very thin] (-1,0) -- (-1,-0.5);\draw[very thin] (-1,-0.5) -- (-1.5,-0.5);\draw[very thin] (-1.5,-0.5) -- (-1,0);
		
		\fill[black] (-1,-1.5) circle (0.8mm) node [right] {\scriptsize \textcolor{black}{$11_{\scalebox{0.8}{2}}$}};
		\fill[black] (-1.5,-1.5) circle (0.8mm) node [left] {\scriptsize \textcolor{black}{$10_{\scalebox{0.8}{2}}$}};
		\fill[black] (-1,-2) circle (0.8mm) node [right] {\scriptsize \textcolor{black}{$12_{\scalebox{0.8}{2}}$}};
		\draw[very thin] (-1,-1.5) -- (-1.5,-1.5);\draw[very thin] (-1.5,-1.5) -- (-1,-2);\draw[very thin] (-1,-2) -- (-1,-1.5);
		
		\fill[black] (-0.5,-1) circle (0.8mm) node [left] {\scriptsize \textcolor{black}{$13_{\scalebox{0.8}{2}}$}};
		\draw[dashed, very thin] (0,0) .. controls(-0.4,-0.5).. (-0.5,-1);\draw[dashed, very thin] (0,-0.5) -- (-0.5,-1);\draw[dashed, 	very thin] (0.5,-0.5) ..controls(0.2,-0.75).. (-0.5,-1);
		\draw[very thin] (-0.5,-1) .. controls(-1.2,-1.25).. (-1.5,-1.5); \draw[very thin] (-0.5,-1) .. controls(-0.48,-1.5).. (-1,-2);
		
		\end{tikzpicture}}\hspace{0.5cm}
		\subfigure[Layer 3]	{
		\begin{tikzpicture}[scale = 1.2]
		\fill[black] (0,0) circle (0.8mm) node [left] {\textcolor{black}{\scriptsize $4_{\scalebox{0.8}{3}}$}};
		\fill[black] (0,-0.5) circle (0.8mm) node [left] {\textcolor{black}{\scriptsize $5_{\scalebox{0.8}{3}}$}};		
		\fill[black] (0.5,-0.5) circle (0.8mm) node [right] {\textcolor{black}{\scriptsize $6_{\scalebox{0.8}{3}}$}};
		\draw[very thin] (0,0) -- (0,-0.5);\draw[very thin] (0,-0.5) -- (0.5,-0.5);\draw[very thin] (0.5,-0.5) -- (0,0);

		\fill[black] (0.5,-1.5) circle (0.8mm) node [right] {\textcolor{black}{\scriptsize $7_{\scalebox{0.8}{3}}$}};
		\fill[black] (0,-1.5) circle (0.8mm) node [left] {\textcolor{black}{\scriptsize $8_{\scalebox{0.8}{3}}$}};
		\fill[black] (0,-2) circle (0.8mm) node [left] {\textcolor{black}{\scriptsize $9_{\scalebox{0.8}{3}}$}};	
		\draw[very thin] (0.5,-1.5) -- (0,-1.5);\draw[very thin] (0,-1.5) -- (0,-2);\draw[very thin] (0,-2) -- (0.5,-1.5);
				
		\fill[black] (-1,0) circle (0.8mm) node [right] {\textcolor{black}{\scriptsize $1_{\scalebox{0.8}{3}}$}};
		\fill[black] (-1,-0.5) circle (0.8mm) node [right] {\textcolor{black}{\scriptsize $2_{\scalebox{0.8}{3}}$}};
		\fill[black] (-1.5,-0.5) circle (0.8mm) node [left] {\textcolor{black}{\scriptsize $3_{\scalebox{0.8}{3}}$}};
		\draw[very thin] (-1,0) -- (-1,-0.5);\draw[very thin] (-1,-0.5) -- (-1.5,-0.5);\draw[very thin] (-1.5,-0.5) -- (-1,0);
		
		\fill[black] (-1,-1.5) circle (0.8mm) node [right] {\textcolor{black}{\scriptsize $11_{\scalebox{0.8}{3}}$}};
		\fill[black] (-1.5,-1.5) circle (0.8mm) node [left] {\textcolor{black}{\scriptsize $10_{\scalebox{0.8}{3}}$}};
		\fill[black] (-1,-2) circle (0.8mm) node [right] {\textcolor{black}{\scriptsize $12_{\scalebox{0.8}{3}}$}};
		\draw[very thin] (-1,-1.5) -- (-1.5,-1.5);\draw[very thin] (-1.5,-1.5) -- (-1,-2);\draw[very thin] (-1,-2) -- (-1,-1.5);
		
		\fill[black] (-0.5,-1) circle (0.8mm) node [above] {\textcolor{black}{\scriptsize $13_{\scalebox{0.8}{3}}$}};
		\draw[dashed,very thin] (-0.5,-1) -- (0,-1.5); \draw[dashed, very thin] (-0.5,-1) ..controls(0.25,-1.25).. (0.5,-1.5); \draw[dashed, very thin] (-0.5,-1) ..controls(-0.45,-1.5).. (0,-2);
		\draw[very thin] (-0.5,-1) .. controls(-1.2,-1.25).. (-1.5,-1.5); \draw[very thin] (-0.5,-1) .. controls(-0.48,-1.5).. (-1,-2);
		\end{tikzpicture}}\hspace{0.5cm}
		
			\subfigure[Partition $\mathcal{C}_1$] { 
		\begin{tikzpicture}[scale = 1.5]
		\foreach \i in {1,...,3}{ 
			\foreach \j in {1,...,3}{ 
				{\pgfmathsetmacro{\x}{0.22*(\i-1)} 
				\pgfmathsetmacro{\y}{-0.17*(\j-1)} 
			 	\fill[RedOrange] (\x,\y) circle (0.8mm);}}}			 	
			\foreach \i in {1,...,3}{ 
				\foreach \j in {4,...,6}{ 
					{\pgfmathsetmacro{\x}{0.22*(\i-1)} 
					\pgfmathsetmacro{\y}{-0.17*(\j-1)} 
			 		\fill[SkyBlue] (\x,\y) circle (0.8mm);}}}	
			 \foreach \i in {1,...,3}{ 
				\foreach \j in {7,...,9}{ 
					{\pgfmathsetmacro{\x}{0.22*(\i-1)} 
					\pgfmathsetmacro{\y}{-0.17*(\j-1)} 
			 		\fill[gray] (\x,\y) circle (0.8mm);}}}	
			 \foreach \i in {1,...,3}{ 
				\foreach \j in {10,...,12}{ 
					{\pgfmathsetmacro{\x}{0.22*(\i-1)} 
					\pgfmathsetmacro{\y}{-0.17*(\j-1)} 
			 		\fill[lightgray] (\x,\y) circle (0.8mm);}}}
			 \foreach \i in {1}{ 
				\foreach \j in {13}{ 
					{\pgfmathsetmacro{\x}{0.22*(\i-1)} 
					\pgfmathsetmacro{\y}{-0.17*(\j-1)} 
			 		\fill[RedOrange] (\x,\y) circle (0.8mm);}}}		
			 \foreach \i in {2}{ 
				\foreach \j in {13}{ 
					{\pgfmathsetmacro{\x}{0.22*(\i-1)} 
					\pgfmathsetmacro{\y}{-0.17*(\j-1)} 
			 		\fill[SkyBlue] (\x,\y) circle (0.8mm);}}}	
			 \foreach \i in {3}{ 
				\foreach \j in {13}{ 
					{\pgfmathsetmacro{\x}{0.22*(\i-1)} 
					\pgfmathsetmacro{\y}{-0.17*(\j-1)} 
			 		\fill[gray] (\x,\y) circle (0.8mm);}}}
			 		\draw[white] (-0.75,-2.2) -- (1.25,-2.2);
			 	\end{tikzpicture}}\hspace{-2.5em}
			\subfigure[Partition $\mathcal{C}_2$] { 
		\begin{tikzpicture}[scale = 1.5]
		\foreach \i in {1,...,3}{ 
			\foreach \j in {1,...,3}{ 
				{\pgfmathsetmacro{\x}{0.22*(\i-1)} 
				\pgfmathsetmacro{\y}{-0.17*(\j-1)} 
			 	\fill[RedOrange] (\x,\y) circle (0.8mm);}}}			 	
			\foreach \i in {1,...,3}{ 
				\foreach \j in {4,...,6}{ 
					{\pgfmathsetmacro{\x}{0.22*(\i-1)} 
					\pgfmathsetmacro{\y}{-0.17*(\j-1)} 
			 		\fill[SkyBlue] (\x,\y) circle (0.8mm);}}}	
			 \foreach \i in {1,...,3}{ 
				\foreach \j in {7,...,9}{ 
					{\pgfmathsetmacro{\x}{0.22*(\i-1)} 
					\pgfmathsetmacro{\y}{-0.17*(\j-1)} 
			 		\fill[gray] (\x,\y) circle (0.8mm);}}}	
			 \foreach \i in {1,...,3}{ 
				\foreach \j in {10,...,13}{ 
					{\pgfmathsetmacro{\x}{0.22*(\i-1)} 
					\pgfmathsetmacro{\y}{-0.17*(\j-1)} 
			 		\fill[lightgray] (\x,\y) circle (0.8mm);}}}
			 	\draw[white] (-0.75,-2.2) -- (1.25,-2.2);
			 	\end{tikzpicture}}\hspace{1em}
\hspace{-1.1cm}\vspace{1em}
	\subfigure[Change in multilayer modularity value with respect to partition $\mathcal{C}_1$]{	
		\includegraphics[width = 5cm]{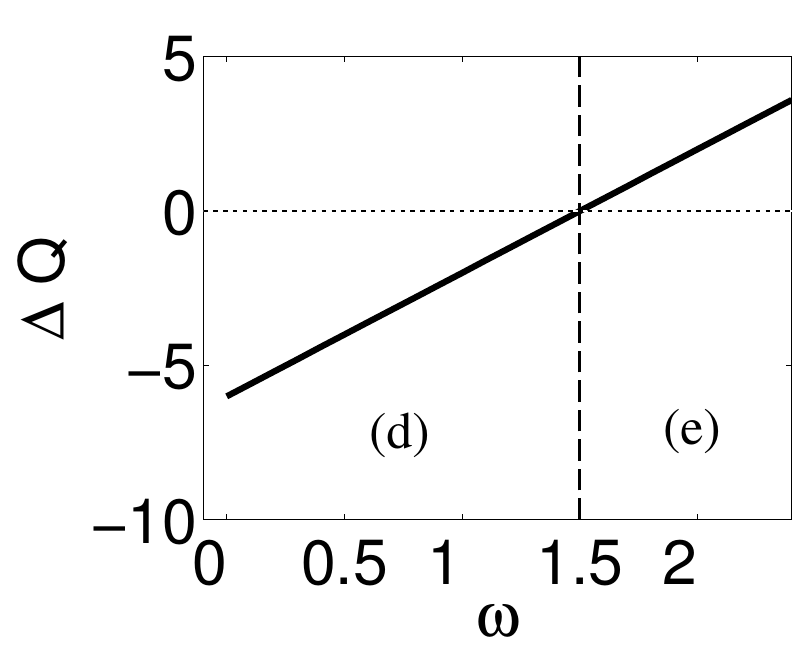}}
\caption{Toy example illustrating the use of ordinal diagonal and uniform  inter-layer coupling for detecting shared connectivity patterns across layers. We consider three layers ($\vert\mathcal{T}\vert = 3$) with thirteen nodes ($N=13$) in each layer. We show the network structures in (a) layer $1$, (b) layer $2$, and (c) layer $3$. Solid lines represent edges present in all three layers and dashed lines represent edges that are only present in one of the layers. Panels (d) and (e) illustrate two different multilayer partitions. In each panel, the $s^{\text{th}}$ column of circles represents the nodes in the $s^{\text{th}}$ layer, which we order $1$ to $13$. We show sets of nodes in the same community using colors in panels (d) and (e). In panel (f), we show the difference between the multilayer modularity value between the partition in panel (e) and the partition in panel (d) for different values of $\omega$. We include the horizontal dotted line to show the point at which the line intercepts the  horizontal axis. The panel labels in the regions defined by the area between two consecutive vertical lines in panel (f) indicate which of the multilayer partitions in panels (d) and (e) has a larger value of multilayer modularity.}
	\label{FIG5.2}
\end{figure}
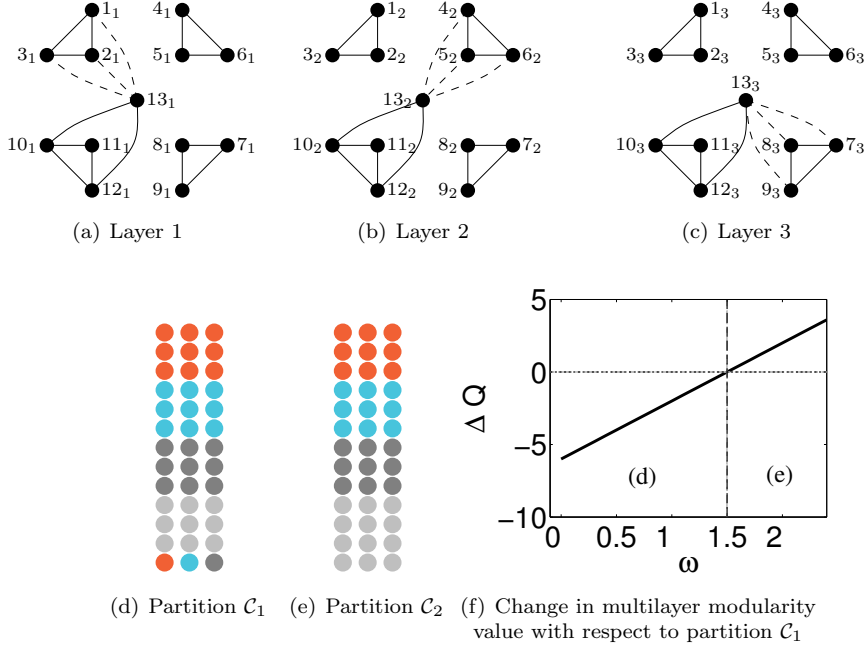

	In Fig.~\ref{FIG5.2}, we consider an unweighted multilayer network with $\vert\mathcal{T}\vert = 3$ layers and $N=13$ nodes in each layer. Every $s^{\text{th}}$ layer contains four 3-node cliques and a node that is connected to each of the three nodes in the $s^{\text{th}}$ clique, and to nodes $10_s$ and $12_s$ in the $4^{\text{th}}$ clique. We show the layers of the multilayer network in panels (a)--(c). We examine its communities using a U null network with a resolution-parameter value of $\gamma=1$. The optimal partition in each layer is unique and is $\{\{1_1,2_1,3_1,13_1\}, \{4_1, 5_1,6_1\},  \{7_1,8_1,9_1\}, \{10_1,11_1,12_1\}\}$ for layer 1, $\{\{1_2,2_2,3_2\}, \\ \{4_2, 5_2,6_2,13_2\}, \{7_2,8_2,9_2\}, \{10_2,11_2, 12_2\}\}$ for layer 2, and $\{\{1_3,2_3,3_3\}, \{4_3, 5_3,6_3\}, \\ \{7_3,8_3,9_3,13_3\}, \{10_3,11_3,12_3\}\}$ for layer 3. We obtain the multilayer partition $\mathcal{C}_1$ in panel (d) by combining these sets such that induced intra-layer partitions are optimal for each layer when $\omega = 0$ and persistence is maximized between layers. The multilayer partition $\mathcal{C}_2$ in panel (e) reflects connectivity patterns that are shared by all layers (i.e., node $13_s$ is with the fourth 3-node clique instead of the $s^{\text{th}}$ 3-node clique); but its intra-layer partitions are not optimal for any layer when $\omega=0$. By carrying out similar calculations to those in the previous toy example, one can show that when $\omega > 3/2$,\footnote{i.e., when $4\omega + 6[2(1-\langle \p{A}\rangle_{s}) - \langle \p{A}\rangle_{s} - 3(1 - \langle \p{A}\rangle_{s})]>0$, with $\langle \p{A}\rangle_{1} = \langle \p{A}\rangle_{2} = \langle \p{A}\rangle_{3}$ by construction in this example.} the multilayer partition in panel (e) yields a larger modularity value than the multilayer partition in panel (d). We illustrate this result in Fig.~\ref{FIG5.2}(f) by plotting $Q(\mathcal{C}_2\vert\p{\mathcal{B}}) - Q(\mathcal{C}_1\vert\p{\mathcal{B}})$ against $\omega$. This example is a simple illustration of how inter-layer connections can help identify connectivity patterns that are shared across layers.\\

		 
	\subsection{Some properties of multilayer partitions}\label{Subsection5.2}	
	
	We now ask how introducing positive ordinal diagonal and uniform coupling (i.e., $\omega > 0$) alters the set of maximum-modularity partitions of static networks (i.e., the case $\omega = 0$). To clearly differentiate between intra-layer and inter-layer modularity contributions, we denote the quality function by
	\begin{equation*}
		Q(C\vert\p{B_1},\ldots,\p{B}_{\vert\mathcal{T}\vert};\omega) := \sum_{s=1}^{\vert\mathcal{T}\vert}\sum_{i,j = 1}^{N}B_{ijs}\delta(c_{i_s},c_{j_s}) + 2\omega\text{Pers}(C)
	\end{equation*}
	instead of $Q(C\vert\p{\mathcal{B}})$, where $C\in\mathcal{C}$ is a multilayer partition. We assume throughout this section that $\vert\mathcal{T}\vert \geq 2$. Let $\mathcal{C}_{\text{{max}}}(\omega)$ denote the set of optimal partitions for the multilayer modularity-maximization problem \eqref{EQ5.3}, and let $C_{\text{{max}}}^{\omega}$ be an arbitrary partition in $\mathcal{C}_{\text{{max}}}(\omega)$. In the discussion to follow, it will be helpful to recall our assumption that each of the partitions in the set $\mathcal{C}$ contains sets that do not have multiple connected components in the weighted graph with adjacency matrix $\p{\mathcal{B}}$. In particular, this applies to partitions $C_{\text{{max}}}^{\omega}\in\mathcal{C}_{\text{{max}}}(\omega)$. We prove several propositions that hold for an arbitrary choice of the matrices $\p{B}_s$ (for example, if one uses the modularity quality function with a U null network and a resolution parameter value of 1, then $\p{B}_s = \p{A}_s - \langle \p{A}_s\rangle\p{1}_{N}$).
	
	\vspace{0.25cm}
	\begin{proposition} \label{P1}
	$\mathrm{Pers}(C_{\mathrm{max}}^{\omega}) > 0  \Leftrightarrow \omega > 0$\,.
	\end{proposition}
	\vspace{0.25cm}
	
	Proposition \ref{P1} ensures that as soon as (and only when) the value of $\omega$ is strictly positive, the value of persistence of an optimal solution is also positive. To prove this, it suffices to observe that if one rearranges sets in a multilayer partition by combining some of the sets into the same set \emph{without} changing the partitions induced on individual layers, then one only changes the value of persistence in the expression of multilayer modularity. For example, this phenomenon occurs in Fig.~\ref{FIG5.1} when going from the partition in panel (d) to the partition in panel (e).

	 \vspace{0.25cm}
	\begin{proof}	\\
	$\Rightarrow$: We prove the contrapositive. Assume that $\omega = 0$ and consider a multilayer partition $C$ such that $\text{Pers}(C) > 0$. The partition $C$ contains at least one set with multiple connected components (because $\text{Pers}(C)>0$ and nodes in different layers are not connected), and $C$ is not optimal by our assumption that global optima do not contain sets with multiple connected components.
	\\ \\
	$\Leftarrow$: Assume that $\omega > 0$ and consider a multilayer partition $C$ such that $\text{Pers}(C) = 0$. We will show that $C$ is not optimal. Let $i_r$ be an arbitrary node in $\{1_1,\ldots,N_1;\ldots;1_{\vert\mathcal{T}\vert},\ldots,N_{\vert\mathcal{T}\vert}\}$, and let $C_{i_r}$ denote the set in $C$ that contains $i_r$. Let $C^{'}$ be the partition obtained from $C$ by combining all sets that contain $i_s$, for some $s$, into one set:
	\begin{displaymath}
		C^{'} = \bigg( C\setminus \bigcup_{s=1}^{\vert\mathcal{T}\vert}\{C_{i_s}\}\bigg) \cup \bigg\{\bigcup_{s=1}^{\vert\mathcal{T}\vert}C_{i_s}\bigg\}\,,
	\end{displaymath}
	where $C_{i_s}$ denotes the set in $C$ that contains $i_s$.
	Consequently,
	\begin{displaymath}
		Q(C^{'}\vert \p{B}_1,\ldots,\p{B}_{\vert\mathcal{T}\vert}; \omega) \geq Q(C\vert \p{B}_1,\ldots,\p{B}_{\vert\mathcal{T}\vert}; \omega) + 2\omega(\vert\mathcal{T}\vert-1)\,,
	\end{displaymath}
	so $C$ is not optimal. 
	(Note that $2\omega(\vert\mathcal{T}\vert-1)$ is strictly positive for $\omega > 0$ because we have assumed that $\vert\mathcal{T}\vert\geq 2$.)
	\end{proof}

	\vspace{0.25cm}
	\begin{proposition}\label{P2} 
	\begin{equation*}
		\text{If } C_l\vert_r = \varnothing \text{ for some } r\in\{1,\ldots,\vert\mathcal{T}\vert-1\}, \text{ then } C_l\vert_s = \varnothing \text{ for all } s>r\,,
	\end{equation*}
	where $C_l\in C_{\mathrm{max}}^{\omega}$ and $C_{\mathrm{max}}^{\omega} \in \mathcal{C}_{\mathrm{max}}({\omega})$.
	\end{proposition}
	\vspace{0.25cm}
	
	Proposition \ref{P2} ensures that if a community becomes empty in a given layer, then it remains empty in all subsequent layers. We omit the proof as this result follows directly from the sparsity pattern of $\p{\mathcal{B}}$ and our assumption that optimal partitions do not contain sets with multiple connected components in the graph with adjacency matrix $\p{\mathcal{B}}$.

	\vspace{0.25cm}
	\begin{proposition}\label{P3}
	$C_{\max}^{\omega}\vert_s = C_{\max}^{\omega}\vert_{s+1} \Leftrightarrow \mathrm{Pers}(C_{\mathrm{max}}^{\omega})\vert_s = N$.
	\end{proposition}
	\vspace{0.25cm}
	
	Proposition \ref{P3} connects the notion of persistence between a pair of layers to the notion of change in community structure within layers. 
	Various numerical experiments that have been performed with ordinal diagonal and uniform inter-layer coupling consist of varying the value of $\omega$ and using information about when nodes change communities between layers as an indication of change in community structure within these layers \cite{Mucha2010, Bassett2011,Bassett2012}. The equivalence relation in Proposition \ref{P3} motivates the use of $\text{Pers}(C)\vert_s$ (or a variant thereof) as an indication of intra-layer change in community structure. 	  
	\vspace{0.25cm} 
	\begin{proof} \\
		$\Leftarrow$: This follows straightforwardly by transitivity of community assignments: if $\delta(c_{j_{s}}, c_{j_{s+1}}) = \delta(c_{i_{s}},c_{i_{s+1}}) = 1$ for all $i,j$, then $\delta(c_{i_s},c_{j_s}) =1$ if and only if $\delta(c_{i_{s+1}},c_{j_{s+1}}) = 1$
	for all $i,j$. (This direction holds for \emph{any} multilayer partition; it need not be optimal.)\newline \\
	$\Rightarrow$: Let $C\in\mathcal{C}$ be a multilayer partition such that $C\vert_s = C\vert_{s+1}$ and $\text{Pers}(C)\vert_s < N$ for some $s\in\{1,\ldots ,\vert\mathcal{T}\vert\}$. We show that $C$ is not optimal.\footnote{Imposing $\text{Pers}(C)\vert_s = N$ by setting $c_{i_{s+1}} = c_{i_s}$ is not sufficient because changing $\text{Pers}(C)\vert_s$ locally can change $\text{Pers}(C)\vert_{s+1}$ or $\text{Pers}(C)\vert_{s-1}$.} Consider a set $C_l\in C$ of nodes such that $C_l\vert_s\neq\varnothing$. If $\delta(c_{i_s},c_{i_{s+1}}) = 1$ (respectively, $\delta(c_{i_s},c_{i_{s+1}}) = 0$) for some $i_s\in C_l\vert_s$, then $\delta(c_{j_s},c_{j_{s+1}}) = 1$ (respectively, $\delta(c_{j_s},c_{j_{s+1}}) = 0$) for all $j_s\in C_l\vert_s$ by transitivity of community assignments and because $C\vert_s = C\vert_{s+1}$ by hypothesis. Because $\text{Pers}(C)\vert_s < N$ by hypothesis, there exists at least one set $C_k\vert_s$ of nodes (where $C_k\in C$) such that $\delta(c_{i_s}, c_{i_{s+1}}) = 0$ for all $i_s\in C_k\vert_s$. Let $C_m\vert_{s+1}$, with $C_m\in C$, denote the set of nodes in layer $s+1$ that contains $i_{s+1}$ for all $i_s\in C_k\vert_s$. Consider the set $\cup_{r\leq s}C_k\vert_r$ of nodes in $C_k$ that are in layers $\{1,\ldots,s\}$ and the set $\cup_{r>s}C_m\vert_r$ of nodes in $C_m$ that are in layers $\{s+1,\ldots,\vert\mathcal{T}\vert\}$. Because $\delta(c_{i_{s}},c_{i_{s+1}}) = 0$ for all $i_s\in C_k\vert_s$, it follows by Proposition \ref{P2} that $C_k = \cup_{r\leq s}C_k\vert_{r}$ and $C_m = \cup_{r>s}C_m\vert_{r}$. Define the partition $C^{'}$ by
	\begin{equation*}
		C^{\prime} = \bigg(C\setminus\big(\{C_k\}\cup \{C_m\}\big)\bigg)\bigcup\bigg(\{C_k\cup C_m\}\bigg)\,.
	\end{equation*}
	This partition satisfies $C^{\prime}\vert_r = C\vert_r$ for all $r\in\{1,\ldots,\vert\mathcal{T}\vert\}$,  $\text{Pers}(C^{\prime})\vert_r = \text{Pers}(C)\vert_r$ for all $r\neq s$, and $\text{Pers}(C^{\prime})\vert_s > \text{Pers}(C)\vert_s$. It follows that $Q(C^{'}\vert\p{B}_1,\ldots,\p{B}_{\vert\mathcal{T}\vert};\omega) > Q(C\vert\p{B}_1,\ldots,\p{B}_{\vert\mathcal{T}\vert};\omega)$ and $C$ is not optimal. 
	\end{proof}	
	\vspace{0.25cm}	
		
	Propositions \ref{P1}, \ref{P2}, and \ref{P3} apply to an optimal partition obtained with any positive value of $\omega$. The next two propositions concern the existence of ``boundary'' values for $\omega$.

		\vspace{0.25cm}
		\begin{proposition}\label{P4}
		There exists $\omega_0 > 0$ such that
		\begin{equation*}
		\text{if } \omega < \omega_0\,, \text{ then } \bigcup_{s=1}^{\vert\mathcal{T}\vert}C_\mathrm{max}^{\omega}\vert_s \in \mathcal{C}_{\mathrm{max}}(0)\,.
		\end{equation*}
		Moreover, we show that $\omega_0 = \Delta Q/\big[2N(\vert\mathcal{T}\vert-1)\big]$, where $\Delta Q$ is the difference between $Q(C_{\mathrm{max}}^{0}\vert\p{B}_1,\ldots,\p{B}_{\vert\mathcal{T}\vert}; 0)$ and the second largest value of $Q(C\vert\p{B}_1,\ldots,\p{B}_{\vert\mathcal{T}\vert}; 0)$ among the partitions of $\mathcal{C}$. 
		\end{proposition}\\
	
Proposition \ref{P4} reinforces the idea of thinking of $\omega$ as the cost of breaking static community structure within layers in favor of larger values of persistence across layers. It demonstrates that there is a positive value of inter-layer coupling such that for any smaller coupling, multilayer modularity maximization only gives more information than single-layer modularity maximization in that it identifies the set of partitions in $\mathcal{C}_\mathrm{max}(0)$ with largest persistence. The proof of this property relies on the fact that the set of possible modularity values for a given modularity matrix is finite. 	
	\vspace{0.25cm}
		\begin{proof}\\
	Let $C$ be an arbitrary partition such that $\cup_{s=1}^{\vert\mathcal{T}\vert}C\vert_s\notin\mathcal{C}_{\mathrm{max}}(0)$. We will show that there exists a value $\omega_0$ of the inter-layer coupling parameter $\omega$ such that $C$ is never optimal for any inter-layer coupling less than $\omega_0$. Given a sequence of single-layer modularity matrices $\{\p{B}_1,\ldots,\p{B}_{\vert\mathcal{T}\vert}\}$, the set of possible multilayer modularity values for a fixed value of $\omega>0$ is finite and is given by 
	\begin{displaymath}
		\mathcal{Q}_{\omega} = \left\{Q(C\vert\p{B}_1,\ldots,\p{B}_{\vert\mathcal{T}\vert}; \omega)\vert C\in\mathcal{C}\right\}\,,
	\end{displaymath}
	where $C$ is a multilayer partition. Let $Q_0^1 = \max \mathcal{Q}_{0}, Q_0^2 = \max \mathcal{Q}_{0}\setminus\{Q_0^1\}$, and $\Delta Q = Q_0^1 - Q_0^2 > 0.$
	By hypothesis,
	\begin{equation*}
		\,Q(C\vert \p{B}_1,\ldots,\p{B}_{\vert\mathcal{T}\vert};0) < Q(C_{\mathrm{max}}^{0}\vert \p{B}_1,\ldots,\p{B}_{\vert\mathcal{T}\vert};0)\,,
	\end{equation*}
	where $C_{\mathrm{max}}^{0}\in \mathcal{C}_{\mathrm{max}}(0)$. Furthermore, by definition of persistence, it follows that
	\begin{equation}
		\,Q(C\vert \p{B}_1,\ldots,\p{B}_{\vert\mathcal{T}\vert};\omega) \leq Q_0^2 + 2\omega N(\vert\mathcal{T}\vert-1)
		\label{EQ5.4}
	\end{equation}
	for all values of $\omega$. By choosing $\omega < \omega_0$, with $\omega_0 = \Delta Q/\big[2N(\vert\mathcal{T}\vert-1)\big]$, we obtain
	\begin{equation*}
		\,Q(C\vert \p{B}_1,\ldots,\p{B}_{\vert\mathcal{T}\vert};\omega) \leq Q_0^2 + 2\omega N(\vert\mathcal{T}\vert-1) < Q_0^2 + \Delta Q = Q_1^0\,,
	\end{equation*}
	so $C$ is not optimal for any inter-layer coupling below $\omega_0$.
	\end{proof}
	\vspace{0.25cm}
	
	Clearly, $\omega_0 = \Delta Q/\big[2N(\vert\mathcal{T}\vert-1)\big]$ is not an upper bound for the set $\big\{\omega \in\mathbb{R}^{+} : \bigcup_{s=1}^{\vert\mathcal{T}\vert}C_\mathrm{max}^{\omega}\vert_s \in \mathcal{C}_{\mathrm{max}}(0)\big\}$,\footnote{For example, one could replace $N(\vert\mathcal{T}\vert-1)$ in \eqref{P4} by $N(\vert\mathcal{T}\vert-1) - \text{Pers}(\mathcal{C}_{\mathrm{max}}(0))$, where $\text{Pers}(\mathcal{C}_{\mathrm{max}}(0))$ denotes the maximum value of persistence that one can obtain by combining sets in each partition of $\mathcal{C}_{\mathrm{max}}(0)$ without changing the partitions induced on individual layers. Proposition \ref{P4} still holds if one takes $\omega_0 = \Delta Q/\big[2\big(N(\vert\mathcal{T}\vert-1) - \text{Pers}(\mathcal{C}_{\mathrm{max}}(0))\big)\big]$, where $\Delta Q/\big[2\big(N(\vert\mathcal{T}\vert-1) - \text{Pers}(\mathcal{C}_{\mathrm{max}}(0))\big)\big] > \Delta Q/2N(\vert\mathcal{T}\vert-1)$, because $\text{Pers}(\mathcal{C}_{\mathrm{max}}(0)) \geq \vert\mathcal{T}\vert - 1$ in any multilayer network.} but our main concern is that the smallest upper bound of this set is \emph{not} $0$. (In fact, we have shown that it must be at least as large as $\Delta Q/\big[2N(\vert\mathcal{T}\vert-1)\big]>0$.)

	\vspace{0.25cm}
	\begin{proposition}\label{P5}
	There exists $\omega_{\infty} > 0$ such that
		\begin{equation*}
		\text{if } \omega > \omega_\infty\,, \text{ then } \mathrm{Pers}(C_{\mathrm{max}}^{\omega})\vert_s = N \text{ for all } s\in\{1,\ldots,\vert\mathcal{T}\vert\}\,.
		\end{equation*}
		Moreover, we show that $\omega_{\infty} = \vert\mathcal{T}\vert N^2\big[\mathrm{max}(\p{\mathcal{B}}_{\mathrm{diag}}) - \mathrm{min}(\p{\mathcal{B}}_{\mathrm{diag}})\big]/2$, where $\mathrm{max}(\p{\mathcal{B}}_{\mathrm{diag}}) = \max_{ijs}B_{ijs}$ and $\mathrm{min}(\p{\mathcal{B}}_{\mathrm{diag}}) = \min_{ijs}B_{ijs}$.
	\end{proposition}\\
	
	Proposition \ref{P5} implies that a sufficiently large value of inter-layer coupling $\omega$ guarantees that $C_{\mathrm{max}}^{\omega}\vert_s$ remains the same across layers (by Proposition \ref{P3}). The proof of this proposition is similar to the proof of Proposition \ref{P4}.
	\vspace{0.25cm}

	\begin{proof}\\
	Let $C$ be an arbitrary partition of a multilayer network such that $\text{Pers}(C)\vert_s < N$ for some $s\in\{1,\ldots,\vert\mathcal{T}\vert\}$. We show that there exists a value $\omega_\infty>0$ of the inter-layer coupling parameter $\omega$ such that $C$ is never optimal for $\omega > \omega_\infty$. We first rewrite the quality function as
	\begin{displaymath}
		Q(C\vert \p{B}_1,\ldots,\p{B}_{\vert\mathcal{T}\vert};\omega) = \beta_1 + 2\omega(N(\vert\mathcal{T}\vert-1) - A)\,,
	\end{displaymath}
	where $\beta_1 = \sum_{s=1}^{\vert\mathcal{T}\vert}\sum_{i,j = 1}^{N}B_{ijs}\delta(c_{i_s},c_{j_s})$ and $A \geq 1$ because $\text{Pers}(C) < N(\vert\mathcal{T}\vert-1)$ by assumption. Now consider the set of values on the diagonal blocks of the multilayer modularity matrix $\p{\mathcal{B}}$:
	\begin{equation*}
		\p{\mathcal{B}}_{\text{diag}} = \big\{B_{ijs}\vert i,j\in\{1,\dots,N\}, s\in\{1,\ldots,\vert\mathcal{T}\vert\}\big\}\,,
	\end{equation*}
	and let $\mathrm{max}(\p{\mathcal{B}}_{\text{diag}})$ and $\mathrm{min}(\p{\mathcal{B}}_{\text{diag}})$, respectively, denote the maximum and minimum values of the set $\p{\mathcal{B}}_{\text{diag}}$. Without loss of generality,\footnote{If $\mathrm{min}(\p{\mathcal{B}}_{\text{diag}})$ and $\text{max}(\p{\mathcal{B}}_{\text{diag}})$ have the same sign, then every diagonal block of $\p{\mathcal{B}}$ either has all positive entries or all negative entries. In both cases, an optimal partition $C_{\mathrm{max}}^{\omega}$ has maximal persistence for any value of $\omega > 0$, because $C_{\text{max}}^{\omega}\vert_s$ is given either by a single community or by $N$ singleton communities for all $s$. Consequently, by Proposition \ref{P3}, $\text{Pers}(C_{\mathrm{max}}^{\omega})\vert_s = N$ for all $s$.} we assume that $\mathrm{min}(\p{\mathcal{B}}_{\text{diag}}) < 0$ and $\text{max}(\p{\mathcal{B}}_{\text{diag}}) > 0$. Let $C^{\prime}$ be any multilayer partition with a maximal value of persistence.
	It follows that
	\begin{displaymath}
		Q(C^{\prime}\vert \p{B}_1,\ldots,\p{B}_{\vert\mathcal{T}\vert}; \omega) = \beta_2 + 2\omega N(\vert\mathcal{T}\vert-1)
	\end{displaymath} 
	for some $\beta_2\in\mathbb{R}$. Because $A\geq 1$, choosing 
	\begin{displaymath}
		2\omega > \vert\mathcal{T}\vert N^2\big[\mathrm{max}(\p{\mathcal{B}}_{\text{diag}}) - \mathrm{min}(\p{\mathcal{B}}_{\text{diag}})\big] \geq \beta_1 - \beta_2 
	\end{displaymath}
	ensures that $C^{\prime}$ yields a larger value of multilayer modularity than $C$ for any $\beta_1$ and for all $A\in\{1,\ldots, N(\vert\mathcal{T}\vert-1)\}$. 
	\end{proof}
	\vspace{0.25cm}

	The following proposition follows directly from Proposition \ref{P5}.

	\vspace{0.25cm}
	\begin{proposition}\label{P5b}
	There exists $\omega_{\infty} > 0$ such that
	\begin{displaymath}
	\text{For all } r\in\{1,\ldots,\vert\mathcal{T}\vert\},	C_{\mathrm{max}}^{\omega}\vert_r \text{ is a solution of } \max_{C\in\mathcal{C}}Q\left(C\vert \sum_{s=1}^{\vert\mathcal{T}\vert}\p{B}_s\right)
	\end{displaymath}
	for all $\omega > \omega_{\infty}$.
	\end{proposition}
	\vspace{0.25cm}
	
	Propositions \ref{P5} and \ref{P5b} imply the existence of a ``boundary value'' for $\omega$ above which single-layer partitions induced by optimal multilayer partitions (1) are the same on all layers and (2) are optimal solutions for the single-layer modularity-maximization problem defined on the mean modularity matrix. 
	\vspace{0.25cm}
	\begin{proof}\\
	Suppose that $\omega > \omega_\infty$, where $\omega_\infty$ is as defined in Proposition \ref{P5}, and let $C_{\mathrm{max}}^{\omega}\in\mathcal{C}_{\mathrm{max}}({\omega})$. By Proposition \ref{P5}, $\text{Pers}(C_{\mathrm{max}}^{\omega}) = N(\vert\mathcal{T}\vert-1)$, and community assignments in $C_{\mathrm{max}}^{\omega}$ are the same across layers. Consequently, for $\omega > \omega_{\infty}$
	\begin{align*}
		C^{*} &= \argmax_{C\in\mathcal{C}}\left(\sum_{s=1}^{\vert\mathcal{T}\vert}\sum_{i,j}^{N}B_{ijs}\delta(c_{i_s},c_{j_s}) + 2\omega \text{Pers}(C)\right)\,,\\
	\Leftrightarrow C^{*} &= \argmax_{C\in\mathcal{C}}\left(\sum_{s=1}^{\vert\mathcal{T}\vert}\sum_{i,j=1}^{N}B_{ijs}\delta(c_{i},c_{j}) + 2\omega N(\vert\mathcal{T}\vert-1)\right)\,,\\
	\Leftrightarrow C^{*} &= \argmax_{C\in\mathcal{C}}\left(\sum_{i,j}^{N}\Bigg(\sum_{s=1}^{\vert\mathcal{T}\vert}B_{ijs}\Bigg)\delta(c_{i},c_{j})\right)\,,
	\end{align*}
	where $c_i$ denotes the community assignment of node $i$ in all layers.
	\end{proof}
	\vspace{0.25cm}

	The next two propositions formalize the intuition that an optimal multilayer partition measures a trade-off between static community structure within layers (i.e., intra-layer modularity) and persistence of community structure across layers. 
	
	\vspace{0.25cm}
	\begin{proposition}\label{P6}
	Let $\omega_1 > \omega_2 > 0$. For all $C_{\mathrm{max}}^{\omega_2} \in\mathcal{C}_{\mathrm{max}}(\omega_2)$, one of the following two conditions must hold:
	\begin{align*}
		&(1)\quad C_{\mathrm{max}}^{\omega_2}\in\mathcal{C}_{\mathrm{max}}(\omega_1)\,,\\
	\text{or } &(2)\quad\mathrm{Pers}(C_{\mathrm{max}}^{\omega_2}) < \mathrm{Pers}(C_{\mathrm{max}}^{\omega_1}) \text{ for all } C_{\mathrm{max}}^{\omega_1}\in\mathcal{C}_{\mathrm{max}}(\omega_1)\,.
	\end{align*}
	\end{proposition}
	\vspace{0.25cm}
	\begin{proof}\\
	Let $C_{\mathrm{max}}^{\omega_2}\in\mathcal{C}_{\mathrm{max}}(\omega_2)$. If $C_{\mathrm{max}}^{\omega_2}\in\mathcal{C}_{\mathrm{max}}(\omega_1)$, then condition (1) is satisfied. Suppose that $C_{\mathrm{max}}^{\omega_2}\notin\mathcal{C}_{\mathrm{max}}(\omega_1)$, and assume that $\text{Pers}(C_{\mathrm{max}}^{\omega_2}) \geq \text{Pers}(C_{\mathrm{max}}^{\omega_1})$ for some $C_{\mathrm{max}}^{\omega_1}\in\mathcal{C}_{\mathrm{max}}(\omega_1)$. By definition of optimality, $C_{\mathrm{max}}^{\omega_2}\notin\mathcal{C}_{\mathrm{max}}(\omega_1)$ implies that 
	\begin{equation}\label{above}
		Q(C_{\mathrm{max}}^{\omega_2}\vert\p{B}_1,\ldots,\p{B}_{\vert\mathcal{T}\vert}; \omega_1) < Q(C_{\mathrm{max}}^{\omega_1}\vert\p{B}_1,\ldots,\p{B}_{\vert\mathcal{T}\vert}; \omega_1)\,,
	\end{equation}
	where $\omega_1 > \omega_2$ by hypothesis. By writing 
	\begin{displaymath}
	Q(C_{\mathrm{max}}^{\omega_k}\vert\p{B}_1,\ldots,\p{B}_{\vert\mathcal{T}\vert}; \omega_{k^{\prime}}) = \sum_{s=1}^{\vert\mathcal{T}\vert}\sum_{i,j=1}^{N} B_{ijs}\delta(c_{i_s}^{\omega_k},c_{j_s}^{\omega_k}) + 2\omega_{k^{\prime}}\text{Pers}(C_{\mathrm{max}}^{\omega_k})\,,
	\end{displaymath}
	where $c_{i_s}^{\omega_k}$ is the community assignment of node $i_s$ in $C_{\mathrm{max}}^{\omega_k}$ and $k,k^{\prime}\in\{1,2\}$; and by substituting $\omega_1$ by $\omega_2 + \Delta$ for some $\Delta > 0$, one can show that the inequality \eqref{above} implies 
	\begin{displaymath}
	Q(C_{\mathrm{max}}^{\omega_2}\vert\p{B}_1,\ldots,\p{B}_{\vert\mathcal{T}\vert}; \omega_2) < Q(C_{\mathrm{max}}^{\omega_1}\vert\p{B}_1,\ldots,\p{B}_{\vert\mathcal{T}\vert}; \omega_2)\,,
	\end{displaymath}
	 which contradicts the optimality of $C_\mathrm{max}^{\omega_2}$.
	\end{proof}
	\vspace{0.25cm}
	
	One can similarly prove the following proposition. 
	\vspace{0.25cm}	
	\begin{proposition}\label{P6b}
	Let $\omega_1 > \omega_2 > 0$. For all $C_{\mathrm{max}}^{\omega_2} \in\mathcal{C}_{\mathrm{max}}(\omega_2)$, one of the following two conditions must hold:
	\begin{align*}
		&(1)\quad C_{\mathrm{max}}^{\omega_2}\in\mathcal{C}_{\mathrm{max}}(\omega_1)\,,\\
	\text{or } &(2)\quad Q(C_{\mathrm{max}}^{\omega_2}\vert \p{B}_1,\ldots,\p{B}_{\vert\mathcal{T}\vert}; 0) > Q(C_{\mathrm{max}}^{\omega_1}\vert \p{B}_1,\ldots,\p{B}_{\vert\mathcal{T}\vert}; 0) \text{ for all } C_{\mathrm{max}}^{\omega_1}\in\mathcal{C}_{\mathrm{max}}(\omega_1)\,.
	\end{align*}
	\end{proposition}
		\vspace{0.25cm}
	\begin{proof}\\
	Let $C_{\text{max}}^{\omega_2}\in\mathcal{C}_{\text{max}}(\omega_2)$. If $C_{\text{max}}^{\omega_2}\in\mathcal{C}_{\text{max}}(\omega_1)$, then condition (1) is satisfied. Suppose that $C_{\text{max}}^{\omega_2}\notin\mathcal{C}_{\text{max}}(\omega_1)$, and assume that 
\begin{equation}
	\quad Q(C_{\text{max}}^{\omega_2}\vert \p{B}_1,\ldots,\p{B}_{\vert\mathcal{T}\vert}; 0) \leq Q(C_{\text{max}}^{\omega_1}\vert \p{B}_1,\ldots,\p{B}_{\vert\mathcal{T}\vert}; 0)
	\label{assumption}
\end{equation}
for some $C_{\text{max}}^{\omega_1}\in\mathcal{C}_{\text{max}}(\omega_1)$. By definition of optimality, $C_{\text{max}}^{\omega_2}\notin\mathcal{C}_{\text{max}}(\omega_1)$ implies that 
	\begin{equation}
		Q(C_{\text{max}}^{\omega_2}\vert\p{B}_1,\ldots,\p{B}_{\vert\mathcal{T}\vert}; \omega_1) < Q(C_{\text{max}}^{\omega_1}\vert\p{B}_1,\ldots,\p{B}_{\vert\mathcal{T}\vert}; \omega_1)\,,
		\label{opt2}
	\end{equation}
	where $\omega_1 > \omega_2$ by hypothesis. By writing 
	\begin{displaymath}
	Q(C_{\text{max}}^{\omega_k}\vert\p{B}_1,\ldots,\p{B}_{\vert\mathcal{T}\vert}; \omega_1) = Q(C_{\text{max}}^{\omega_k}\vert\p{B}_1,\ldots,\p{B}_{\vert\mathcal{T}\vert}; 0) + 2\omega_1\text{Pers}(C_{\text{max}}^{\omega_k}),
	\end{displaymath}
	where $k\in\{1,2\}$ and by using \eqref{assumption}, one can show that \eqref{opt2} implies that
	\begin{displaymath}
	Q(C_{\text{max}}^{\omega_2}\vert\p{B}_1,\ldots,\p{B}_{\vert\mathcal{T}\vert}; \omega_2) < Q(C_{\text{max}}^{\omega_1}\vert\p{B}_1,\ldots,\p{B}_{\vert\mathcal{T}\vert}; \omega_2).
	\end{displaymath}
	for all $\omega_2 < \omega_1$. This contradicts the optimality of $C_{\text{max}}^{\omega_2}$.
	\end{proof}
	\vspace{0.25cm}
	
	To develop intuition for Propositions \ref{P6} and \ref{P6b}, it is helpful to think of a multilayer quality function $Q(C\vert\p{B}_1,\ldots,\p{B}_{\vert\mathcal{T}\vert}; \omega)$ for a given partition $C$ as a linear function of $\omega$ that crosses the vertical axis at $Q(C\vert\p{B}_1,\ldots,\p{B}_{\vert\mathcal{T}\vert}; 0)$ with a slope $\text{Pers}(C)$ (see, for example, the last panels of Figs.~\ref{FIG5.1} and \ref{FIG5.2}).

	 The next three corollaries follow straightforwardly from Propositions \ref{P6} and \ref{P6b}. The first states that the largest achievable value of persistence for an optimal partition obtained with a given value of inter-layer coupling is a non-decreasing function in $\omega$. The second states that the largest achievable value of intra-layer modularity for an optimal partition obtained with a given value of inter-layer coupling is a non-increasing function in $\omega$. The third property states that if two distinct values of $\omega$ have the same set of optimal partitions, then this set is also optimal for all intermediate values.

	\vspace{0.25cm}
	\begin{corollary}\label{P7}
	Let $\omega_1 > \omega_2 > 0$. It follows that 
	\begin{equation*}
		\mathrm{Pers}(\mathcal{C}_{\mathrm{max}}(\omega_1)) \geq \mathrm{Pers}(\mathcal{C}_{\mathrm{max}}(\omega_2))\,,
	\end{equation*}
	where $\mathrm{Pers}(\mathcal{C}_{\mathrm{max}}(\omega)) := \max \big\{\mathrm{Pers}(C_{\mathrm{max}}^{\omega}), C_{\mathrm{max}}^{\omega}\in\mathcal{C}_{\mathrm{max}}(\omega)\big\}$.
	\end{corollary}
	
	\vspace{0.25cm}
	\begin{corollary}\label{P8}
	Let $\omega_1 > \omega_2 > 0$. It follows that
	\begin{equation*}
		Q\left(\mathcal{C}_{\mathrm{max}}(\omega_1)\right)\vert \p{B_1},\ldots,\p{B}_{\vert\mathcal{T}\vert};0) \leq Q\left(\mathcal{C}_{\mathrm{max}}(\omega_2)\right)\vert \p{B_1},\ldots,\p{B}_{\vert\mathcal{T}\vert};0)\,,
	\end{equation*}
	where $Q(\mathcal{C}_{\mathrm{max}}(\omega)\vert \p{B_1},\ldots,\p{B}_{\vert\mathcal{T}\vert};0) := \max \big\{Q(C_{\mathrm{max}}^{\omega}\vert \p{B_1},\ldots,\p{B}_{\vert\mathcal{T}\vert};0), C_{\mathrm{max}}^{\omega}\in\mathcal{C}_{\mathrm{max}}(\omega)\big\}.$
	\end{corollary}
	
	\vspace{0.25cm}
	\begin{corollary}\label{P9}
	Assume that $\mathcal{C}_{\mathrm{max}}(\omega_1) = \mathcal{C}_{\mathrm{max}}(\omega_2)$ for $\omega_1 > \omega_2 > 0$. It follows that
	\begin{equation*}
		\mathcal{C}_{\mathrm{max}}(\omega_1) = \mathcal{C}_{\mathrm{max}}(\omega) = \mathcal{C}_{\mathrm{max}}(\omega_2) \text{ for all } \omega \in (\omega_2,\omega_1)\,.
	\end{equation*}
	\end{corollary}
	\vspace{0.25cm}

	One can extend the proofs of Propositions \ref{P1}--\ref{P6b} so that they apply for inter-layer coupling that is uniform between each pair of contiguous layers but may differ from pair to pair. In other words, one can obtain similar results for the maximization problem 
\begin{equation*}
		\max_{C\in\mathcal{C}}\left[\sum_{s=1}^{\vert\mathcal{T}\vert}\sum_{i,j = 1}^{N}B_{ijs}\delta(c_{i_s},c_{j_s}) + 2\sum_{s=1}^{\vert\mathcal{T}\vert-1}\omega_s\text{Pers}(C)\vert_s\,\right].
\end{equation*}	
Propositions \ref{P1}--\ref{P5b} trivially extend to this case, and one can extend Propositions  \ref{P6}--\ref{P6b} if (for example) one assumes that $\omega^{(1)}_s >\omega^{(2)}_s > 0$ for all $s\in\{1,\ldots,\vert\mathcal{T}\vert-1\}$, where $\p{\omega}^{(1)}$ and $\p{\omega}^{(2)}$ are  $(\vert\mathcal{T}\vert-1)$-dimensional vectors. (For example, one could set $\p{\omega}^{(2)} = \omega\p{\omega}^{(1)}$ and vary $\omega > 0$.)

			
	\subsection{Computational issues}\label{Subsection5.3}
	
	We now examine issues that can arise when using the Louvain heuristic (see Section \ref{Subsection2.2}) to maximize multilayer modularity \eqref{EQ3.3}.
	
		
	\subsubsection{Underemphasis of persistence}

	\begin{figure}[t]
			\centering
			\subfigure[center][Layer 1]{
		  \includegraphics[width = 2.5cm]{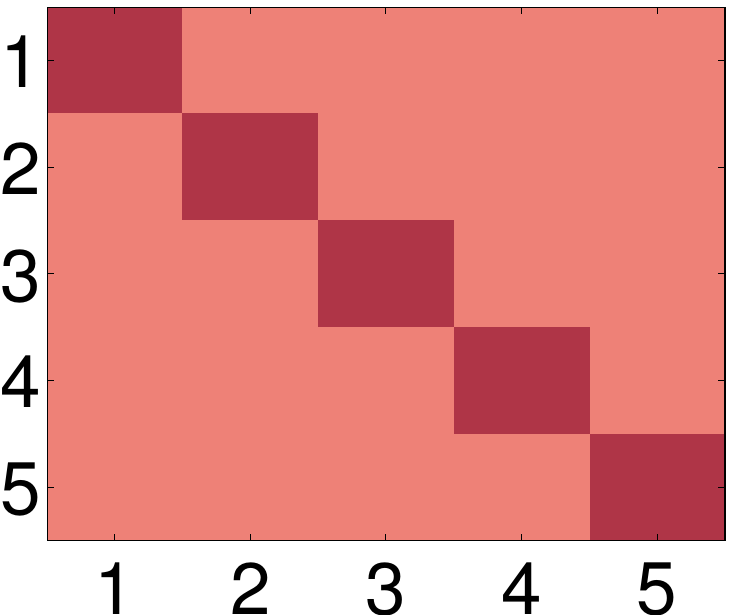}}	
		  \subfigure[Layer 2]{
		  \includegraphics[width = 2.5cm]{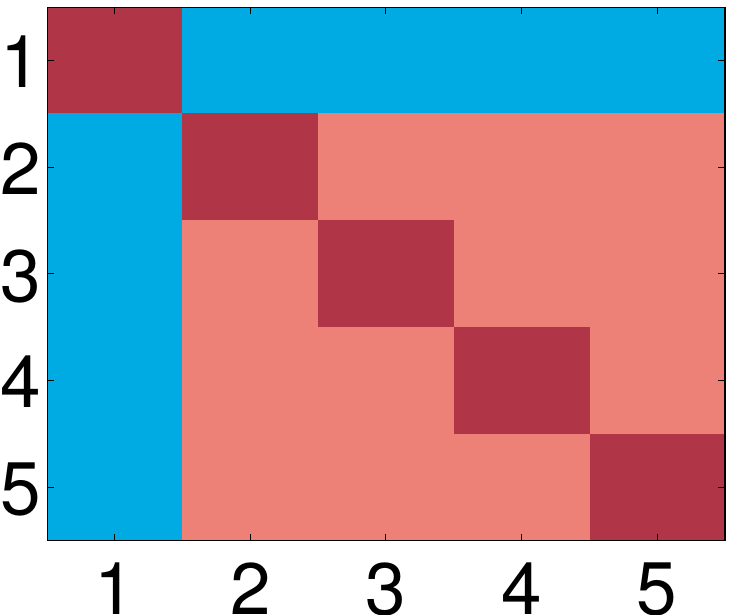}}
		  \subfigure[Layer 3]{
		  \includegraphics[width = 2.5cm]{figs2/FIG53a.pdf}}
		  
		  \includegraphics[width = 5cm]{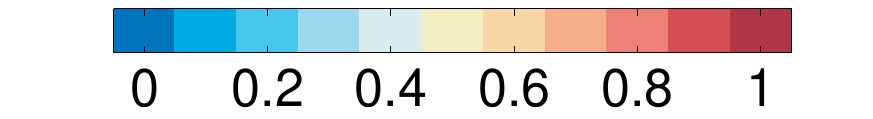}
		  
		  \subfigure[Multilayer partition before post-processing]{
		  \includegraphics[width = 3.2cm]{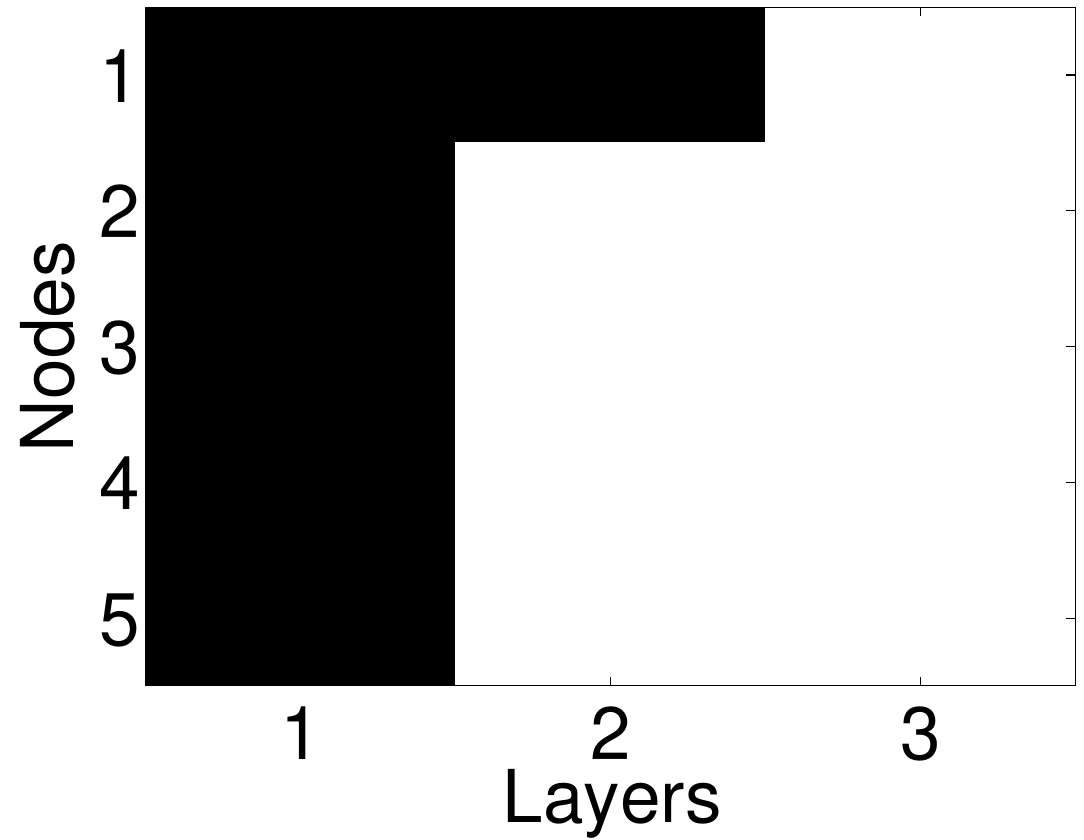}}\hspace{2em}
		  \subfigure[Multilayer partition after post-processing]{
		  \includegraphics[width = 3.2cm]{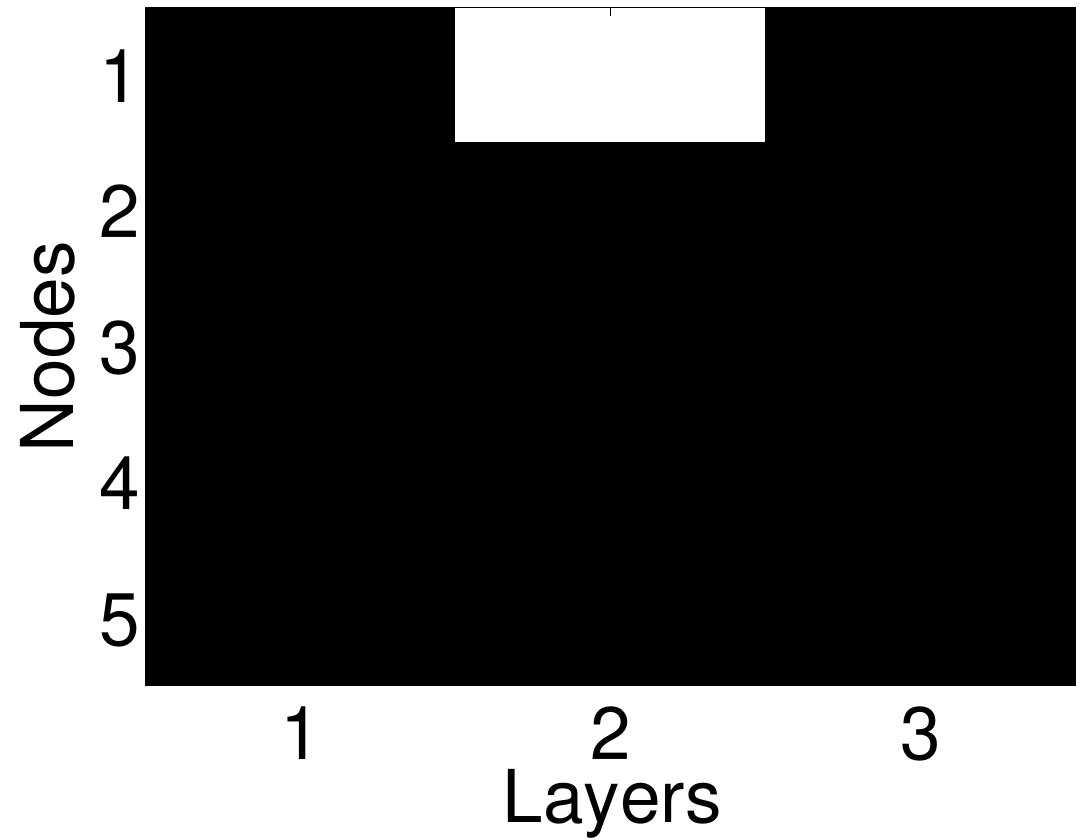}}\hspace{-2em}
		  
		  \caption{Toy example illustrating the effect of post-processing on a multilayer partition by increasing multilayer modularity via community-assignment swaps that increase the value of persistence but do not change intra-layer partitions. The colors in panels (a)--(c) scale with the entries of the adjacency matrix. Panel (d) (respectively, panel (e)) represents the output multilayer partition obtained with Louvain before (respectively, after) post-processing. The horizontal axis represents the layers, and the vertical axis represents the nodes. The shading in panels (d,e) represents the community assignments of nodes in each layer.} 
	  \label{FIG5.3}
	\end{figure}

	Consider the example network in Fig.~\ref{FIG5.3}, which is a 3-layer network that has 5 nodes in each layer. Suppose that all nodes are strongly connected to each other in layers 1 and 3, and that the edge weight between node $1_2$ and nodes $\{2_2,3_2,4_2,5_2\}$ is smaller in layer $2$ than the edge weight between node $1_s$ and nodes $\{2_s,3_s,4_s,5_s\}$ when $s=1,3$. We use the uniform null network with $\gamma = 0.5$ and set $\omega = 0.1$. This produces a multilayer modularity matrix in which all the single-layer modularity entries $B_{ijs}$ except those of node $1_2$ are positive and exceed the value of inter-layer coupling. Suppose that one loops over the nodes ordered from $1$ to $N\vert\mathcal{T}\vert$ in phase 1 of the Louvain heuristic. The initial partition consists of $N\vert\mathcal{T}\vert$ singletons, and each node is then moved to the set that maximally increases modularity. The partition at the end of phase 1 is $\{\{1_1,2_1,3_1,4_1,5_1,1_2\}, \{2_2,3_2,4_2,5_2\}, \{1_3,2_3,3_3,4_3,5_3\}\}$. In phase 2, the second and third sets merge to form a single set,\footnote{Note that combining the first and second set into a single set decreases modularity because the value of inter-layer coupling is too small to compensate for the decrease in intra-layer contributions to modularity.} and the Louvain heuristic gets trapped in a local optimum in which the smaller set of nodes (i.e., $\{1_1\}$) remains in the same community across layers 1 and 2 and the larger set of nodes (i.e., $\{2_1,3_1,4_1,5_1\}$) changes community. We show this multilayer partition in Fig.~\ref{FIG5.3}(d). Repeating this experiment $1000$ times using a randomized node order at the start of each iteration of phase 1  of the Louvain heuristic yields the same multilayer partition. 
	One can modify this multilayer partition to obtain a new partition with a larger value of multilayer modularity by increasing the value of persistence across layers without changing intra-layer partitions (we use this idea in the proof of Proposition \ref{P1}). We show an example of this situation in Fig.~\ref{FIG5.3}(e). 
		
	In Fig.~\ref{FIG5.3}(d), we illustrate the above issue visually via abrupt changes in colors between layers. (These are more noticeable in larger networks.) Such changes are misleading because they imply a strong decrease in persistence that might not be accompanied by a significant change in intra-layer partitions.  In Fig.~\ref{FIG5.3}(d), for example, the intra-layer partitions differ in the community assignment of only a single node. To mitigate this problem, we apply a post-processing function to all output partitions that maximizes persistence between layers without changing the partitions that are induced on each layer. We thereby produce a partition with a larger value of multilayer modularity. In our post-processing, we relabel the community assignments of nodes in each layer such that (1) the number of nodes that remain in the same community between consecutive layers is maximized and (2) the partition induced on each layer by the original multilayer partition is unchanged. By reformulating the problem of maximizing persistence for a given set of intra-layer partitions as a weighted bipartite matching problem between consecutive layers, one can implement our post-processing procedure using the \emph{Hungarian Algorithm} \cite{Edmonds1972, Munkres1957, Kuhn1955}.

	
	
	\subsubsection{Abrupt drop in the number of intra-layer merges}
	
\begin{figure}
		\center
			\subfigure[Mean number of intra-layer merges after the first completion of phase 1 of Louvain]{\includegraphics[width = 3.5cm]{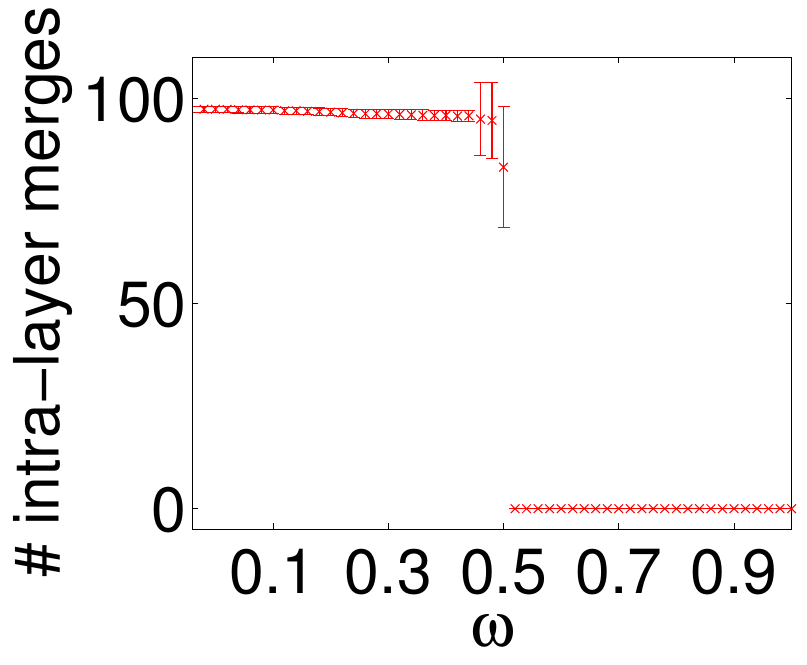}}\hspace{2cm}
			\subfigure[Mean number of intra-layer merges after the first completion of phase 1 of LouvainRand]{\includegraphics[width = 3.5cm]{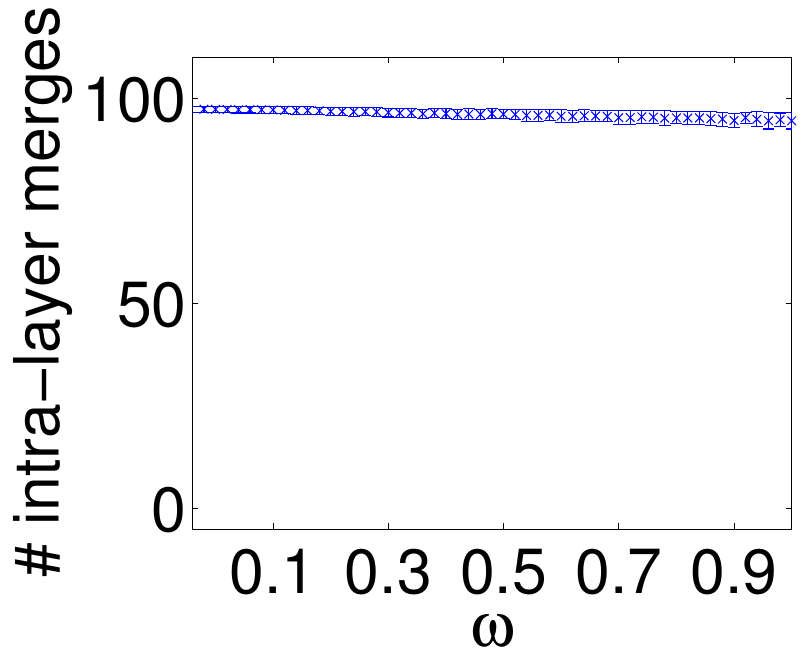}}
			
			\center
			\vspace{-0.35cm}\subfigure[Mean number of nodes that change community assignment between layers in output partition of Louvain]{\includegraphics[width = 6cm]{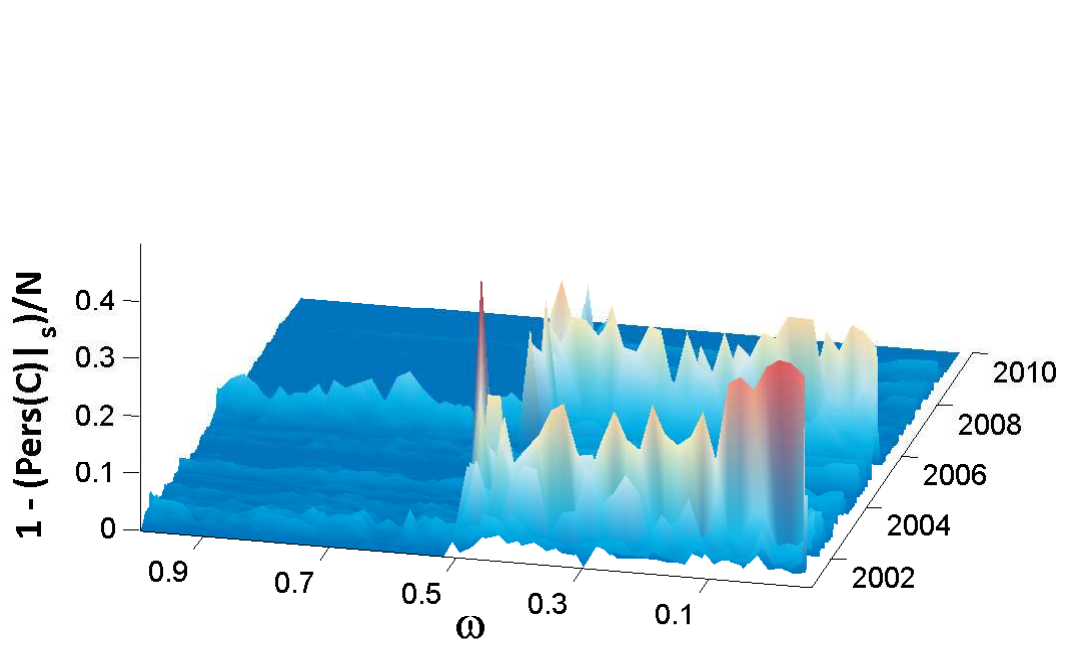}}\hspace{2em}
			\subfigure[Mean number of nodes that change community assignment between layers in output partition of LouvainRand]{\includegraphics[width = 6cm]{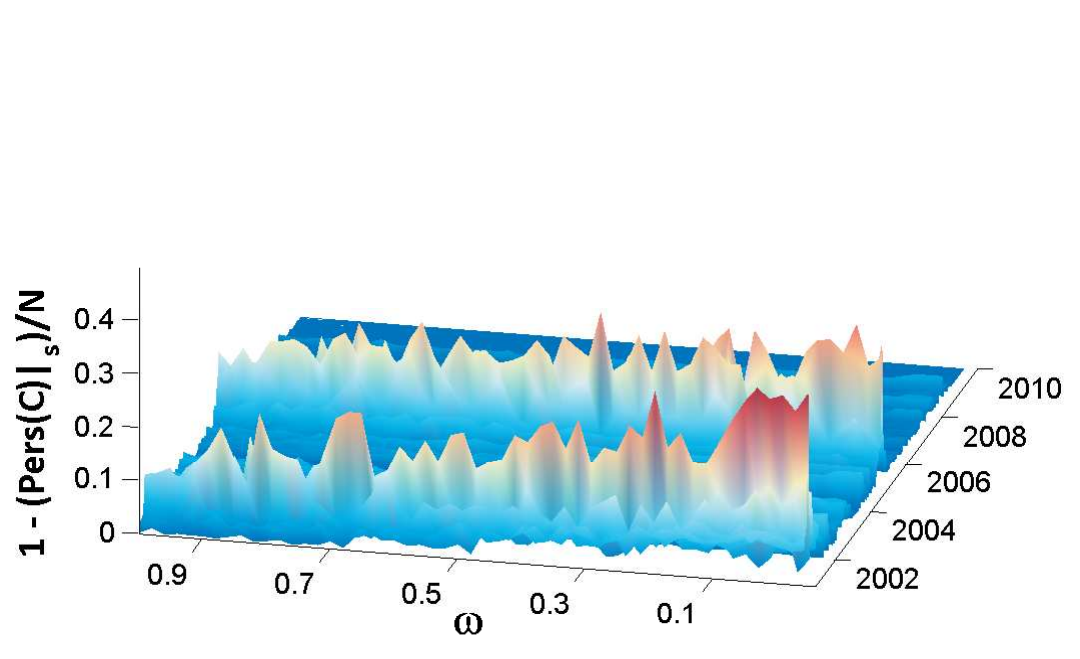}}
					
			\caption{Comparison between the Louvain and LouvainRand algorithms. The sample of inter-layer coupling values is the set $\{0,0.02,\dots,0.98,1\}$ with a discretization step of $0.02$ between each pair of consecutive values. (a,b) The number of nodes $n_{\text{intra}}$ that have been merged with at least one node from the same layer after the first completion of phase 1 of (a) the Louvain heuristic and (b) the LouvainRand heuristic. For each heuristic, we average $n_{\text{intra}}$ over $\vert\mathcal{T}\vert = 238$ layers and $100$ iterations. The error bars in panels (a,b) indicate standard deviations. (c,d) The value of $1 - \text{Pers}(C)\vert_s/N$ averaged over $100$ runs of (c) the Louvain heuristic and (d) the LouvainRand heuristic after convergence of the algorithms to a local optimum.	
			}
			\label{FIG5.4}
		\end{figure}

	The Louvain heuristic faces a second problem in multilayer networks. When the value of inter-layer coupling satisfies
	\begin{equation}
		\omega > \mathrm{max}(\p{\mathcal{B}}_{\text{diag}})\,,
		\label{EQ5.5}
	\end{equation}
	where $\p{\mathcal{B}}_{\text{diag}}$ is the set of values on the diagonal blocks of $\p{\mathcal{B}}$ defined in equation \eqref{EQ5.4}, the inter-layer contributions to multilayer modularity are larger than the intra-layer contributions for all pairs of nodes. Consequently, only inter-layer merges occur during the first completion of phase 1 of the Louvain heuristic. In Fig.\ref{FIG5.4}(a), we illustrate this phenomenon using the {\sc MultiAssetClasses} data set. The mean number of intra-layer merges drops from roughly $N=98$ (almost every node contains at least one other node from the same layer in its community) to $0$. For $\omega$ values larger than $\mathrm{max}(\p{\mathcal{B}}_{\text{diag}})$, every set at the end of the first completion of phase 1 only contains copies of each node in different layers and, in particular, does not contain nodes from the same layer. This can yield abrupt changes in the partitions induced on individual layers of the output multilayer partition.

	In Fig.~\ref{FIG5.4}(c), we show an example using the {\sc MultiAssetClasses} data set of how the above issue can lead to an abrupt change in a quantitative measure computed from a multilayer output partition obtained with the Louvain heuristic. We note that the mean size of sets (averaged over $100$ runs) after the first completion of phase 1 of the Louvain algorithm for $\omega > \mathrm{max}(\p{\mathcal{B}}_{\text{diag}})$ is relatively small for {\sc MultiAssetClasses}. (The mean is 3 nodes per set, and the maximum possible number of nodes per set is $\vert\mathcal{T}\vert = 238$, because each of these sets only contains copies of the same node when $\omega > \mathrm{max}(\p{\mathcal{B}}_{\text{diag}})$.) Nevertheless, as $\omega$ increases past $\omega = \mathrm{max}(\p{\mathcal{B}}_{\text{diag}})$ there is a sudden drop in the value of $(1 - \text{Pers}(C)\vert_s/N)$ between consecutive layers in the output partition at $\omega = \mathrm{max}(\p{\mathcal{B}}_{\text{diag}})$ [see Fig.~\ref{FIG5.4}(c)]. Nonzero values of $(1 - \text{Pers}(C)\vert_s/N)$ indicate that community assignments have changed between layers $s$ and $s+1$ (by Proposition \ref{P3}). Roughly speaking, Fig.~\ref{FIG5.4}(c) suggests that when $\omega > \mathrm{max}(\p{\mathcal{B}}_{\text{diag}})$, one abruptly moves away from the situation $\omega \approx \omega_0$ to a scenario that is closer to $\omega \approx \omega_{\infty}$.\footnote{For another instance in which an abrupt transition occurs as one increases $\omega$, see the discussion in Ref.~\cite{Radicchi2013} on the change in eigenvalues of a multilayer Laplacian as a function of $\omega$.}
		
	The above phenomenon manifests when the values of inter-layer coupling are large relative to the entries of $\p{\mathcal{B}}_{\text{diag}}$. In the correlation multilayer networks that we consider (or in unweighted multilayer networks), entries of the adjacency matrix satisfy $\vert A_{ijs}\vert \leq 1$. Assuming that one uses the modularity quality function on each layer and that $P_{ijs} \geq 0$ (e.g., $P_{ijs} = \langle \p{A}_s\rangle$), this implies that
	\begin{equation*}
		\mathrm{max}(\p{\mathcal{B}}_{\text{diag}}) \leq 1 \text{ for all } \gamma\in[\gamma^-,\gamma^+]\,.
	\end{equation*}
	For networks in which the modularity cost of changing intra-layer partitions in favor of persistence is large in comparison to the values of $\mathcal{B}_{\text{diag}}$, it may be desirable to use $\omega > 1$  to gain insight into multilayer community structure. (For example, this occurs in both toy examples of Section \ref{Subsection5.1}.)

	To mitigate this problem, we change the condition for merging nodes in the Louvain heuristic. Instead of moving a node to a community that maximally increases modularity, we move a node to a community  chosen uniformly at random from those that increase modularity. We call this heuristic \emph{LouvainRand} \cite{NetWiki}, and we illustrate the results of using it in Figs.~\ref{FIG5.4}(b,d). Although LouvainRand can increase the output variability (by increasing the search space of the optimization process), it seems to mitigate the problem for multilayer networks with ordinal diagonal and uniform coupling.


		
	\subsection{Multilayer community structure in asset correlation networks}\label{Subsection5.4}
	
	\begin{figure}[t!]
		\center
		\subfigure[Persistence versus $\omega$ for $\gamma =1$]{\includegraphics[width = 4cm]{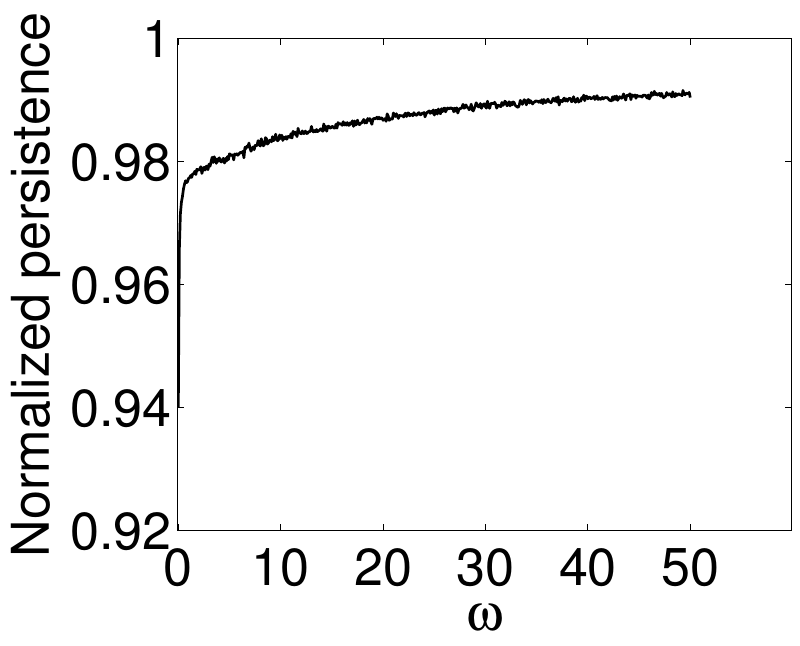}}\hspace{5em}
		\subfigure[Intra-layer contributions versus $\omega$ for $\gamma =1$]{\includegraphics[width = 4cm]{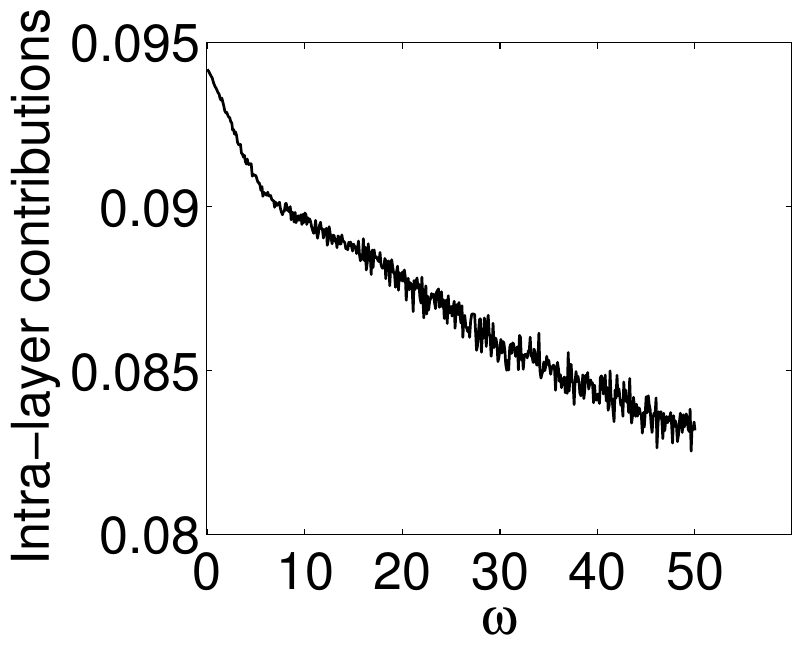}}		
		\subfigure[Output partition for $(\gamma,\omega) = (1,0.1)$]{\includegraphics[width = 4cm]{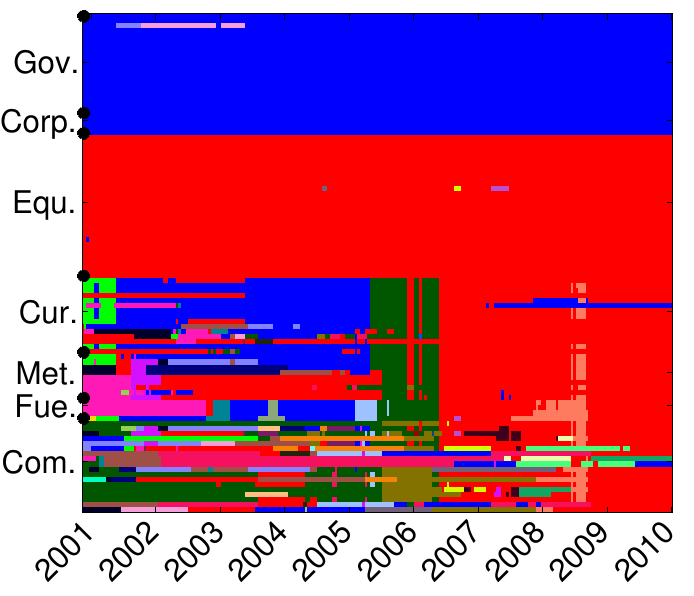}}\hspace{1em}
		\subfigure[Persistence between all pairs of layers]{\includegraphics[width = 4cm]{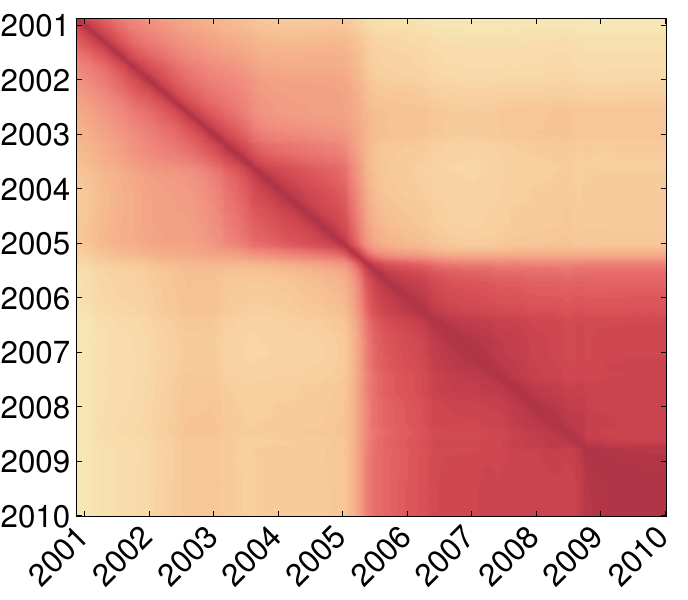}}\hspace{1em}
		\subfigure[Mean co-classification index between all pairs of nodes over all layers]{\includegraphics[width = 3.9cm]{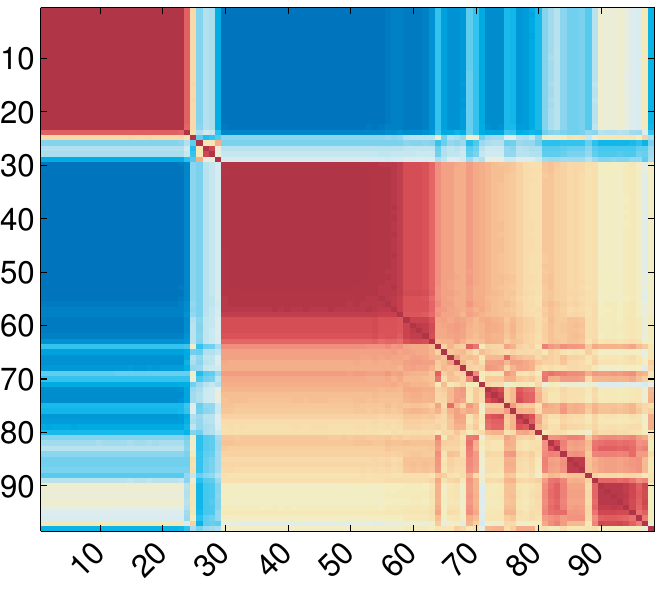}}
			
		\hspace{4.5cm}\includegraphics[width = 7cm]{figs2/colorbar.pdf}
		
		\caption{Numerical experiments with {\sc MultiAssetClasses}. We sample the set of inter-layer edge weights uniformly from the interval $[0, 50]$ with a discretization step of $0.1$ (so there are 501 values of $\omega$ in total), and we use the uniform null network (i.e., $P_{ijs} = \langle \p{A}_s\rangle$) with $\gamma = 1$. (a) The persistence normalized by $N(\vert\mathcal{T}\vert-1)$ for each value of $\omega$ averaged over 20 runs of LouvainRand. (b) The intra-layer modularity $\sum_{s=1}^{\vert\mathcal{T}\vert}\sum_{i,j=1}^{N}B_{ijs}\delta(c_{i_s},c_{j_s})$ normalized by $\sum_{s=1}^{\vert\mathcal{T}\vert}\sum_{i,j=1}^{N} A_{ijs}$ for each value of $\omega$ averaged over 20 runs of LouvainRand. (c) Sample output multilayer partition. Each point on the horizontal axis represents a single time window, and each position on the vertical axis is an asset. We order the assets by asset class, and the colors represent communities. (d Association matrix of normalized persistence values between all pairs of layers averaged over all values of $\omega\in[0, 50]$ in our sample and $20$ runs for each value. The normalized persistence between a pair of layers $\{s,r\}$ is $\sum_{i=1}^{N}\delta(c_{is},c_{ir})/N$. (e) Association matrix indicating the co-classification of nodes averaged over the set of partitions induced on each layer for each value of $\omega$ and 20 runs of LouvainRand.
		}
		\label{FIG5.5}
	\end{figure}

	In this section, we show the results of computational experiments in which we fix the value of the resolution parameter $\gamma$ and vary the value of inter-layer coupling $\omega$. We use the uniform null network (i.e., $P_{ijs} = \langle \p{A}\rangle_s$) and set $\gamma = 1$. We use the LouvainRand heuristic to identify multilayer partitions and apply our post-processing procedure that increases persistence without changing partitions induced on individual layers to all output multilayer partitions. We showed in Proposition \ref{P5} that for $2\omega > 2\omega_{\infty} = \vert\mathcal{T}\vert N^2\big[\mathrm{max}(\p{\mathcal{B}}_\text{diag}) - \mathrm{min}(\p{\mathcal{B}}_\text{diag})\big]$, the set $\mathcal{C}_{\mathrm{max}}(\omega)$ of global optima no longer changes and every optimal partition in this set has maximal persistence.\footnote{Note that there can also be smaller values of $\omega_{\infty}$ for which this is true; in other words, we did not show that $\omega_\infty$ is the smallest lower bound of the set $\{w: \text{Pers}(C_{\mathrm{max}}^{\omega}) = N(\vert\mathcal{T}\vert-1) \text{ for all } C_{\mathrm{max}}^{\omega}\in\mathcal{C}_{\mathrm{max}}(\omega)\}$.} In our example, $\vert\mathcal{T}\vert N^2\big[\mathrm{max}(\p{\mathcal{B}}_\text{diag}) - \mathrm{min}(\p{\mathcal{B}}_\text{diag})\big]\leq 2\vert\mathcal{T}\vert N^2$, with $N=98$ and $\vert\mathcal{T}\vert = 238$.  However, for the purposes of the present paper, we take the set $\{0,0.1,\ldots,49.9,50\}$ with a discretization step of $0.1$ between consecutive values (giving $501$ values in total) as our sample of $\omega$ values.

	In agreement with the properties derived in Propositions \ref{P6} and \ref{P6b}, we observe in Fig.~\ref{FIG5.5}(a) that normalized persistence (given by $\text{Pers}(C)/[N(\vert\mathcal{T}\vert-1)]$) tends to be larger for larger values of inter-layer coupling, and in Fig.~\ref{FIG5.5}(b) that intra-layer modularity (which we normalize by $\sum_{s=1}^{\vert\mathcal{T}\vert}(\p{1}^{T}\p{A}_s\p{1})$) tend to be smaller for larger values of inter-layer coupling. The increase of persistence and the decrease of intra-layer modularity need not be monotonic, because we are a finding a set of local optima for each value of $\omega$ rather than the set of global optima.

	In Fig.~\ref{FIG5.5}(c), we show a sample output of the multilayer partition (which contains 35 communities). (See Section \ref{datasets} for our definitions of the asset-class abbreviations.) Some of the changes in community structure correspond to known events (e.g., Lehman Bankruptcy in September 2008 [marked by an increase in the size of the equity asset class]). Observe that the two largest communities are the ones that contain the government bond assets and the equity assets. In particular, the community that contains equities becomes noticeably larger between 2006 and 2007, and again towards the end of 2008 [after the pink streak between 2008 and 2009 in Fig.~\ref{FIG5.5}(d)]. For larger values of the resolution parameter $\gamma$, this community instead becomes noticeably larger only in 2008. (By inspecting the correlation matrices, one can check that the increase in correlation between equities and other assets is greater in 2008 than in 2006.)
	
	In Fig.~\ref{FIG5.5}(d), we show the matrix of mean values of persistence between all pairs of layers. The $(s,r)^{\text{th}}$ entry is the term $\sum_{i=1}^{N}\delta(c_{i_s},c_{i_{r}})/N$, where $s,r\in\{1,\ldots,\vert\mathcal{T}\vert\}$ need not be from consecutive layers, averaged over all values of $\omega\in[0,50]$ in our sample, and multiple runs for each value of $\omega$. Instead of only plotting $\text{Pers}(C)\vert_s$ for consecutive layers, Fig.~\ref{FIG5.5} gives some indication as to whether nodes change communities between layers $s$ and $s+1$ to join a community that contains a copy of some of these nodes from another time layer (i.e., $\sum_{i=1}^{N}\delta(c_{i_{s+1}},c_{i_{r}}) \neq 0$ for some $r$) or to join a community that does not contain a copy of these nodes in any other time layer (i.e., $\sum_{i=1}^{N}\delta(c_{i_{s+1}},c_{i_{r}}) = 0$ for all $r$). Figure \ref{FIG5.5} also gives some insight into whether there are sets of consecutive layers across which persistence values remain relatively large. This may shed light on when connectivity patterns change in a multilayer network. As indicated by the values on the color scale, the values of persistence in Fig.~\ref{FIG5.5}(d) remain relatively high (which can be explained in part by the fact that equities and bonds remain in the same community across almost all layers, and these constitute roughly 50$\%$ of the node sample). The most noticeable separation into diagonal blocks in the middle of Fig.~\ref{FIG5.5}(d) corresponds to the change in Fig.~\ref{FIG5.5}(c) between 2005 and 2006, at which various currencies, metals, and fuels separate from the bond community (blue) to form a dark green community. The smaller diagonal block at the bottom right of Fig.~\ref{FIG5.5}(d) corresponds to the change in Fig.~\ref{FIG5.5}(c) after the Lehman Bankruptcy between 2008 and 2009 [after the pink streak in Fig.~\ref{FIG5.5}(c)].

	In Fig.~\ref{FIG5.5}(e), we show the co-classification index of nodes in partitions induced on individual layers, which we average over layers, all values of $\omega\in[0,50]$ in our sample, and multiple runs for each value of $\omega$ (we reorder the nodes to emphasize diagonal blocks in the association matrix). This figure yields insight into what sets of nodes belong to the same community across layers for increasing values of $\omega$. This may shed light on connectivity patterns that are shared across layers. Unsurprisingly, the first diagonal block corresponds primarily to bond assets and the second diagonal block mainly corresponds to equity assets. Figures \ref{FIG5.5}(d,e) complement each other: at a given $\gamma$ resolution, the former gives an idea about \emph{when} community structure has changed, and the latter gives an idea about how it has changed.

		 

	\section{Conclusions}\label{Section6} 
		
	Modularity maximization in temporal multilayer networks is a clustering technique that produces a time-evolving partition of nodes. We have investigated two questions that arise when using this method: (1) the role of null networks in modularity maximization, and (2) the effect of inter-layer edges on the multilayer modularity-maximization problem. We demonstrated that one must be cautious in interpreting communities obtained with a null network in which the distribution of expected edge weights is sample dependent. Furthermore, we showed that an optimal partition in multilayer modularity maximization reflects a trade-off between static community structure within layers and persistence of community structure across layers. One can try to exploit this in practice to detect changes in connectivity patterns and shared connectivity in a time-dependent network. Our results and observations hold for any choice of single-layer networks within layers; we discussed correlation networks in detail to use them as illustrative examples. 
	
	At the heart of modularity maximization is a comparison between what one anticipates and what one observes. The ability to specify what is anticipated is a desirable (albeit under-exploited) feature of modularity maximization, because one can explicitly adapt it for different applications \cite{Expert2011,marta2014,Bassett2012,granular2014}. By defining a null model as a probability distribution over the space of adjacency matrices and a null network as the expected adjacency matrix under the specified distribution, we highlight the important point that the \emph{same} null network can correspond to \emph{different} null models; this is not something that has been appreciated properly in the literature. Moreover, one needs to be very careful with one's choice of null network because it determines what one regards as densely connected in a network: different choices in general yield different communities.  As we illustrated in Section \ref{Section4} for financial correlation networks, this choice can have a large impact on results, and can lead to misleading conclusions. 
	
	In Section \ref{Section5}, we proved several properties that describe the effect of ordinal diagonal and uniform inter-layer coupling on multilayer modularity maximization, or more generally, on any maximization problem that can be cast in the form \eqref{EQ3.4}. Although our theoretical results do not necessarily apply to the local optima that one attains in practice, they do provide useful guidelines for how to interpret the outcome of a computational heuristic for maximizing modularity: if a multilayer partition is inconsistent with one of the proven properties, then it must be an artifact of the heuristic and not a feature of the quality function.

	To further examine multilayer modularity maximization, we defined a measure that we called \emph{persistence} to quantify how much community assignments change in time in a multilayer partition. For zero inter-layer coupling, the value of persistence is $0$, and it achieves a maximum finite value for sufficiently large inter-layer coupling. We showed that the highest achievable value of persistence for an optimal partition obtained with a given value of inter-layer coupling $\omega$ is a non-decreasing function in $\omega$. Similarly, the highest achievable value of intra-layer contributions to the quality function for an optimal partition obtained with a given value of inter-layer coupling $\omega$ is a non-increasing function in $\omega$. The notion of persistence makes it possible to measure this trade-off between static community structure within layers and temporal persistence across layers. We illustrated this trade-off in our numerical experiments.

	Finally, we showed that the Louvain heuristic can pose two issues when applied to multilayer networks with ordinal diagonal and uniform coupling. These can produce misleading values of persistence (or other quantitative measures of a multilayer partition) and can cause one to draw false conclusions about temporal changes in community structure in a network. We proposed ways to mitigate these problems and showed several numerical experiments on real data as illustrations. To further interpret these results, one needs to investigate more closely how the increase in persistence and the decrease in intra-layer contributions to the quality function actually manifest in a multilayer partition  between the ``boundary cases'' $\omega_0$ and $\omega_{\infty}$ (see Propositions~\ref{P4} and ~\ref{P5}). This may help to identify an interval of $\omega$ values in which the trade-off between persistence and intra-layer modularity yields the most insights.


\section*{Acknowledgements}

MB acknowledges a CASE studentship award from the EPSRC (BK/10/41). MB and MAP were supported by
FET-Proactive project PLEXMATH (FP7-ICT-2011-8; grant \#317614) funded by the European 
Commission.  MAP and MB also received support from the James S. McDonnell Foundation (\#220020177), and MAP was also supported by the EPSRC (EP/J001759/1). We thank HSBC Bank for providing the time-series data of financial assets. We thank Alex Arenas, Mihai Cucuringu, Sergio G\'omez, Lucas Jeub, and three anonymous referees for helpful comments.


		


		\end{document}